\documentclass[journal,twocolumn]{IEEEtran}
\ifCLASSINFOpdf
  \usepackage[pdftex]{graphicx}
\else
  \usepackage[dvips]{graphicx}
\fi

\usepackage{epsfig}
\usepackage{color}
\usepackage{amssymb}
\usepackage{amstext}
\usepackage{amsmath}
\usepackage{amsthm}
\usepackage{multicol}
\usepackage{pslatex}
\usepackage{float}
\usepackage[small]{caption}
\usepackage{subcaption}
\usepackage[update,prepend]{epstopdf}
\usepackage{lmodern}
\usepackage{cite}
\usepackage{balance}
\usepackage{listings}
\usepackage{hyperref}
\def\postbreak{\raisebox{0ex}[0ex][0ex]{\ensuremath{\hookrightarrow\space}}}
\def\sign{\mbox{sign\,}}

\begin{document}

\title{Charge pump phase-locked loop with phase-frequency detector: closed form mathematical model}

\author{
    Kuznetsov~N.$^{\,a,b,c}$, Yuldashev~M.$^{\,a}$, Yuldashev~R.$^{\,a}$,\\
  Blagov~M.$^{\,a,b}$, Kudryashova~E.$^{\,a}$, Kuznetsova~O.$^{\,a}$, Mokaev~T$^{\,a}$.
\thanks{
($^a$) Faculty of Mathematics and Mechanics,
Saint-Petersburg State University, Russia;
($^b$) Dept. of Mathematical Information Technology,
University of Jyv\"{a}skyl\"{a}, Finland;
($^c$) Institute for Problems in Mechanical Engineering of Russian Academy of Science, Russia;
(corresponding author email: nkuznetsov239@gmail.com).
This paper is an extended version of \cite{kuznetsov2018comment}
}
}

\maketitle

\begin{abstract}                
  Charge pump phase-locked loop with phase-frequency detector (CP-PLL)
  is an electrical circuit,
  widely used in digital systems  for frequency synthesis
  and synchronization of the clock signals.
  In this paper a non-linear second-order model of CP-PLL is rigorously derived.
  The obtained model obviates the shortcomings of
  previously known second-order models of CP-PLL.
  Pull-in time is estimated for the obtained second-order CP-PLL.
\end{abstract}

\IEEEpeerreviewmaketitle

\section{Introduction}
\IEEEPARstart{P}{hase}-locked loops (PLLs) are electronic circuits, designed for generation of an electrical signal (voltage), while the frequency is automatically tuned to the frequency of an input (reference, Ref) signal.
Charge pump phase-locked loop with phase-frequency detector (Charge pump PLL, CP-PLL, CPLL)
is widely used in digital systems  for frequency synthesis
and synchronization of the clock signals \cite{Best-2007-book}.
The CP-PLL is able to quickly lock onto the phase of the incoming signal,
achieving low steady-state phase error \cite{Orla-2013-review}.
Important issues in the design of PLL are estimation of the ranges of deviation between oscillators frequencies for which a locked state can be achieved,
the stability analysis of the locked states, and
study of possible transient processes.
The pioneering monographs \cite{Gardner-1966,ShahgildyanL-1966,Viterbi-1966}
 were published in 1966,
a rather comprehensive bibliography \cite{LindseyT-1973} was published in 1973,
and recent surveys \cite{LeonovKYY-2015-TCAS,BestKLYY-2016}.
For the corresponding study of the CP-PLL
F.~Gardner developed in 1980 a linearized model
in vicinity of locked states
\cite{Gardner-1980,Gardner-2005-book}.
Then an approximate discrete-time linear models of the CP-PLL were suggested in \cite{Hein-1988,lu2001discrete}.
However, linear models are essentially limited,
since only local behavior near locked states can be studied.
For the study of non-local behavior and transient processes
some nonlinear second-order models of the CP-PLL were developed in \cite{Paemel-1994,Acco-2004,Orla-2013-review}.

Examples from section~\ref{sec:examples} demonstrate that algorithm and formulas suggested by M. van Paemel in \cite{Paemel-1994}
should be used carefully for simulation even inside allowed area (see Fig.~18 and Fig.~22 in original paper \cite{Paemel-1994}).
While the examples are given for the first time, the main idea of Example 1 was already noticed by P. Acco and O. Feely \cite{acco2003etude,Orla-2012}. 
P. Acco and O. Feely considered only near-locked state, therefore they didn't notice problems with out-of-lock behavior. Example 2 and Example 3 demonstrate problems with out-of-lock behavior, which was not discovered before.

Note that while derivation of non-linear mathematical models for high-order CP-PLL
requires numerical solution of non-linear algebraic equations
or allows to find only approximate solutions
(see, e.g. \cite{Hedayat-1999-3order,Johnson-2005,Wang-2005,Daniels-2008,Guermandi-2011,Brambilla-2012-cyllinder,Shakhtarin-2014,Hedayat-2014-high-order}),
non-linear mathematical models for the second order CP-PLL
can be found in closed-form.
Further, we consider only the second order CP-PLL.
Noise performance and simulation of PLL is discussed in \cite{Best-2007-book,Razavi-1996,meszaros2012does,Rosenkranz-1982,Shahgildyan-1972}.

\section{A model of charge pump phase-locked loop with
phase-frequency detector in the signal space}

\begin{figure*}
  \includegraphics[width=1\linewidth]{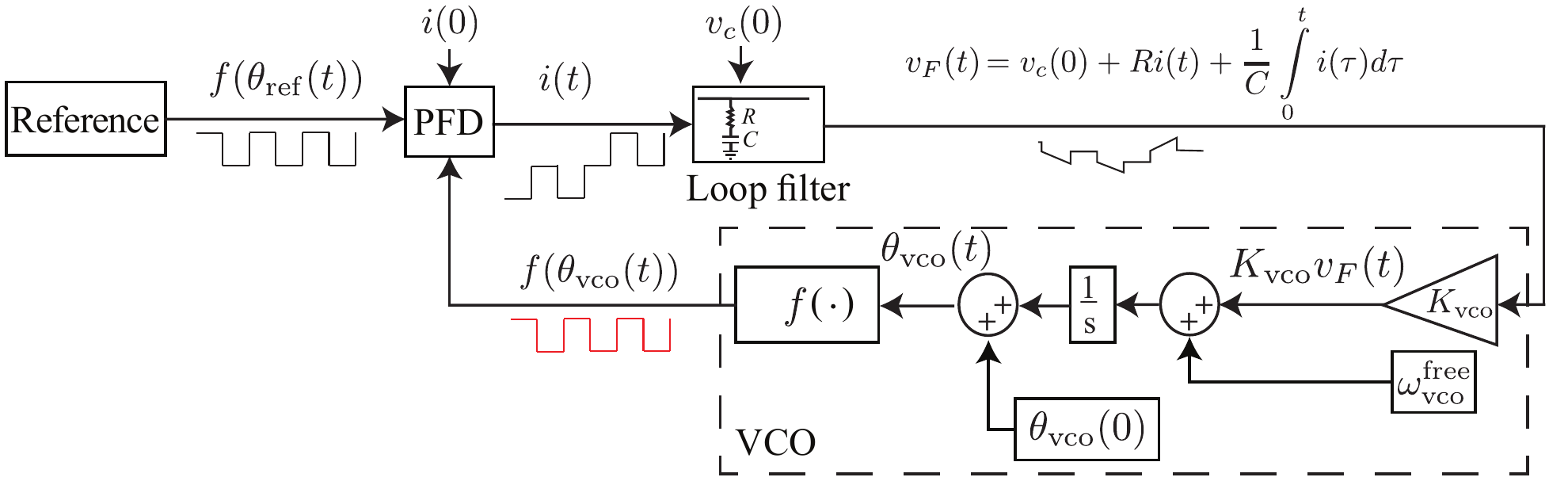}
  \caption{Charge pump phase-locked loop with
phase-frequency detector (Charge pump PLL)}
  \label{pfd_char}
\end{figure*}
\begin{figure}[H]
  \centering
  \includegraphics[width=0.6\linewidth]{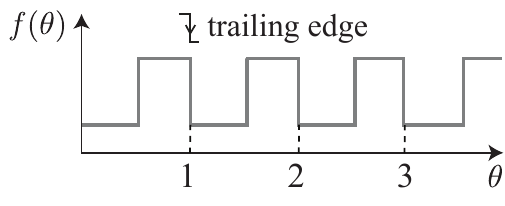}
  \caption{Waveforms of the reference and voltage controlled oscillator (VCO)
signals are periodic functions with period equal to one.
Trailing edges happen at the integer values of corresponding phases.}
\label{f1-f2}
\end{figure}

Consider charge pump phase-locked loop with
phase-frequency detector \cite{Gardner-1980,Gardner-2005-book} on Fig.~\ref{pfd_char}.
Both reference and output of the voltage controlled oscillator (VCO) are square waveform signals, see Fig.~\ref{f1-f2}.
Without loss of generality we suppose that
trailing edges of VCO and reference signals occur when corresponding phase reaches an integer number.
The frequency $\omega_{\rm ref}$ of reference signal (reference frequency)
is usually assumed to be constant:
\begin{equation}\label{omega-ref}
  \theta_{\rm ref}(t) = \omega_{\rm ref}t = \frac{t}{T_{\rm ref}},
\end{equation}
where $T_{\rm ref}$, $\omega_{\rm ref}$  and $\theta_{\rm ref}(t)$
are the period, frequency and phase of reference signal.

\begin{figure}[H]
  \includegraphics[width=0.8\linewidth]{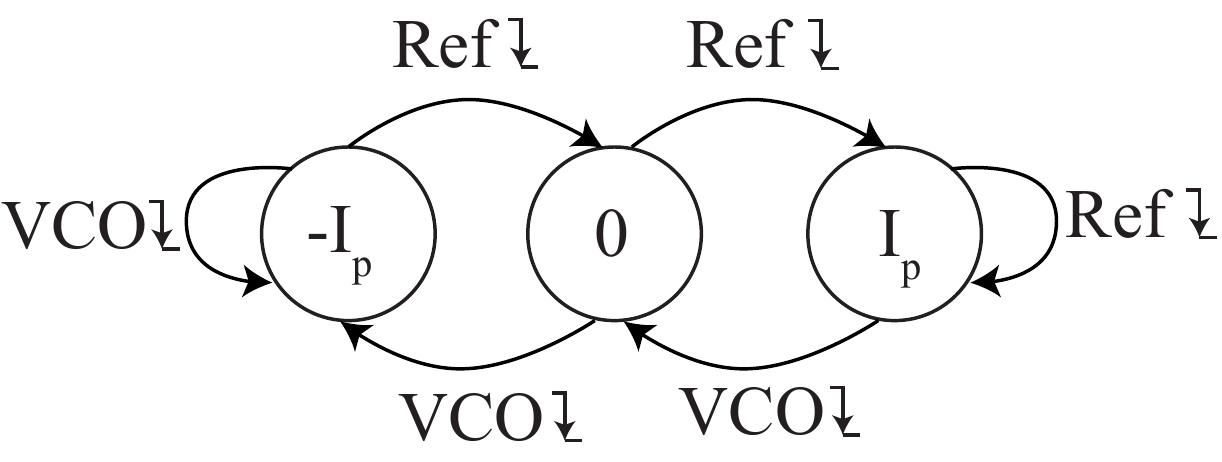}
  \includegraphics[width=0.8\linewidth]{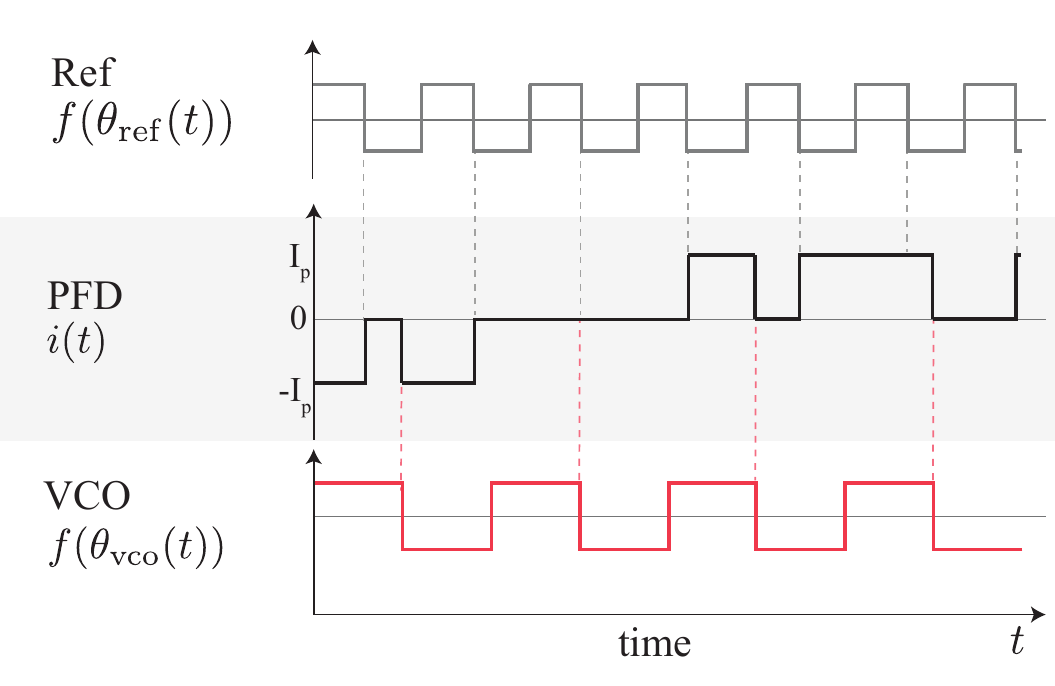}
    \caption{Phase-frequency detector operation.}
  \label{pfd}
\end{figure}

The phase-frequency detector (PFD) is a digital circuit, triggered by the trailing (falling) edges of the reference  Ref and VCO signals. The output signal of PFD $i(t)$
can have only three states (Fig.~\ref{pfd}): 0, $+I_p$, and $-I_p$.
To construct a mathematical model,
we wait for first trailing edge of reference signal and define the corresponding time instance as $t = 0$.
Suppose that before $t=0$ the PFD had a certain constant state $i(0-)$.
A trailing edge of the reference signal forces PFD to switch to higher state,
unless it is already in state $+I_p$.
A trailing edge of the VCO signal forces PFD to switch to lower state,
unless it is already in state $-I_p$.
If both trailing edges happen at the same time then PFD switches to zero.

Thus, $i(0)$ is determined by the values $i(0-)$, $\theta_{\rm vco}(0)$, and $\theta_{\rm ref}(0)$.
Similarly,  $i(t)$
is determined by $i(t-)$, $\theta_{\rm vco}(t)$, and $\theta_{\rm ref}(t)$.
Thus, $i(t)$ is piecewise constant and right-continuous\footnote{approaching any number from the right yields the same value of $i(t)$}.

The relationship between the input current $i(t)$
and the output voltage $v_F(t)$ for a
proportionally integrating (perfect PI) filter
based on  resistor and capacitor
\[
  H(s) = R + \frac{1}{Cs},
\]
is as follows
\begin{equation}
\label{RC-filter}
  \begin{aligned}
    & v_F(t) = v_c(0) + Ri(t) + \frac{1}{C}\int\limits_0^t i(\tau)d\tau,
  \end{aligned}
\end{equation}
where $R>0$ is a resistance, $C>0$ is a capacitance,
and $v_c(t)=v_c(0) + \tfrac{1}{C}\int\limits_0^t i(\tau)d\tau$ is a capacitor charge.

The control signal $v_F(t)$ adjusts the VCO frequency:
\begin{equation} \label{vco first}
   \begin{aligned}
      &
        \dot\theta_{\rm vco}(t) =
         \omega_{\rm vco}(t) =
         \omega_{\rm vco}^{\text{free}} + K_{\rm vco}v_F(t),
   \end{aligned}
\end{equation}
where $\omega_{\rm vco}^{\text{free}}$ is the VCO free-running (quiescent) frequency
(i.e. for $v_F(t)\equiv 0$), $K_{\rm vco}$
is the VCO gain (sensivity),  and $\theta_{\rm vco}(t)$ is the VCO phase.
Further we assume that\footnote{VCO overload is considered in the end of this paper.}
\begin{equation}
\label{vco overload}
\begin{aligned}
  & \dot\theta_{\rm vco}(t)=\omega_{\rm vco}(t) >0.
  \\
\end{aligned}
\end{equation}

If the frequency of VCO $\omega_{\rm vco}(t)$
is much higher than the frequency of reference signal $\omega_{\rm ref}$,
then trailing edges of VCO forces PFD to be in the lower state $-I_p$ most of the time.
In this case the output of loop filter $v_F(t)$ is negative.
Negative filter output, in turn, decrease the VCO frequency
to match the reference frequency.
Similarly, if the VCO frequency is much lower than the reference frequency,
the filter output becomes mostly positive, increasing the VCO frequency.
If the VCO and reference frequencies are close to each other,
then the transient process may be more complicated.
In this case the CP-PLL either tends to a locked state
or to an unwanted oscillation.

From \eqref{omega-ref}, \eqref{RC-filter}, and \eqref{vco first},
for given $i(0-)$ and $\omega_{\rm ref}$ 
we obtain a \emph{continuous time nonlinear mathematical model of CP-PLL}
described by differential equations
\begin{equation}\label{signal space}
\begin{aligned}
  & \dot v_c(t) = \tfrac{1}{C}i(t), \\
  & \dot\theta_{\rm vco}(t)  =
        \omega_{\rm vco}^{\text{free}} + K_{\rm vco}
        \left(
          Ri(t)
          + v_c(t)
        \right), \\
\end{aligned}
\end{equation}
with discontinuous piecewise constant nonlinearity
\[
  i(t) = i\big(i(t-), \omega_{\rm ref}, \theta_{\rm vco}(t)\big)
\]
and initial conditions $\big(v_c(0), \theta_{\rm vco}(0)\big)$.



\section{Derivation of discrete time CP-PLL model}
Here we derive discrete time model of the CP-PLL in the following form
\begin{equation}
  \begin{aligned}
    & \tau_{k+1} = \varphi(\tau_k,v_k),\\
    & v_{k+1} = v(\tau_k,v_k).
  \end{aligned}
\end{equation}
First, let us define the state variables $\tau_k$, $v_k$.

\begin{figure}[H]
\centering
  \includegraphics[width=\linewidth]{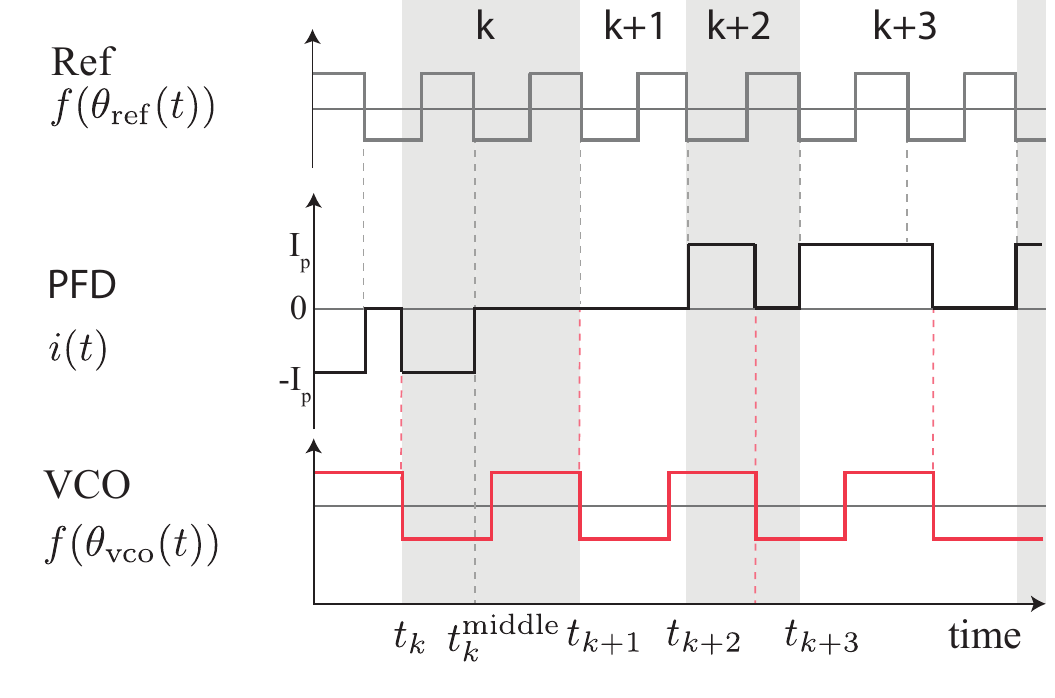}
  \caption{Explanation of $t_k$ and $t_k^{\rm middle}$.}
  \label{fig states}
\end{figure}

Let $t_0 = 0$.
Denote by $t_0^{\rm middle}$ the first instant of time such that the PFD output becomes equal to zero\footnote{Remark that the PFD output $i(t)$ always returns to zero
from non-zero state at certain time.
If $i(t_0)=-1$, then the first Ref trailing edge returns the PFD output to zero.
If $i(t_0)=1$, then the VCO frequency is increasing
until the first VCO trailing edge returns the PFD output to zero.};
if $i(0)=0$ then $t_0^{\rm middle}=0$.
Then we wait until the first trailing edge of the VCO or Ref
and denote corresponding moment of time by $t_1$.
Continuing in a similar way, one obtains
increasing sequences $\{t_k\}$ and $\{t_k^{\rm middle}\}$ for $k=0,1,2...$ (see Fig.~\ref{fig states}).

\begin{figure}[H]
\centering
  \includegraphics[width=\linewidth]{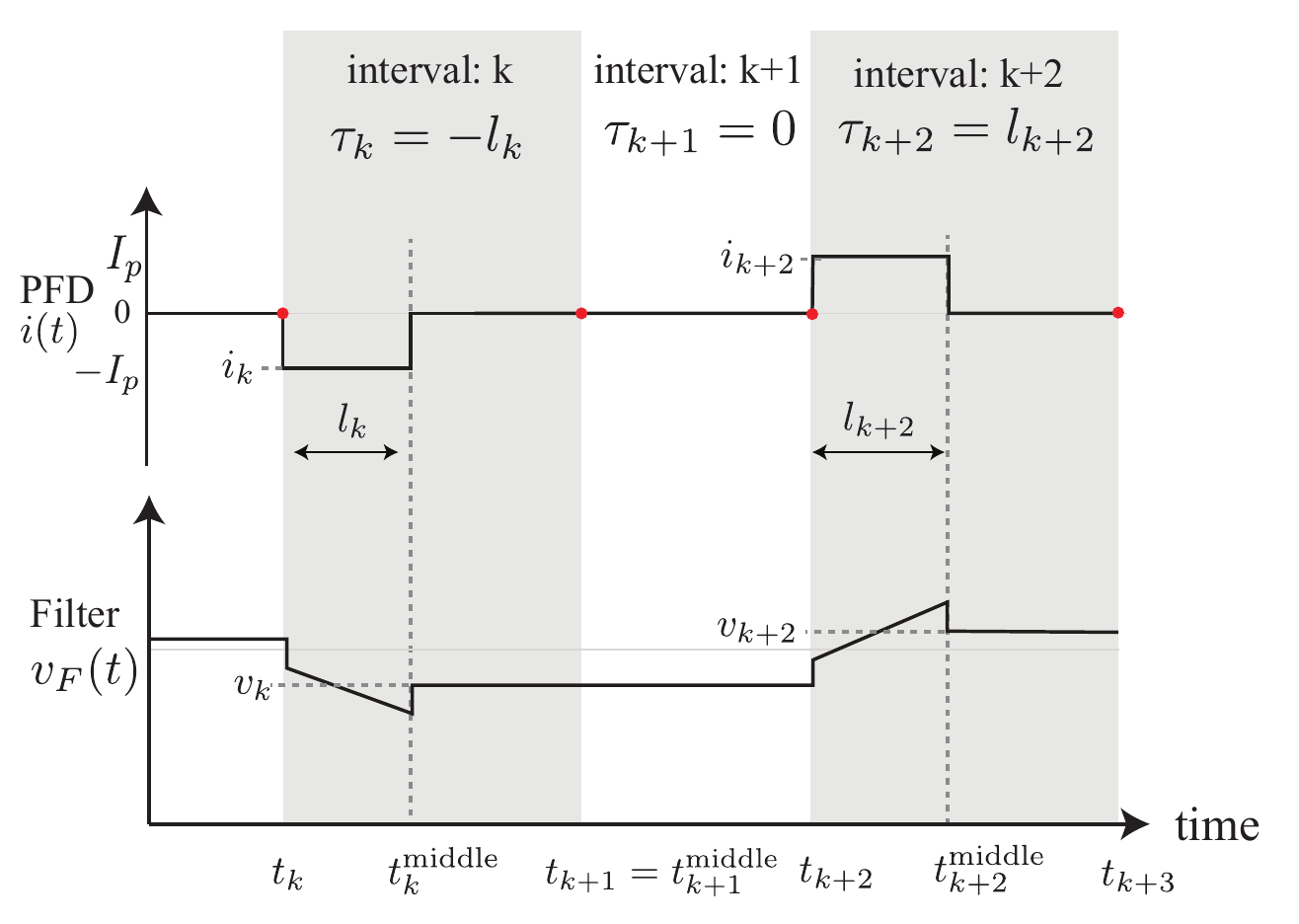}
  \caption{Definition of discrete states $\tau_k$ and $v_k$. $l_k$ is a PFD pulse width.}
  \label{fig states tau v}
\end{figure}

Let $t_k < t_k^{\rm middle}$.
Then for $t \in [t_k,t_k^{\rm middle})$ $\sign(i(t))$ is a non zero constant ($\pm1$).
Denote by $\tau_k$ the PFD pulse width (length of time interval,
where PFD output is non-zero constant),
multiplied by the sign of PFD output (see Fig.~\ref{fig states tau v}):
\begin{equation}
\label{tau k def}
\begin{aligned}
  & \tau_k = \left(t_k^{\rm middle} - t_k\right)\sign(i(t)),
  & t \in [t_k,t_k^{\rm middle}), \\
  &  \tau_k = 0 & t_k=t_k^{\rm middle}.
\end{aligned}
\end{equation}

If the VCO trailing edge hits before the Ref trailing edge,
then $\tau_k < 0$ and in the opposite case we have $\tau_k > 0$.
Thus, $\tau_k$ shows how one signal lags behind another.

From \eqref{RC-filter} it follows that
the zero output of PFD $i(t) \equiv 0$ on the interval $(t_k^{\rm middle},t_{k+1})$
implies a constant filter output.
Denote this constant by $v_k$:
\begin{equation}
\begin{aligned}
  & v_F(t) \equiv v_k, \quad t \in [t_k^{\rm middle},t_{k+1}).
\end{aligned}
\end{equation}
\noindent Finally, from \eqref{RC-filter} we get
\begin{equation}
\label{v_F_v_k}
\begin{aligned}
  & v_F(t) = \left\{
    \begin{array}{ll}
      v_k, & t \in [t_{k}^{\rm middle},t_{k+1}),
      \\
      v_k \pm RI_p \pm \frac{I_p}{C}(t - t_{k+1}), & t \in [t_{k+1}, t_{k+1}^{\rm middle}).
    \end{array}\right.
\end{aligned}
\end{equation}
Combining \eqref{vco first} and \eqref{v_F_v_k} we obtain
\begin{equation}
\label{omega-vco-formula}
\begin{aligned}
  & \omega_{\rm vco}(t) = \left\{
    \begin{array}{l}
      \omega_{\rm vco}^{\rm free} + K_{\rm vco}v_k,
        \quad
        t \in [t_{k}^{\rm middle},t_{k+1}),
      \\
      \omega_{\rm vco}^{\rm free} + K_{\rm vco}
          \left(
            v_k \pm RI_p \pm \frac{I_p}{C}(t - t_{k+1})
          \right),
        \\
        \quad
        \quad
        \quad
         t \in [t_{k+1}, t_{k+1}^{\rm middle}),
    \end{array}\right.
\end{aligned}
\end{equation}
  where the sign $-$ or $+$ in the last equation
  corresponds to $\sign(i(t))$.


By \eqref{v_F_v_k} and \eqref{tau k def}
the value of $v_{k + 1}$ can be expressed via $\tau_{k+1}$ and $v_k$:
\begin{equation}
  \label{vk solution}
      v_{k+1} = v_k+\frac{I_p}{C}\tau_{k+1}.
\end{equation}

\begin{figure}[H]
\centering
  \includegraphics[width=\linewidth]{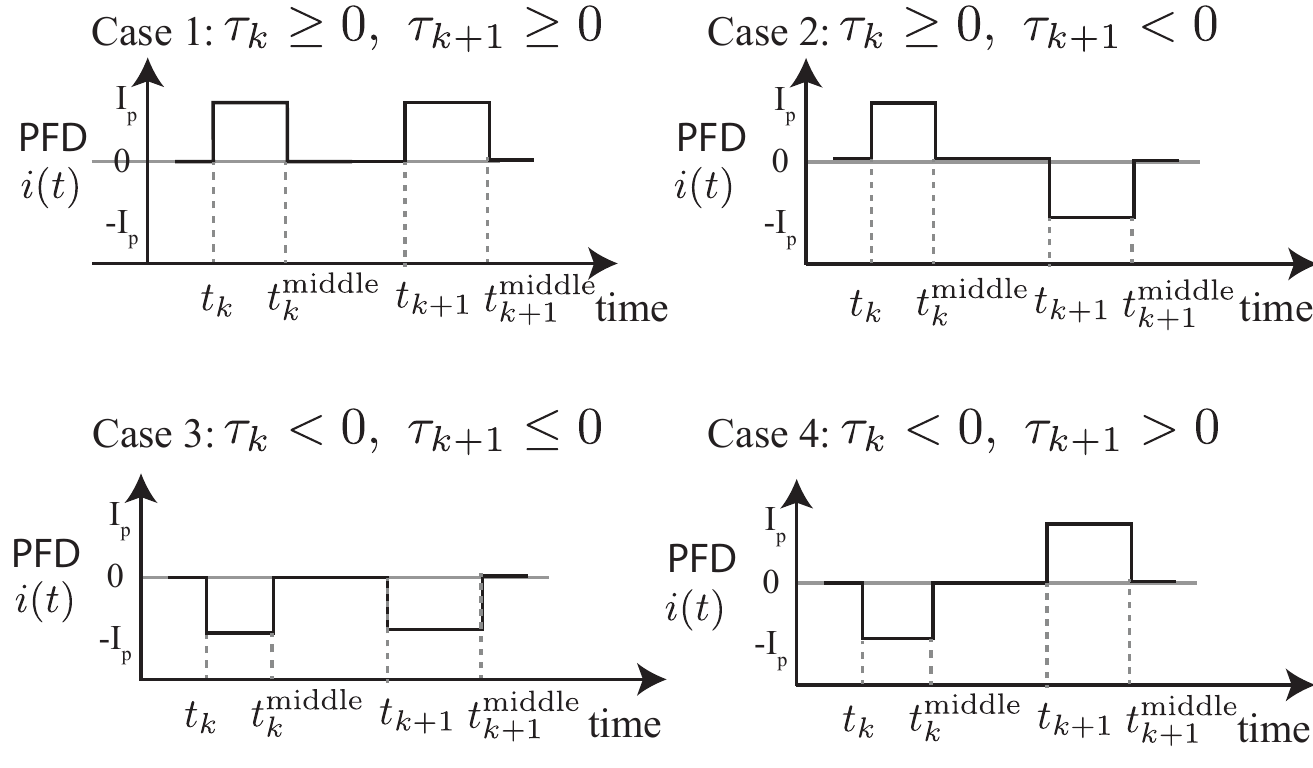}
  \caption{Four cases of PFD states transitions}
  \label{fig-4-cases}
\end{figure}

To find $\tau_{k+1}$ one needs to
consider four possible cases of PFD transitions (see Fig.~\ref{fig-4-cases}):
\begin{enumerate}
  \item $\tau_k \geq 0$, $\tau_{k+1} \geq 0$;
  \item $\tau_k \geq 0$, $\tau_{k+1} < 0$;
  \item $\tau_k < 0$, $\tau_{k+1} \leq 0$;
  \item $\tau_k < 0$, $\tau_{k+1} > 0$.
\end{enumerate}


{\bf Case 1: } $\tau_k\geq 0$ and $\tau_{k+1} \geq 0$
(see Fig.~\ref{fig:case1:1-5}).

\begin{figure}[H]
  \begin{subfigure}[b]{0.49\textwidth}
    \includegraphics[width=\linewidth]{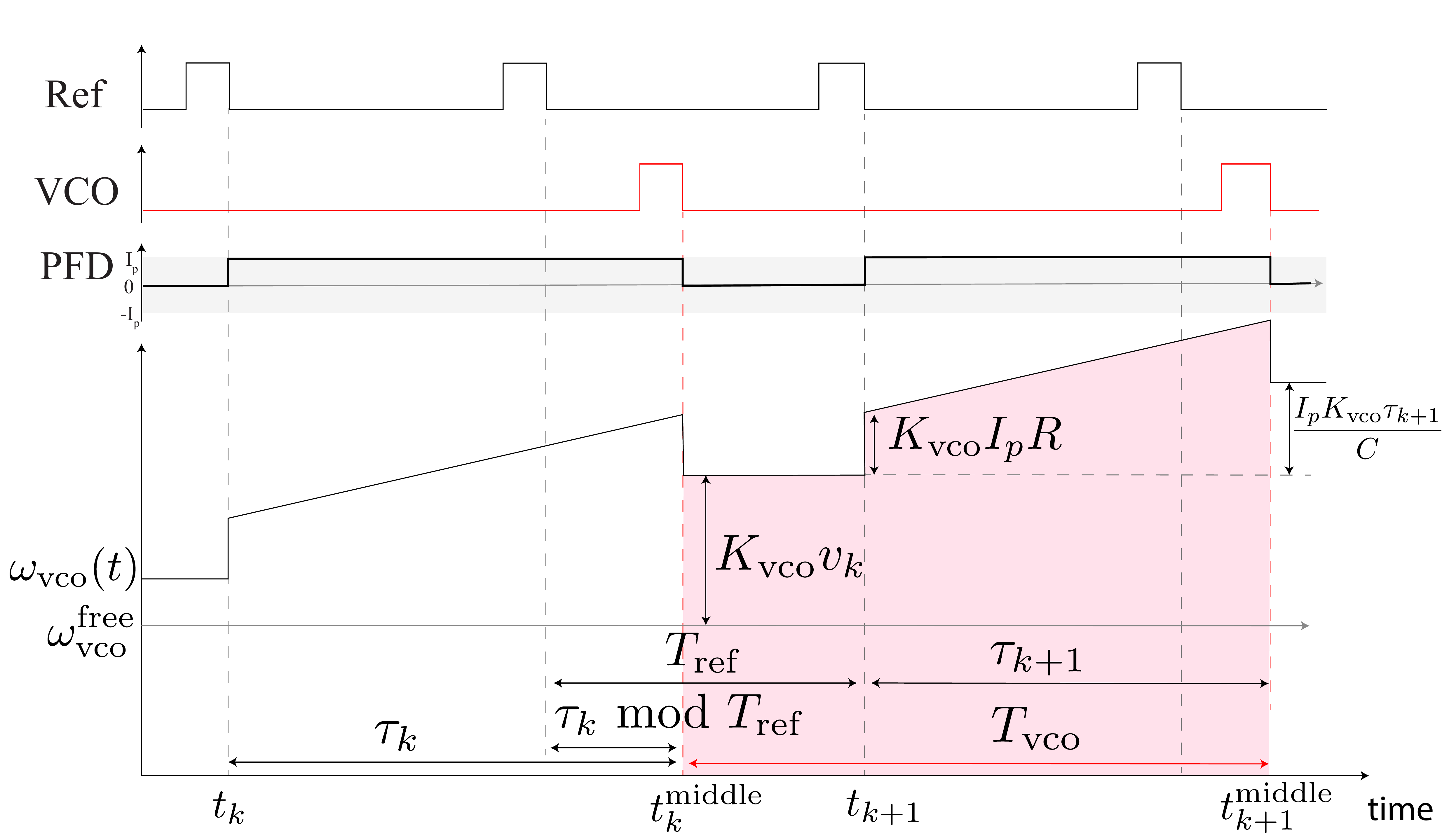}
    \caption{Case 1\_1: general case, Reference cycle slipping; $\tau_k \geq 0$, $\tau_{k+1} \geq 0$}.
    \label{fig case1_0}
  \end{subfigure}
  \caption{
   Subcases of the Case 1. Integral of the VCO frequency $\omega_{\rm vco}$ over
   the VCO period $T_{\rm vco}$
   is pink subgraph area (grey in black/white).
  The integral is equal to $1$ according
  to the PFD switching law and definition of time intervals.
  }
  \label{fig:case1:1-5}
\end{figure}

First, rewrite requirement $\tau_{k+1} \geq 0$ in terms of
$\tau_k$ and $v_k$.
Since $\omega_{\rm vco}(t)>0$,
condition $\tau_{k+1}>0$ takes the form
\begin{equation}
\begin{aligned}
&  \theta_{\rm vco}(t_{k+1}) -
   \theta_{\rm vco}(t_{k}^{\rm middle}) \leq 1.
\end{aligned}
\end{equation}
By \eqref{omega-vco-formula} we get
\begin{equation}
\begin{aligned}
&  \theta_{\rm vco}(t_{k+1}) -
   \theta_{\rm vco}(t_{k}^{\rm middle})
   =
   \int\limits_{t_k^{\rm middle}}^{t_{k+1}}
    \omega_{\rm vco}(\tau)d\tau =
  \\
  &
 = (T_{\rm ref} - (\tau_k\ {\rm mod}\ T_{\rm ref}))
\left(
  \omega_{\rm vco}^{\text{free}} + K_{\rm vco}v_k
\right).
\end{aligned}
\end{equation}
Thus, condition $\tau_{k+1} \geq 0$
can be expressed via $v_k$ and $\tau_k$ in the following way\footnote{
  Here $(a\text{ mod }b)$ is the remainder after division of $a$ by $b$, where $a$ is the dividend and $b$ is the divisor. This function is often called the modulo operation.}:
\begin{equation}
\label{case1 requierment}
\begin{aligned}
& (T_{\rm ref} - (\tau_k\ {\rm mod}\ T_{\rm ref}))
\left(
  \omega_{\rm vco}^{\text{free}} + K_{\rm vco}v_k
\right) \leq 1.
\end{aligned}
\end{equation}

Now find $\tau_{k+1}$.
VCO trailing edges appeared twice on time interval $[t_k,t_{k+2})$: at $t=t_k^{\rm middle}$ and at $t=t_{k+1}^{\rm middle}$.
Thus, we get
\begin{equation}
\begin{aligned}
& \theta_{\rm vco}(t_{k+1}^{\rm middle}) - \theta_{\rm vco}(t_k^{\rm middle}) = 1.
\end{aligned}
\end{equation}
By \eqref{vco first} it is equivalent to
\begin{equation}
\label{case-1-phasing}
\begin{aligned}
  & \int\limits_{t_{k}^{\rm middle}}^{t_{k+1}}
    \omega_{\rm vco}(t)dt +
    \int\limits_{t_{k+1}}^{t_{k+1}^{\rm middle}}
    \omega_{\rm vco}(t)dt
    = 1.
\end{aligned}
\end{equation}

By definition of $\tau_k$ \eqref{tau k def}
we get (see Fig.~\ref{fig case1_0})
\begin{equation}
  \label{case-1-timing}
  \begin{aligned}
    & t_{k+1} = t_k^{\rm middle}- (\tau_k\text{ mod } T_{\rm ref}) + T_{\rm ref},
    \\
    & t_{k+1}^{\rm middle} = t_{k+1} + \tau_{k+1}.
  \end{aligned}
\end{equation}

Substituting \eqref{case-1-timing} and \eqref{omega-vco-formula} into \eqref{case-1-phasing} and calculating the integral (i.e.
shaded area in Fig.~\ref{fig case1_0})
we get the following relation for $\tau_{k+1}$:
\begin{equation}
\label{gz-gz-eq}
\begin{aligned}
& (T_{\rm ref} - (\tau_k\ {\rm mod}\ T_{\rm ref}) + \tau_{k+1})
\left(
  \omega_{\rm vco}^{\text{free}} + K_{\rm vco}v_k
\right)
+
\\
&
+
K_{\rm vco}I_p\left(R \tau_{k+1} + \frac{1}{2C}\tau_{k+1}^2\right)
 = 1.
\end{aligned}
\end{equation}
Equation \eqref{gz-gz-eq} is quadratic with discriminant
\begin{equation}
\begin{aligned}
& b(v(k))^2 - 4ac(\tau_k,v_k),
\\
\end{aligned}
\end{equation}
where
\begin{equation}
\label{abc}
\begin{aligned}
& a = \frac{K_{\rm vco}I_p}{2C},
\\
&
  b = b(v_k) = \omega_{\rm vco}^{\text{free}} + K_{\rm vco}v_k
    + K_{\rm vco}I_pR,
\\
&
  c = c(\tau_k,v_k) = (T_{\rm ref} - (\tau_k\ {\rm mod}\ T_{\rm ref}))
\left(
  \omega_{\rm vco}^{\text{free}} + K_{\rm vco}v_k
\right) - 1.
\end{aligned}
\end{equation}
Relation \eqref{case1 requierment} is equivalent to $c \leq 0$,
which means that the discriminant is non-negative.
Therefore, equation \eqref{gz-gz-eq} has exactly one positive solution:
\begin{equation}
\begin{aligned}
& \frac{-b + \sqrt{b^2 - 4ac}}{2a}.
\end{aligned}
\end{equation}

{\bf Case 2: }$\tau_k\geq 0$ and  $\tau_{k+1} < 0$ (Fig.~\ref{fig:case2:1-4}).
The first VCO edge appears at $t=t_k^{\rm middle}$,
and the second VCO edge appears at $t=t_{k+1}$.

\begin{figure}[H]
  \begin{subfigure}[b]{0.49\textwidth}
    \includegraphics[width=\linewidth]{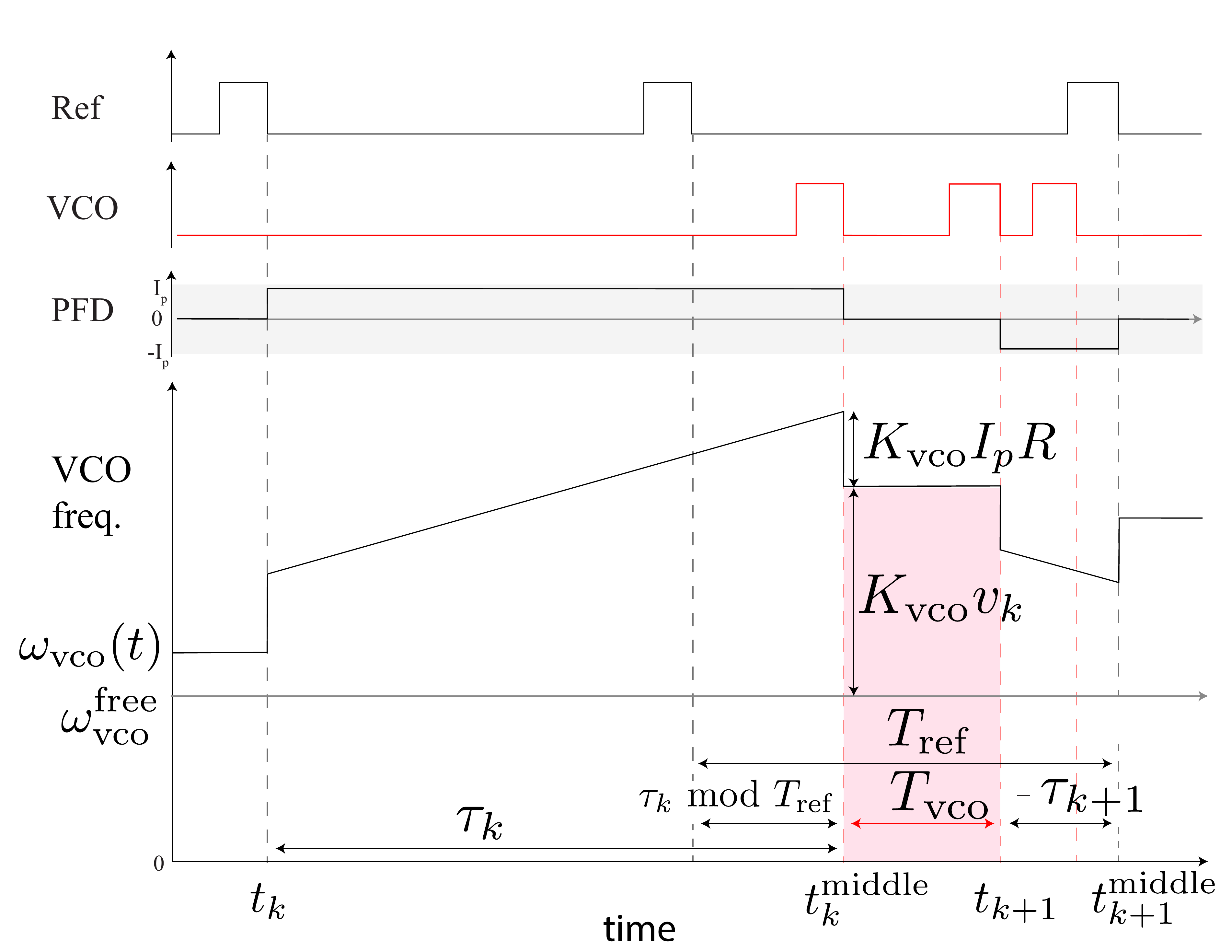}
    \caption{Case 2\_1: general case, Reference and VCO cycle slipping; $\tau_k \geq 0$, $\tau_{k+1} < 0$}
    \label{fig case2_1}
  \end{subfigure}
  \caption{
   Subcases of the Case 2.
   Integral of the VCO frequency $\omega_{\rm vco}$ over
   the VCO period $T_{\rm vco}$
   is pink subgraph area (grey in black/white).
   The integral is equal to $1$ according
   to the PFD switching law and definition of time intervals.
  }
  \label{fig:case2:1-4}
\end{figure}
Similarly to the previous case, we can derive a relation for $\tau_{k+1}$
by integrating VCO frequency $\omega_{\rm vco}(t)$ over its period
$[t_k^{\rm middle}, t_{k+1}]$:
\[
  \begin{aligned}
    & \left(
      T_{\rm ref} + \tau_{k+1}-(\tau_k \text{ mod } T_{\rm ref})
    \right)\left(
  \omega_{\rm vco}^{\text{free}} + K_{\rm vco}v_k
  \right)
    = 1.
  \end{aligned}
\]
Therefore
\begin{equation}
\label{case2 solution}
\begin{aligned}
& \tau_{k+1} =
\frac{1}{\omega_{\rm vco}^{\text{free}} + K_{\rm vco}v_k}
- T_{\rm ref} + (\tau_k \text{ mod } T_{\rm ref}).
\end{aligned}
\end{equation}

{\bf Case 3: } $\tau_k < 0$  and $\tau_{k+1} \leq 0$ (see Fig.~\ref{fig case3}).
\begin{figure}[H]
  \begin{subfigure}[b]{0.49\textwidth}
  \centering
    \includegraphics[width=\linewidth]{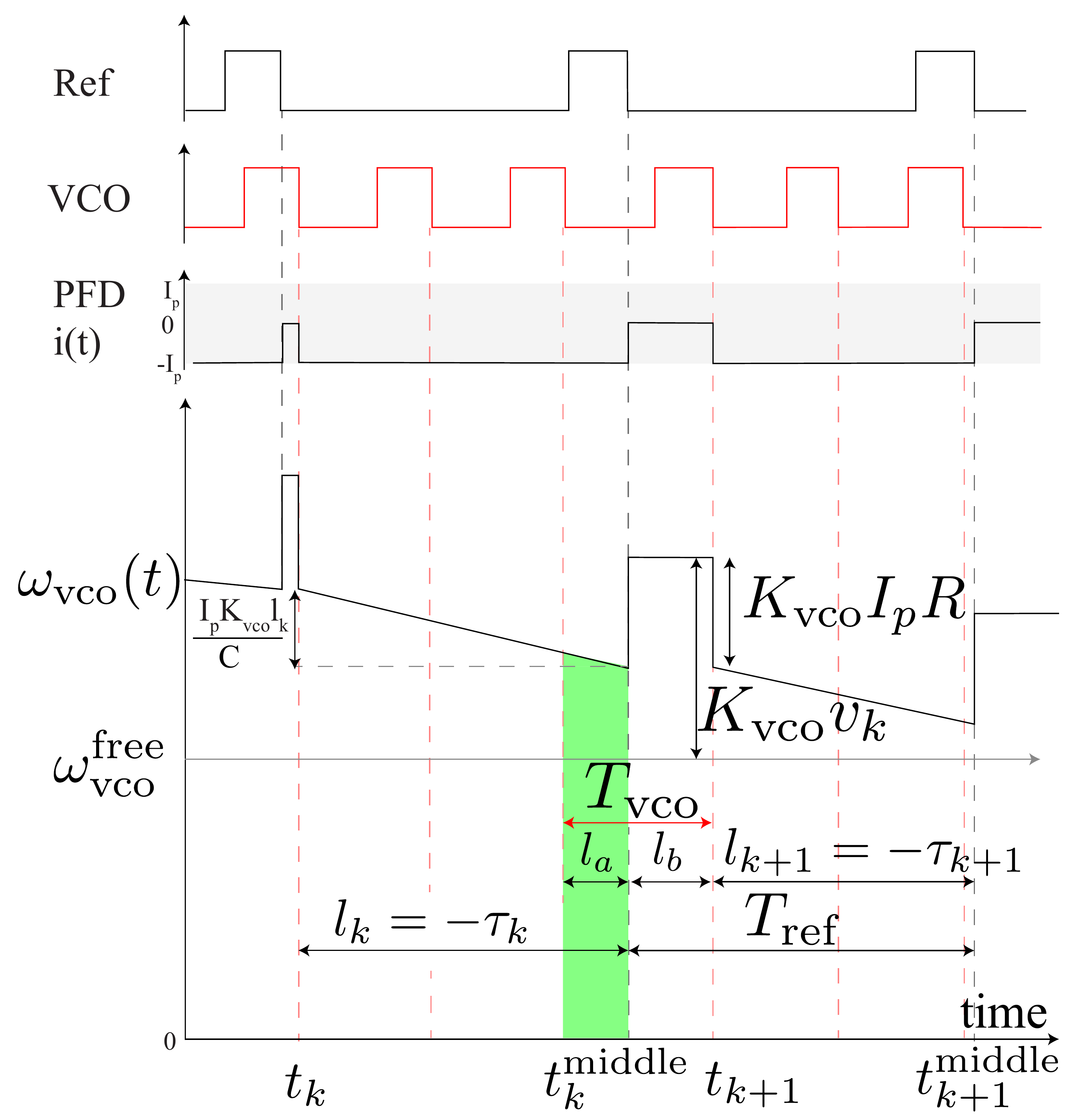}
    \caption{Case 3\_1: general case, VCO cycle slipping; $\tau_k < 0, \quad \tau_{k+1} < 0$}
    \label{fig case3_1}
  \end{subfigure}
  \caption{
   Subcases of the Case 3.
   Integral of the VCO frequency $\omega_{\rm vco}$ over
   the VCO period $T_{\rm vco}$
   is pink subgraph area (grey in black/white).
   The integral is equal to $1$ according
   to the PFD switching law and definition of time intervals.
  }
  \label{fig case3}
\end{figure}

First, we determine the sign of $\tau_{k+1}$.
To do that we need to find out which of the signals (VCO or reference)
reaches its period first after $t_{k}^{\rm middle}$,
i.e. to check if $l_b  = t_{k+1} - t_{k}^{\rm middle} \leq T_{\rm ref}$.
By \eqref{omega-vco-formula} for the time interval $l_b$
we get (see Fig.~\ref{fig case3})
\begin{equation}
\label{case-3-phasing1}
      l_b = l_b(v_k) = \frac{S_{l_b}}{K_{\rm vco}v_k+\omega_{\rm vco}^{\text{free}}},
\end{equation}
where
\begin{equation}
    S_{l_b} = \int\limits_{t_k^{\rm middle}}^{t_{k+1}}\omega_{\rm vco}(t)dt
     = 1-S_{l_a}.
\end{equation}
Here $S_{l_a}$ (green area in Fig.~\ref{fig case3})
is computed as a fractional part of the subgraph area corresponding to $l_k=-\tau_k$:
\begin{equation}
\label{case-3-phasing2}
\begin{aligned}
    & S_{l_a} = S_{l_a}(\tau_k, v_k) = S_{l_k}\text{ mod }1,\\
    & S_{l_k} = S_{l_k}(\tau_k, v_k) =
       \\
       &
       \quad = \left(
        K_{\rm vco}v_k - I_pRK_{\rm vco} +\omega_{\rm vco}^{\text{free}}
      \right)l_k + K_{\rm vco}I_p\frac{l_k^2}{2C}.
\end{aligned}
\end{equation}
Finally, we get
\begin{equation}
  \label{case3 solution}
  \begin{aligned}
    & \tau_{k+1} = \frac{1-S_{l_a}}{K_{\rm vco}v_k+\omega_{\rm vco}^{\text{free}}} -T_{\rm ref}.
  \end{aligned}
\end{equation}

{\bf Case 4: } $\tau_k<0$  and $\tau_{k+1}>0$ (see Fig.~\ref{fig:case4:1-3}).
\begin{figure}[H]
  \begin{subfigure}[b]{0.49\textwidth}
  \centering
    \includegraphics[width=\linewidth]{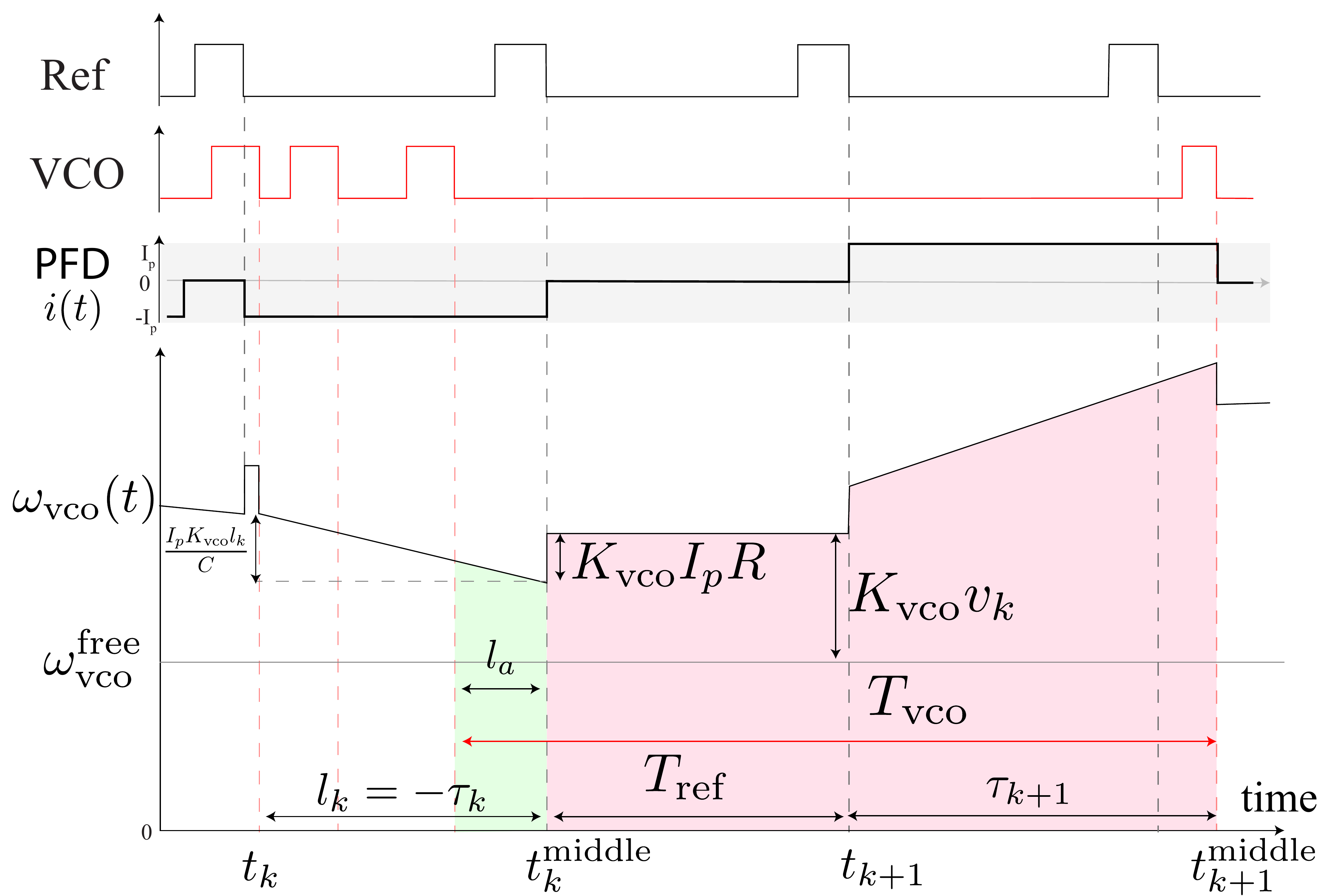}
    \caption{Case 4\_1: general case, VCO and Reference cycle slipping; $\tau_k < 0, \quad \tau_{k+1} > 0$}
    \label{fig case4_1}
  \end{subfigure}
  \caption{
   Subcases of the Case 4.
   Integral of the VCO frequency $\omega_{\rm vco}$ over
   the VCO period $T_{\rm vco}$
   is pink subgraph area (grey in black/white).
   The integral is equal to $1$ according
   to the PFD switching law and definition of time intervals.
   }
  \label{fig:case4:1-3}
\end{figure}
Similar to Case~3 condition $\tau_{k+1} > 0$ is equivalent to
\begin{equation}
\label{case-4-cond}
\begin{aligned}
& l_b > T_{\rm ref},
\end{aligned}
\end{equation}
where $l_b = t_{k+1}^{\rm middle} - t_{k}^{\rm middle}$
can be computed in the following way
\begin{equation}
  \begin{aligned}
    & l_b = l_b(\tau_k,v_k) = \frac{1 - S_{l_a}}{K_{\rm vco}v_k+\omega_{\rm vco}^{\text{free}}},\\
    & S_{l_a} = S_{l_a}(\tau_k,v_k) = S_{l_k}\text{ mod }1,\\
    & S_{l_k} = S_{l_k}(\tau_k,v_k) =
    \\
    & \quad -\left(
    K_{\rm vco}v_k - I_pRK_{\rm vco} +\omega_{\rm vco}^{\text{free}}
    \right)\tau_k
    + K_{\rm vco}I_p\frac{\tau_k^2}{2C}.
  \end{aligned}
\end{equation}
Now we can compute the VCO phase corresponding to its full period:
\begin{equation}
  \begin{aligned}
    & S_{l_a}+S_{T_{\rm ref}}+S_{\tau_{k+1}} = 1,\\
    & S_{T_{\rm ref}} = S_{T_{\rm ref}}(v_k) = T_{\rm ref}(K_{\rm vco}v_k+\omega_{\rm vco}^{\text{free}}),\\
    & S_{\tau_{k+1}} = S_{\tau_{k+1}}(\tau_{k+1},v_k) =
          \tau_{k+1}\left(
              K_{\rm vco}v_k+\omega_{\rm vco}^{\text{free}}
              +I_pRK_{\rm vco}
          \right)
          +
          \\
      & \quad\quad\quad\quad
      + \frac{I_pK_{\rm vco}\tau^2(k+1)}{2C}.
  \end{aligned}
\end{equation}
From \eqref{case-4-cond} there is only one positive solution $\tau_{k+1}$ of quadratic equation
\begin{equation}
\label{case4 solution}
\begin{aligned}
& \tau_{k+1} = \frac{-b + \sqrt{b^2 - 4ad}}{2a},
\\
& a = \frac{K_{\rm vco}I_p}{2C},
\\
&
  b = b(v_k) = \omega_{\rm vco}^{\text{free}} + K_{\rm vco}v_k
    + K_{\rm vco}I_pR,
\\
&
  d = b(\tau_k,v_k) =S_{l_a}+S_{T_{\rm ref}} - 1,
  \\
\end{aligned}
\end{equation}



\section{Corrected full discrete model of CP-PLL}

Combining equations for all four cases we obtain 
a \emph{discrete time nonlinear mathematical model of CP-PLL}
\begin{equation}
  \label{vk-tauk-system}
  \begin{aligned}
    & v_{k+1} = v_k+\frac{I_p}{C}\tau_{k+1}(\tau_k,v_k),\\
    & k = 0,1,2,...
  \end{aligned}
\end{equation}
with initial conditions $v_0$ and $\tau_0$.
Function $\tau_{k+1}(\tau_k,v_k)$ is defined by the following equation
\begin{equation}
\label{complete-model}
  \begin{aligned}
    & \tau_{k+1}(\tau_k,v_k) =
    \left\{\begin{array}{l}
      \frac{-b + \sqrt{b^2 - 4ac}}{2a},
      \ \quad \tau_k \geq 0, \quad c \leq 0,
      \\
      \\
      \frac{1}{\omega_{\rm vco}^{\text{free}} + K_{\rm vco}v_k}
- T_{\rm ref} + (\tau_k \text{ mod } T_{\rm ref}),
      \\
      \quad\quad\quad\quad\quad\quad\quad
       \tau_k \geq 0, \quad c > 0,
      \\
      \\
      l_b-T_{\rm ref},
      \quad\quad\quad
       \tau_k < 0, \quad l_b \leq T_{\rm ref},
      \\
      \\
      \frac{-b + \sqrt{b^2 - 4ad}}{2a},
      \quad \tau_k < 0, \quad l_b > T_{\rm ref},
    \end{array}
    \right.
    \\
      \end{aligned}
\end{equation}
\begin{equation}
  \begin{aligned}
    & a = \frac{K_{\rm vco}I_p}{2C},
    \\
    &
      b = b(v_k) = \omega_{\rm vco}^{\text{free}} + K_{\rm vco}v_k
        + K_{\rm vco}I_pR,
    \\
    &
      c = c(\tau_k,v_k) =
    \\
    &
      \quad (T_{\rm ref} - (\tau_k\ {\rm mod}\ T_{\rm ref}))
\left(
  \omega_{\rm vco}^{\text{free}} + K_{\rm vco}v_k
\right) - 1,\\
    & l_b = l_b(\tau_k,v_k) =\tfrac{1 - S_{l_a}}{K_{\rm vco}v_k+\omega_{\rm vco}^{\text{free}}},\\
    & S_{l_a} = S_{l_a}(\tau_k,v_k) = S_{l_k}\text{ mod }1,\\
    & S_{l_k} = S_{l_k}(\tau_k,v_k) =
    \\
    & \quad -\left(
    K_{\rm vco}v_k - I_pRK_{\rm vco} +\omega_{\rm vco}^{\text{free}}
    \right)\tau_k
    + K_{\rm vco}I_p\tfrac{\tau_k^2}{2C},
    \\
    & d = d(v_k) = S_{l_a}+T_{\rm ref}(K_{\rm vco}v_k+\omega_{\rm vco}^{\text{free}}) - 1.
  \end{aligned}
\end{equation}
Here the initial conditions are the following: $v_0$ is the initial output of the filter,
$\tau_0$ is determined by $\theta_{\rm vco}(0)$ and $i(0)$.

If at some point VCO becomes overloaded (frequency approaches zero)
 one should stop simulation or use another set of equations, based on ideas similar to (34) and (35) in \cite{Paemel-1994}, see section~\ref{sec:vco overload}.

\subsection{Locked states, hold-in range, and pull-in range}
After the synchronization is achieved, i.e. transient process is over,
the loop is said to be in a \emph{locked state}.
For practical purposes, only \emph{asymptotically stable locked states},
in which the loop returns after small perturbations of its state,
are of interest.
CP-PLL is in a {\bf locked state}
if trailing edges of the VCO signal happens near the trailing edges
of the reference signal:
\begin{equation}
\label{locked-state-req}
\begin{aligned}
  & \quad \left|
    \frac{\tau_{k}}{T_{\rm ref}}
    \right| \leq \tau_{\rm lock},
    \quad k = 0,1,2, ...,
\end{aligned}
\end{equation}
where $\tau_{\rm lock}$ is sufficiently small,
and the loop returns in this state after small perturbations of
$(v_k, \tau_k)$.

In a locked state the output of PFD i(t)
can be non-zero
only on short time intervals (shorter than $\tau_{\rm lock}$).
An allowed residual phase difference $\tau_{\rm lock}$
should be in agreement with engineering requirements
for a particular application. The ideal case is $\tau_{\rm lock}=0$.


There is a hypothesis, which has not yet been proven rigorously,
that the hold-in and pull-in ranges\footnote{
The largest symmetric interval (continuous range, without holes) of frequencies
around free-running frequency of the VCO
is called a \emph{hold-in range}
if the loop has a locked state.
If, in addition, the loop
acquires a locked state for any initial state of the loop
and any reference frequency from the interval
then the interval is called a \emph{pull-in range} \cite{KuznetsovLYY-2015-IFAC-Ranges,LeonovKYY-2015-TCAS,BestKLYY-2016,KuznetsovLYY-arxiv2017}.}
coincide and both are infinite if the model has a locked state
(see, e.g. \cite{Acco-2004,Orla-2013-review,Razavi-hb-2016}).

\subsection{Pull-in time}

One of the important characteristics of CP-PLL is how fast it acquires a locked state during
the pull-in process.
Suppose that the CP-PLL is in a locked state with input frequency $\omega_{\rm ref1}$.
Then the reference frequency changes to new frequency $\omega_{\rm ref2}$
from fixed range $[\omega_{\rm ref}^{\rm min},\omega_{\rm ref}^{\rm max}]$.
Since CP-PLL lost it's locked state,
the feedback loop of CP-PLL tunes the VCO to the new input frequency
to acquire new locked state.
Transient process takes time $T_{{\rm ref1} \to {\rm ref2}}$,
and maximum of such possible time intervals is called
a \emph{pull-in time} $T_{\text{pull-in}}$.
The pull-in time can be measured in seconds or the number of cycles of the input frequency.

In practice \eqref{locked-state-req} is hard to check
for all $k$.
Instead for simulation one may check that additionally to small $\tau_k$
frequency of VCO is close to reference frequency:
\begin{equation}
\label{locked-state-req2}
\begin{aligned}
  &  \left|
    \frac{\tau_{k}}{T_{\rm ref}}
    \right| \leq \tau_{\rm lock},
   \\ 
  &  \left| 
      \omega_{\rm vco}^{\rm free} + K_{\rm vco}v_k - \frac{1}{T_{\rm ref}}
    \right| < \omega_{\rm lock}.
\end{aligned}
\end{equation} 

\section{Comparison with original model by M. van Paemel}
\label{sec:examples}
In this section using original notation \cite{Paemel-1994}
we demostrate why existing model has problems
and how proposed model fixes them.

\subsection{Example 1}
Consider the following set of parameters and initial state:
\begin{equation}
\label{ex1}
\begin{aligned}
& R_2 = 0.2;
C = 0.01;
K_v = 20;
I_p = 0.1;
T = 0.125;\\
& \tau(0) = 0.0125;
v(0) = 1.
\end{aligned}
\end{equation}
Calculation of normalized parameters (equations (27)-(28) and (44)-(45) in \cite{Paemel-1994})
\begin{equation}
\label{F_n zeta}
\begin{aligned}
  & K_N = I_pR_2K_vT = 0.05,\\
  & \tau_{2N}=\frac{R_2C}{T} = 0.016,\\
  & F_N = \frac{1}{2\pi}\sqrt{\frac{K_N}{\tau_{2N}}} \approx 0.2813,\\
  & \zeta = \frac{\sqrt{K_N\tau_{2N}}}{2} \approx 0.0141,
\end{aligned}
\end{equation}
shows that parameters \eqref{ex1} correspond to allowed area in Fig.~\ref{fig:allowed-domain} (equations (46)--(47), Fig~18 and Fig.~22 in \cite{Paemel-1994}):
\begin{figure*}
  \centering
  \includegraphics[width=1\linewidth]{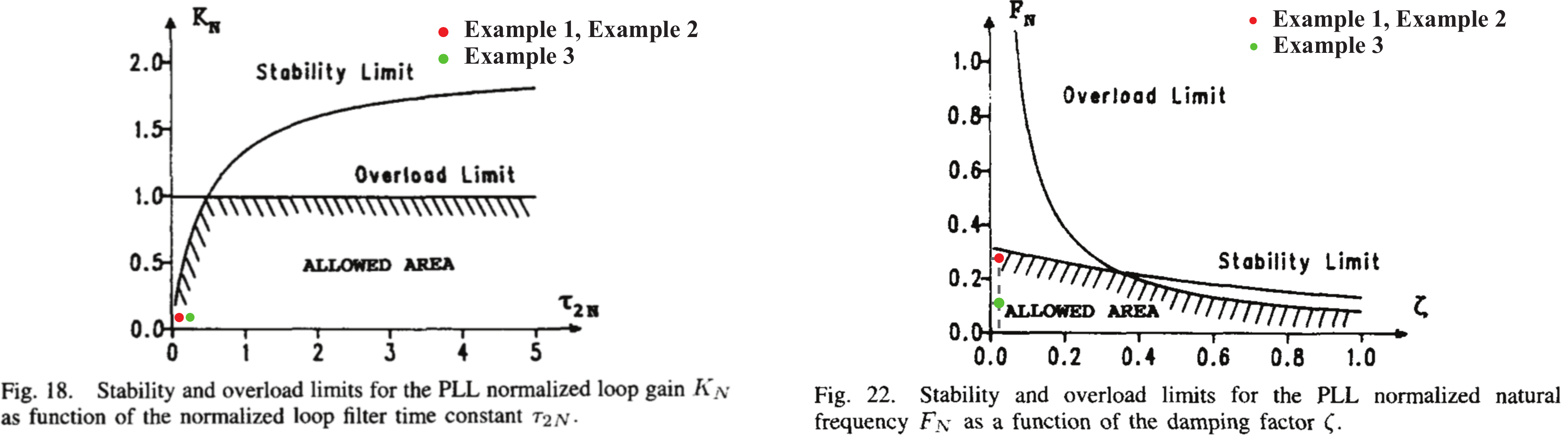}
  \caption{Parameters for Example 1, Example 2, and Example 3 correspond to allowed area (see Fig~18 and Fig.~22 in \cite{Paemel-1994})}
  \label{fig:allowed-domain}
\end{figure*}
\begin{equation}
\label{allowed domain eq}
  \begin{aligned}
    & F_N < \frac{\sqrt{1+\zeta^2}-\zeta}{\pi} \approx 0.3138,
    \\
    & F_N < \frac{1}{4\pi\zeta} \approx 5.6438. 
  \end{aligned}
\end{equation}
\begin{figure}[h]
  \centering
  \includegraphics[width=0.9\linewidth]{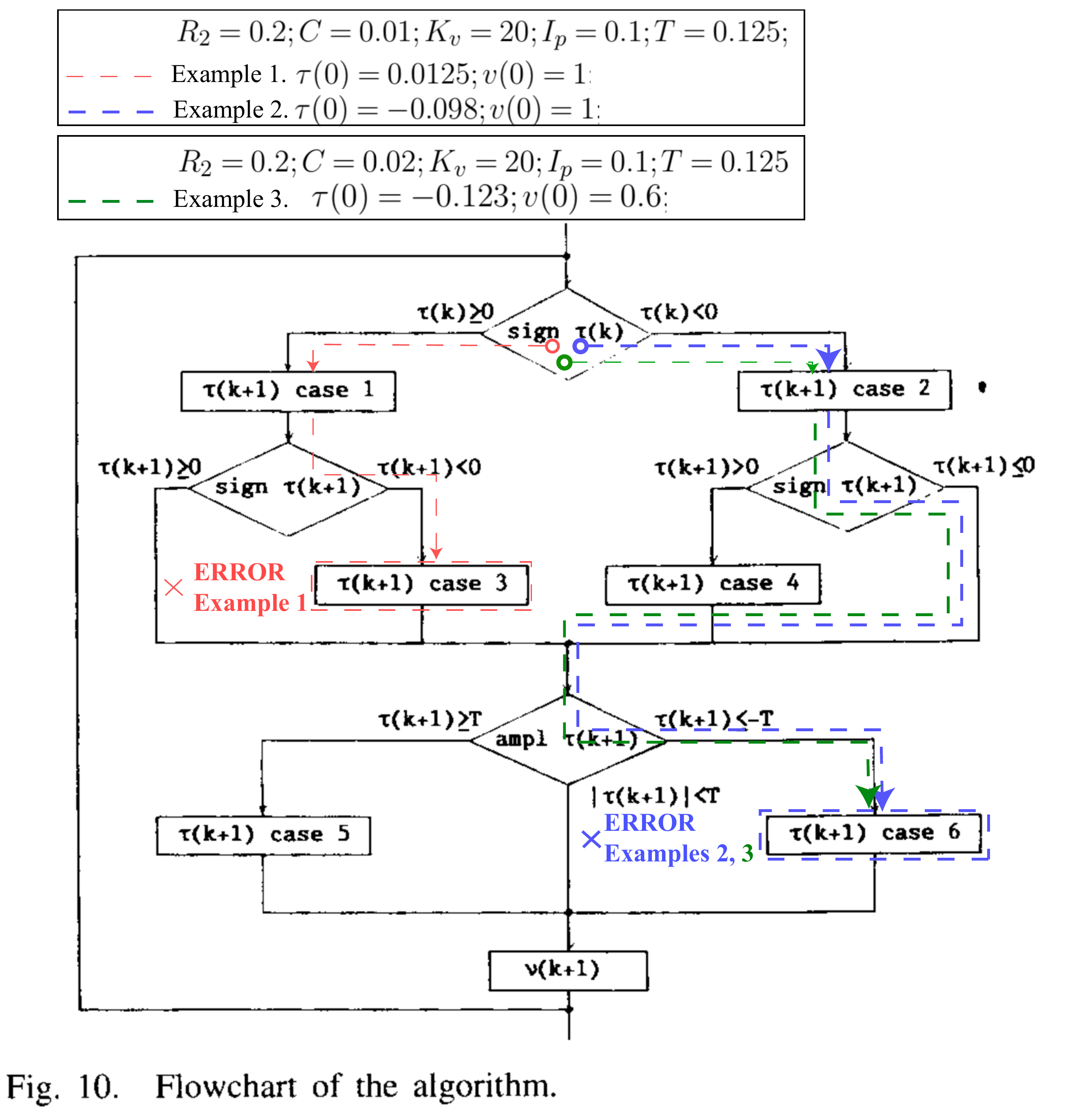}
  \caption{Demonstration of Example 1, Example 2, and Example 3 in the flowchart of the algorithm (see Fig.~10 in \cite{Paemel-1994})}
  \label{fig:paemel-flowchart}
\end{figure}
Now we use the flowchart in Fig.~\ref{fig:paemel-flowchart} (Fig.~10 in \cite{Paemel-1994}) to compute $\tau(1)$ and $v(1)$: since $\tau(0) > 0$ and $\tau(0) < T$,
we proceed to \emph{case 1)} and corresponding relation for $\tau(k+1)$ (equation (7) in \cite{Paemel-1994}):
\begin{equation}
  \label{paemel-case-1}
  \begin{aligned}
    & \tau(k+1)= \frac{-I_pR_2-v(k)}{\frac{I_p}{C}}+
    \\
    & + \frac{\sqrt{(I_p R_2 + v(k))^2 - \frac{2I_p}{C}(v(k)(T - \tau(k)) - \frac{1}{K_v})}}{\frac{I_p}{C}}.
  \end{aligned}
\end{equation}
However, the expression under the square root in \eqref{paemel-case-1} is negative:
\begin{equation}
\begin{aligned}
  & (I_p R_2 + v(0))^2 - \frac{2I_p}{C}(v(0)(T - \tau(0)) - \frac{1}{K_v}) = -0.2096 < 0.
\end{aligned}
\end{equation}
Therefore the algorithm is terminated with error. 

From \eqref{ex1} for our model we have:
\begin{equation}
\label{ex1 correct}
\begin{aligned}
  & \omega_{\rm vco}^{\rm free} = 0,
  \\
  & c =
     (T - (\tau(0)\ {\rm mod}\ T))
     K_v v(0) - 1 = 1.2500,\\
  & \tau(1) = \frac{1}{K_v v(0)}
- T + (\tau(0) \text{ mod } T) = -0.0625,
  \\
  & v(1) = v(0)+\frac{I_p}{C}\tau(1) = 0.3750.
\end{aligned}
\end{equation}
\subsection{Example 2}
Consider the same parameters as in Example 1, but $\tau(0) = -0.098$:
\begin{equation}
\label{ex2}
\begin{aligned}
&
R_2 = 0.2;
C = 0.01;
K_v = 20;
I_p = 0.1;
T = 0.125;\\
& \tau(0) = -0.098;
\quad v(0) = 1.
\end{aligned}
\end{equation}
In this case \eqref{F_n zeta}, \eqref{allowed domain eq}, and Fig.~\ref{fig:allowed-domain} are the same as in Example 1, i.e. we are in the ``allowed area''.
Now we compute $\tau(1)$ and $v(1)$  following the flowchart in Fig.~\ref{fig:paemel-flowchart}: since $\tau(0) < 0$
we proceed to \emph{case 2)} and corresponding equation of $\tau(k+1)$ (equation (9) in \cite{Paemel-1994}):
\begin{equation}
  \label{paemel-case-2}
  \begin{aligned}
    & \tau(1) 
      = \frac{\frac{1}{K_v}-I_pR_2\tau(0)-\frac{I_p\tau(0)^2}{2C}}{v(0)}
        -T+\tau(0) 
      = -0.21906,\\
    & -0.2191 < -T = -0.125.
  \end{aligned}
\end{equation}
This fact indicates cycle-slipping (out of lock).
According to the flowchart in Fig.~\ref{fig:paemel-flowchart} (see Fig.~10 in \cite{Paemel-1994}), we should proceed to \emph{case 6)}
and recalculate $\tau(1)$.
First step of case 6) is to calculate $t_1, t_2, t_3,...$ (equations (16) and (17) in \cite{Paemel-1994}):
\begin{equation}
  \label{case-6-n}
  \begin{aligned}
    & t_n = \frac{v_{n-1}-I_pR_2-\sqrt{(v_{n-1}-I_pR_2)^2-2\frac{I_p}{C}\cdot \frac{1}{K_v}}}{\frac{I_p}{C}},\\
    &v_n = v_{n-1}-\frac{I_p}{C}t_n,\\
    &v_0 = v(k-1).
  \end{aligned}
\end{equation}
Since $k = 0$, then
\begin{equation}
  \begin{aligned}
    & t_1 = \frac{v_{0}-I_pR_2-\sqrt{(v_{0}-I_pR_2)^2-2\frac{I_p}{C}\cdot \frac{1}{K_v}}}{\frac{I_p}{C}},\\
    &v_1 = v_{0}-\frac{I_p}{C}t_1,\\
    &v_0 = v(-1).
  \end{aligned}
\end{equation}
However, $v(-1)$ doesn't make sense and algorithm terminates with error.
Even if we suppose that it is a typo and $v_0 = v(0)$, then relation under the square root become negative:
\begin{equation}
\begin{aligned}
  & (v(0) - I_p R_2)^2 - 2\frac{I_p}{CK_v} = -0.0396 < 0.
\end{aligned}
\end{equation}
In both cases the algorithm is terminated with error. 
Note, that modification of case 2) corresponding to VCO overload 
(equation (35) in \cite{Paemel-1994}) can not be applied here, 
since $v(0) > I_p R_2$ (no overload) and $v(1)$ is not computed yet because of the error.

Corrected model successfully detects overload without any error:
\begin{equation}
 \begin{aligned}
   & v(1)+\frac{\omega_{\rm vco}^{\rm free}}{K_{\rm vco}}-I_p R_2 \approx -0.2106 < 0
   \\
 \end{aligned}
 \end{equation}

\subsection{Example 3}
Consider parameters:
\begin{equation}
\label{ex3}
\begin{aligned}
& \tau(0) = -0.123;
\quad v(0) = 0.6,\\
&
R_2 = 0.2;
C = 0.02;
K_v = 20;
I_p = 0.1;
T = 0.125.
\end{aligned}
\end{equation}
Similar to \eqref{F_n zeta} and \eqref{allowed domain eq}
\begin{equation}
\begin{aligned}
  & K_N = 0.05,
   \quad \tau_{2N} = 0.032,
  \\
  & F_N \approx 0.1989,
    \quad  \zeta = 0.02,
\end{aligned}
\end{equation}
\begin{equation}
\label{allowed domain eq3}
  \begin{aligned}
    & F_N < \frac{\sqrt{1+\zeta^2}-\zeta}{\pi} \approx 0.3120,
    \\
    & F_N < \frac{1}{4\pi\zeta} \approx 3.9789,
  \end{aligned}
\end{equation}
parameters \eqref{ex3} correspond to allowed area in Fig.~\ref{fig:allowed-domain} (equations (46)-(47), Fig.~18 and Fig.~22 in \cite{Paemel-1994}).

Now we compute $\tau(1)$ and $v(1)$  following the flowchart in Fig.~\ref{fig:paemel-flowchart}: since $\tau(0) < 0$
one proceeds to \emph{case 2)} and corresponding equation for computing $\tau(k+1)$ (equation (9) in \cite{Paemel-1994}):
\begin{equation}
\label{paemel-case-2}
  \begin{aligned}
    & \tau(1) = \frac{\frac{1}{K_v}-I_pR_2\tau(0)-\frac{I_p\tau(0)^2}{2C}}{v(0)}
      -T+\tau(0)
      \approx -0.224,\\
    & -0.224 < -T = -0.125.
  \end{aligned}
\end{equation}
The last inequality indicates cycle-slipping (out of lock).
According to the flowchart in Fig.~\ref{fig:paemel-flowchart} (see Fig.~10 in \cite{Paemel-1994}), one proceeds to \emph{case 6)}
and recalculates $\tau(1)$.
First step of \emph{case 6)} is to calculate $t_1, t_2, t_3,...$ using \eqref{case-6-n} (see equations (16) and (17) in \cite{Paemel-1994}) until $t_1+t_2+\ldots+t_n>|\tau(0)|$.
Even if we suppose $v(-1)=v(0)-\frac{I_p}{C}\tau(0)$, we get
\begin{equation}
  \begin{aligned}
    & t_1 = 0.0463,\ v_1 = 1.215;\\
    & t_2 = 0.0618,\ v_2 = 0.983;\\
    & t_1+t_2=0.1081<|\tau(0)|=0.123.
  \end{aligned}
\end{equation}
However, $t_3$ can not be computed, because the relation under the square root in \eqref{case-6-n} is negative:
\begin{equation}
  \begin{aligned}
    & (v_2-I_pR_2)^2-2\frac{I_p}{C}\cdot \frac{1}{K_v} \approx -0.0726.
  \end{aligned}
\end{equation}

Corrected model gives 
\begin{equation}
\begin{aligned}
  & \tau(1) = -0.0569,
  \quad v(1) = 0.3153.
  \\
\end{aligned}
\end{equation}






\section{Mathematical simplification of the discrete model of CP-PLL}
Follow the ideas from \cite{acco2003etude,Orla-2013-review},
the number of parameters in \eqref{complete-model} can be reduced
to just two ($\alpha$ and $\beta$)
\begin{equation}
\label{alpha-beta-eqs}
  \begin{aligned}
    & \tau^{\alpha\beta}_k = \frac{\tau_k}{T_{\rm ref}},\\
    & \omega^{\alpha\beta}_k=T_{\rm ref}
      \left(
        \omega_{\rm vco}^{\text{free}} + K_{\rm vco}v_k
      \right) - 1,\\
    & \alpha = K_{\rm vco}I_pT_{\rm ref}R,\\
    & \beta = \frac{K_{\rm vco}I_pT_{\rm ref}^2}{2C}.
  \end{aligned}
\end{equation}
Then
\begin{equation}
  \label{vk-tauk-system}
  \begin{aligned}
    & \omega^{\alpha\beta}_{k+1} =  \omega^{\alpha\beta}_k +2\beta\tau^{\alpha\beta}_{k+1}(\tau^{\alpha\beta}_k,\omega^{\alpha\beta}_k),
    \\
    &
     k = 0,1,2,...
  \end{aligned}
\end{equation}
with initial conditions $\tau^{\alpha\beta}(0)$ and $\omega^{\alpha\beta}(0)$.
Function $\tau^{\alpha\beta}_{k+1}(\tau^{\alpha\beta}_k,\omega^{\alpha\beta}_k) $ is defined by the following equation
\begin{equation}
\label{complete-model-ab}
  \begin{aligned}
    & \tau^{\alpha\beta}_{k+1}(\tau^{\alpha\beta}_k,\omega^{\alpha\beta}_k) =
    \\
    &
    = \left\{\begin{array}{l}
      \frac{-b_n + \sqrt{b_n^2 - 4\beta c}}{2\beta},
      \ \quad \tau^{\alpha\beta}_k \geq 0, \quad c \leq 0,
      \\
      \\
      \frac{1}{\omega^{\alpha\beta}_k + 1}
      -1 + (\tau^{\alpha\beta}_k\text{ mod }1),
      \\
      \quad\quad\quad\quad\quad\quad\quad
       \tau^{\alpha\beta}_k \geq 0, \quad c > 0,
      \\
      \\
      l_{bn}-1,
      \quad\quad\quad
       \tau^{\alpha\beta}_k < 0, \quad l_{bn} \leq 1,
      \\
      \\
      \frac{-b_n + \sqrt{b_n^2 - 4\beta d}}{2\beta},
      \quad \tau^{\alpha\beta}_k < 0, \quad l_{bn} > 1,
    \end{array}
    \right.
    \\
    &
      b_n = b_n(\omega^{\alpha\beta}_k) = \omega^{\alpha\beta}_k
    + \alpha + 1,
    \\
    &
      c = c(\tau^{\alpha\beta}_k,\omega^{\alpha\beta}_k) =
      \\
    & \quad (1 - (\tau^{\alpha\beta}_k\text{ mod }1))
(\omega^{\alpha\beta}_k +1) - 1,
    \\
    & l_{bn} = l_{bn}(\tau^{\alpha\beta}_k,\omega^{\alpha\beta}_k) = \frac{1 - (S_{l_k}\text{ mod }1)}{\omega^{\alpha\beta}_k + 1},\\
    & S_{l_k} = S_{lk}(\tau^{\alpha\beta}_k,\omega^{\alpha\beta}_k) =
    \\
    & \quad
    -\left(
      \omega^{\alpha\beta}_k - \alpha + 1
    \right)\tau^{\alpha\beta}_k
    + \beta\tau^2_n(k),\\
    & d = d(\tau^{\alpha\beta}_k,\omega^{\alpha\beta}_k) = (S_{l_k}\text{ mod 1}) + \omega^{\alpha\beta}_k.
  \end{aligned}
\end{equation}
This change of variables allows to reduce parameter space
and simplify design of PFD PLLs.

Moreover, checking requirement \eqref{locked-state-req}
for all $k$ is impossible in practice during simulation.
Instead one can check that values of
$\omega_{\alpha\beta}$ and $\tau_{\alpha\beta}$ are small,
which indicate
that frequencies of VCO and input signal are close
and trailing edges happen almost at the same time.
It means that the loop is in a locked state.

\subsection{Locked states and periodic solutions}
By \eqref{alpha-beta-eqs}, there is only one stationary point
\begin{equation}
\begin{aligned}
  & \omega^{\alpha\beta}_k = \omega^{\alpha\beta}_{k+1} \equiv 0,
  \\
  & \tau^{\alpha\beta}_k = \tau^{\alpha\beta}_{k+1} \equiv 0.
\end{aligned}
\end{equation}
which is a locked state if it is locally stable.
Period-$2$ points
\begin{equation}
\begin{aligned}
  & \omega^{\alpha\beta}_{k+2} = \omega^{\alpha\beta}_k,
  \\
  & \tau^{\alpha\beta}_{k+2} = \tau^{\alpha\beta}_k,
\end{aligned}
\end{equation}
also can be found by \eqref{alpha-beta-eqs}:
\begin{equation}
\label{eq-periodic}
\begin{aligned}
  & \omega^{\alpha\beta}{k+2} =
  \omega^{\alpha\beta}_{k+1} +2\beta\tau^{\alpha\beta}_{k+2},
  \\
  & =
  \omega^{\alpha\beta}_k + 2\beta\tau^{\alpha\beta}_{k+2} + 2\beta\tau^{\alpha\beta}_{k+1},
  \\
  & \tau^{\alpha\beta}_{k+2} = -\tau^{\alpha\beta}_{k+1}.
\end{aligned}
\end{equation}
From \eqref{eq-periodic} period-2 points correspond to the alternating between Case 2 and Case 4.
In general, period-$P$ points satisfy equation
\begin{equation}
\begin{aligned}
  & \sum\limits_{i = k+1}^{k+P} \tau_i = 0.
  \\
\end{aligned}
\end{equation}
By \eqref{complete-model-ab} it is possible to obtain
exact relations for all periodic solutions.

\section{Comparison of Simulink vs V.Paemel's model vs Corrected model}
Correctness of proposed model was verified by extensive simulation
in Matlab Simulink.
Circuit level model in Matlab Simulink was compared with original
model by V.~Paemel and proposed model.
Full code also can be found in Github repository \url{https://github.com/mir/PFD-simulation}
Here we consider only few examples.
Based on simulation for this set of parameters
all three models produce almost the same results.
\begin{figure*}
  \centering
  \includegraphics[width=1\linewidth]{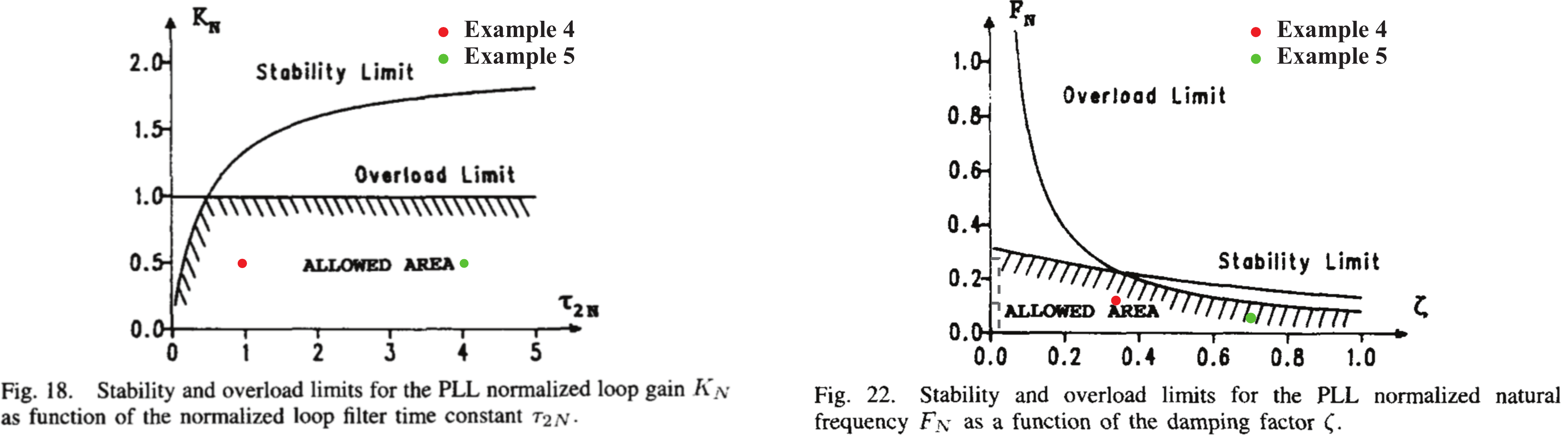}
  \caption{Parameters for Example 5 and Example 6 correspond to allowed area (see Fig~18 and Fig.~22 in \cite{Paemel-1994})}
  \label{fig:allowed-domain}
\end{figure*}

\subsection{Example 5}
\begin{equation}
\label{Simulink parameters ex5}
\begin{aligned}
& \tau(0) = 0;
\quad v(0) = 10,\\
&
R_2 = 1000;
C = 10^{-6};
K_v = 500;
I_p = 10^{-3};
T = 10^{-3}.
\\
& \tau_{2N} = 1;
K_N = 0.5;
F_N = 0.1125;
\zeta =0.3536.
\end{aligned}
\end{equation}
\begin{figure}[H]
  \centering
  \includegraphics[width=1\linewidth]{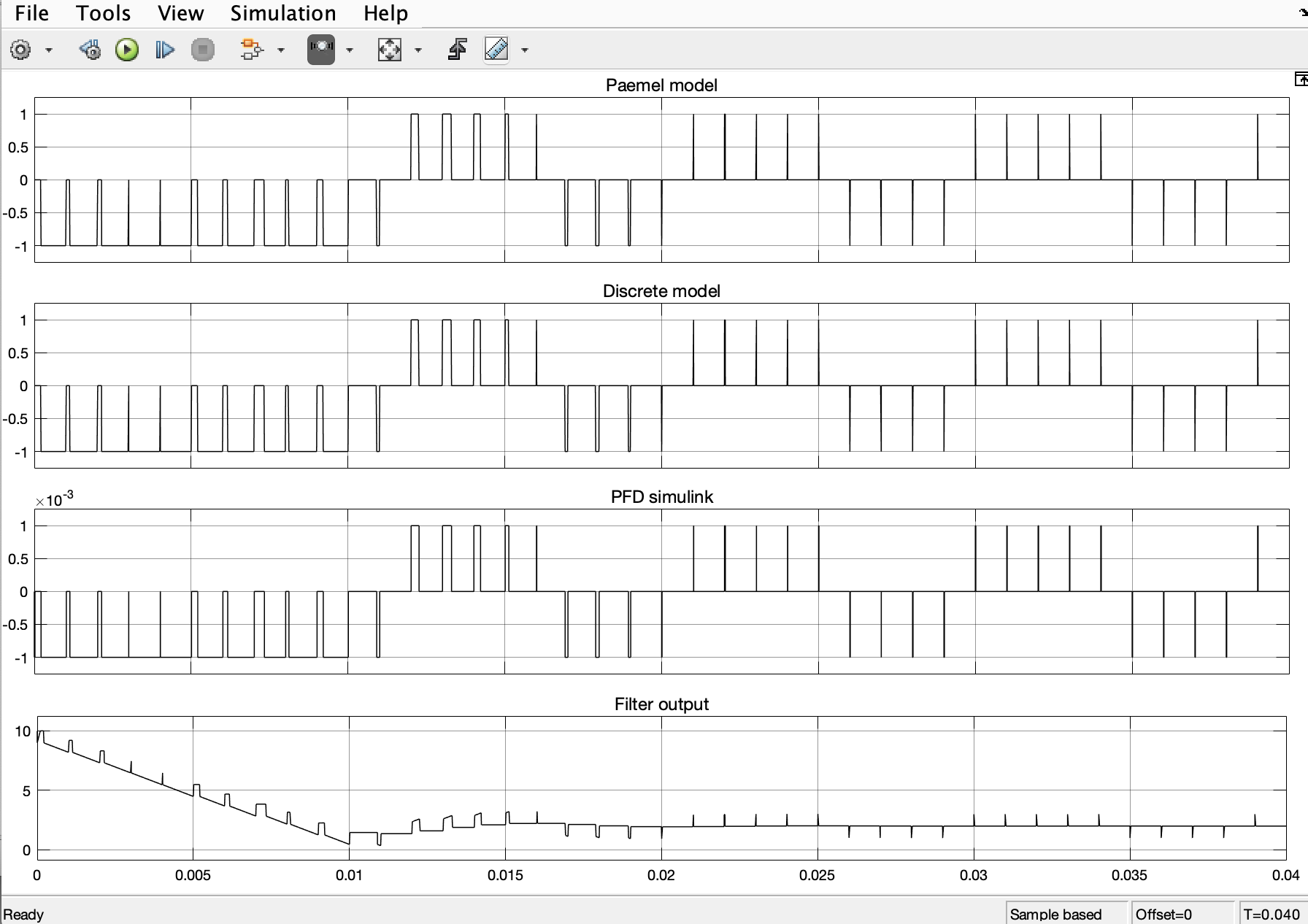}
  \caption{Comparison of PFD outputs of Simulink model (PFD Simulink) vs V.Paemel's model (Paemel model) vs Corrected model (Discrete model).
  Lower subfigure demonstrates output of Loop filter.
  For considered set of parameters 
  ($\tau(0) = 0;
  v(0) = 10;
  R_2 = 1000;
  C = 10^{-6};
  K_v = 500;
  I_p = 10^{-3};
  T = 10^{-3};
  \tau_{2N} = 1;
  K_N = 0.5;
  F_N = 0.1125;
  \zeta =0.3536$)
  all three models produce almost the same results.}
 \label{fig:sim vs paemel vs correcter}
\end{figure}

\subsection{Example 6}
\begin{equation}
\label{Simulink parameters ex6}
\begin{aligned}
& \tau(0) = 0;
v(0) = 100,\\
&
R_2 = 1000;
C = 4\cdot10^{-6};
K_v = 500;
I_p = 10^{-3};
T = 10^{-3}.
\\
& \tau_{2N} = 4;
K_N = 0.5;
F_N = 0.0563;
\zeta =0.7017.
\end{aligned}
\end{equation}

\begin{figure}[H]
  \centering
  \includegraphics[width=1\linewidth]{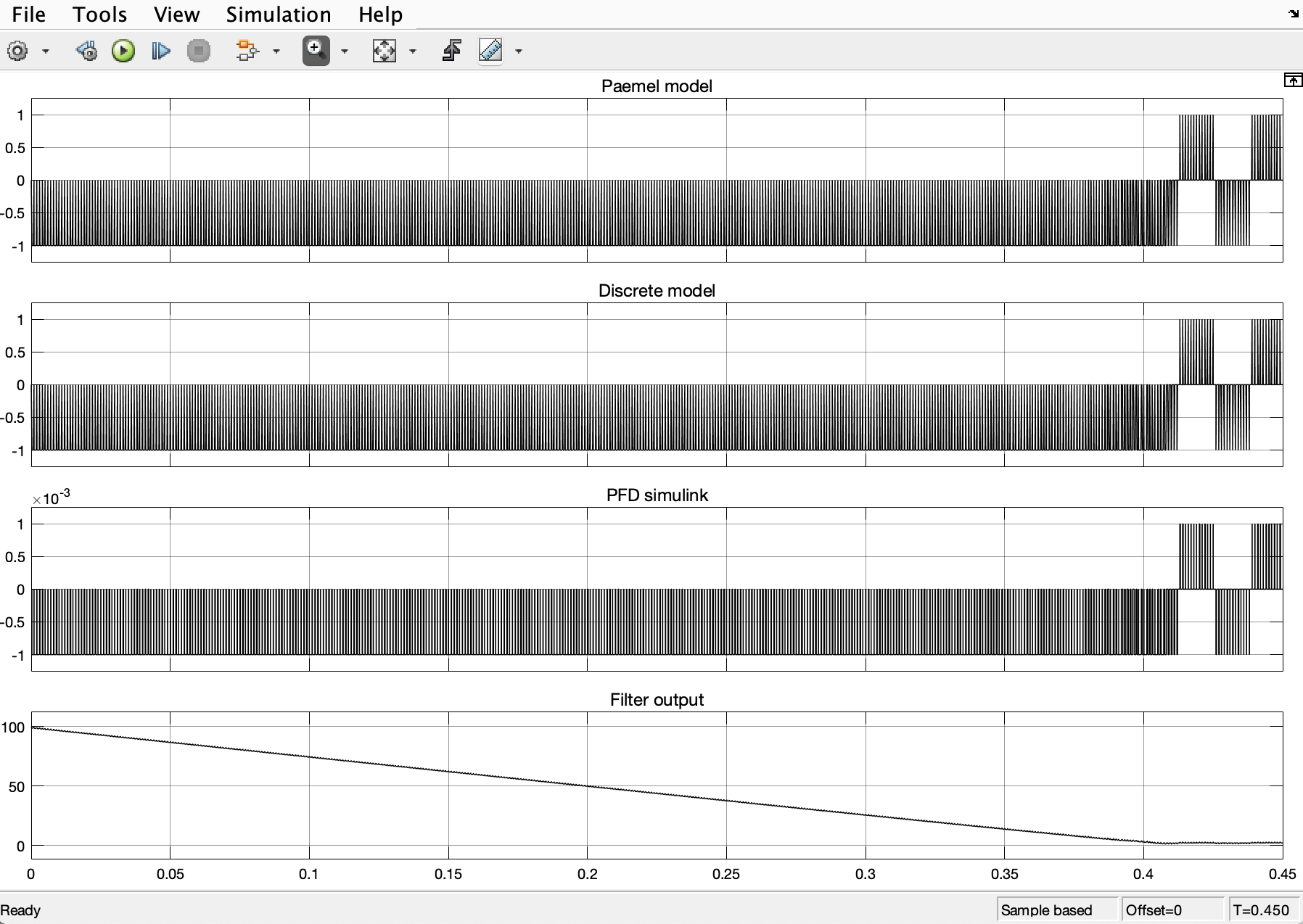}
  \caption{Comparison of PFD outputs of Simulink model (PFD Simulink) vs V.Paemel's model (Paemel model) vs Corrected model (Discrete model).
  Lower subfigure demonstrates output of Loop filter.
  For considered set of parameters 
  ($\tau(0) = 0;
  v(0) = 100;
  R_2 = 1000;
  C = 4\cdot10^{-6};
  K_v = 500;
  I_p = 10^{-3};
  T = 10^{-3};
  \tau_{2N} = 4;
  K_N = 0.5;
  F_N = 0.0563;
  \zeta =0.7071$)
  all three models produce almost the same results.}
 \label{fig:sim vs paemel vs correcter}
\end{figure}

\section{Algorithm for VCO overload}
\label{sec:vco overload}
If filter output $v_F(t)$ is small enough
then VCO equation \eqref{vco first} should be modified\cite{van1994analysis}
to satisfy \eqref{vco overload}:
\begin{equation}
\begin{aligned}
  & \dot\theta_{\rm vco}(t) = \left\{
    \begin{array}{ll}
      \omega_{\rm vco}^{\rm free} + K_{\rm vco} v_F(t), & v_F(t) > - \frac{\omega_{\rm vco}^{\rm free}}{K_{\rm vco}}\\
      0, & v_F(t) \leq - \frac{\omega_{\rm vco}^{\rm free}}{K_{\rm vco}}
    \end{array}
    \right.
  \\
\end{aligned}
\end{equation}
At step $k$ VCO overload can be detected by checking the following conditions:
\begin{equation}
\label{overload eq}
  \begin{aligned}
    &\tau_k>0 \text{ and }
    v_k+\frac{\omega_{\rm vco}^{\rm free}}{K_{\rm vco}}-\frac{I_p}{C}\tau_k<0,\\
    & \tau_k<0 \text{ and }
    v_k+\frac{\omega_{\rm vco}^{\rm free}}{K_{\rm vco}}-I_p R<0.\\
  \end{aligned}
\end{equation}
Following Cases describe how to modify \eqref{complete-model} to take into account VCO overload.
\begin{itemize}
  \item {\bf Case O1}.
   $\tau_k < 0$, $K_{\rm vco}v_k+\omega_{\rm vco}^{\rm free}>0$, $\tau_{k+1}<0$
  \item {\bf Case O2}. 
   $\tau_k < 0$, $K_{\rm vco}v_k+\omega_{\rm vco}^{\rm free}>0$, $\tau_{k+1} \geq 0$
  \item {\bf Case O3}. 
   $\tau_k < 0$, $K_{\rm vco}v_k+\omega_{\rm vco}^{\rm free} \leq 0$, $v_k+\frac{\omega_{\rm vco}^{\rm free}}{K_{\rm VCO}}+I_pR<0$
  \item {\bf Case O4}. 
   $\tau_k<0$, $K_{\rm vco}v_k+\omega_{\rm vco}^{\rm free}\leq0$, $v_k+\frac{\omega_{\rm vco}^{\rm free}}{K_{\rm VCO}}+I_pR \geq 0$
  \item {\bf Case O5}. $\tau_k \geq 0$, $K_{\rm vco}v_k+\omega_{\rm vco}^{\rm free}\leq0$, $\tau_{k+1} > 0$
  \item {\bf Case O5}*. Impossible. $\tau_k \geq 0$, $K_{\rm vco}v_k+\omega_{\rm vco}^{\rm free}\leq0$, $\tau_{k+1} < 0$.
  \item {\bf Case O6}. $\tau_k \geq 0$, $K_{\rm vco}v_k+\omega_{\rm vco}^{\rm free} > 0$, $\tau_{k+1} > 0$,
  \item {\bf Case O7}.  $\tau_k \geq 0$, $K_{\rm vco}v_k+\omega_{\rm vco}^{\rm free} > 0$, $\tau_{k+1} \leq 0$,
\end{itemize}

\subsection*{{\bf Case O1}. $\tau_k<0$,
$K_{\rm vco}v_k+\omega_{\rm vco}^{\rm free}>0$, $\tau_{k+1}<0$ (equivalently $l_b < T_{\rm ref}$)}
\begin{figure}[H]
  \centering
    \includegraphics[width=\linewidth]{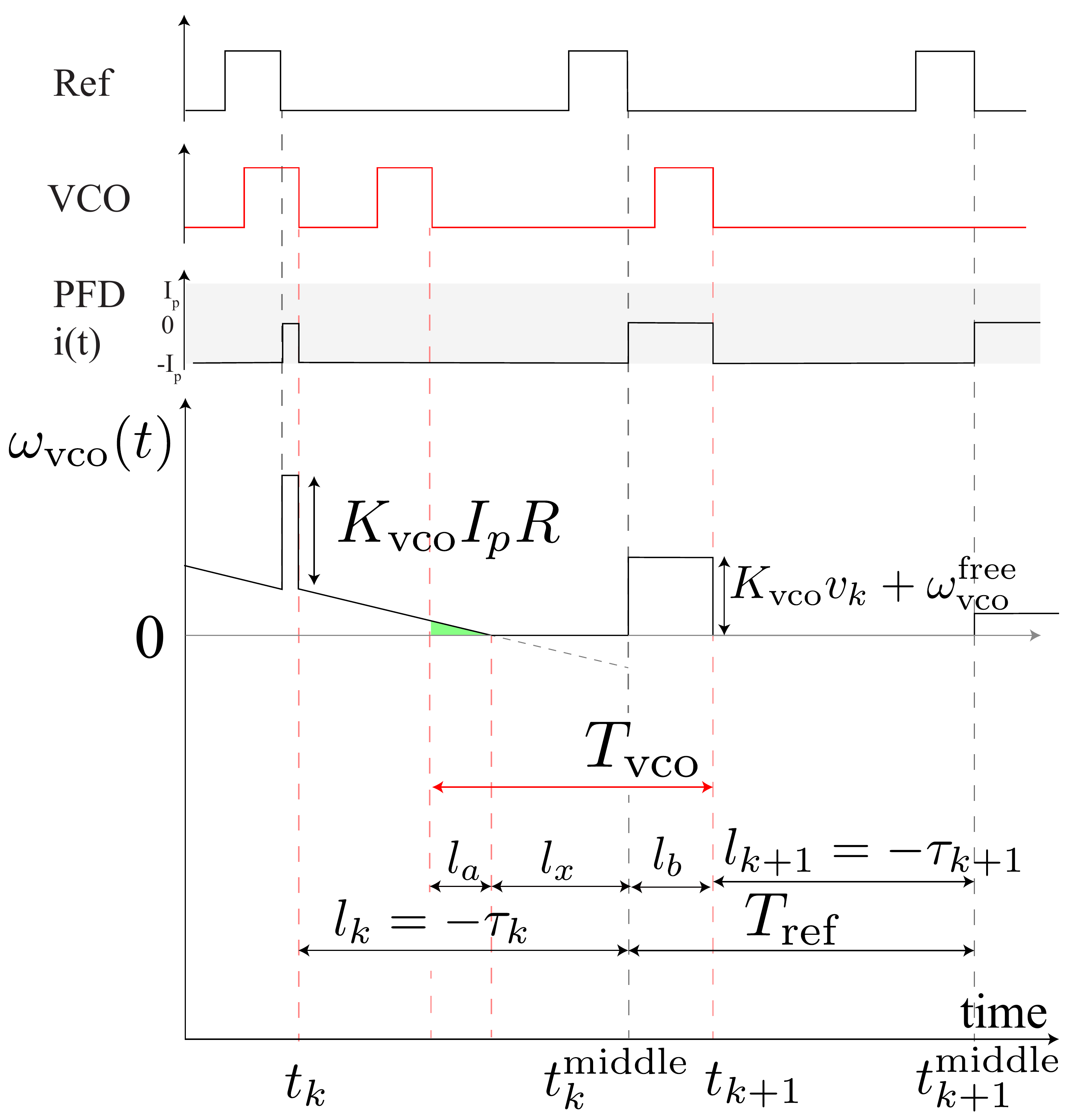}
    \caption{VCO overload case O1. $\tau_k<0$,
$K_{\rm vco}v_k+\omega_{\rm vco}^{\rm free}>0$, $\tau_{k+1}<0$ (equivalently $l_b < T_{\rm ref}$)}
    \label{fig case_overload_1}
\end{figure}
Note, that values $\tau_k$ and $v_k$ can be correctly computed by \eqref{complete-model}. 

In order to compute $\tau_{k+1}$ taking into account VCO overload, one can compute phase of VCO before it's frequency hit zero.
The VCO phase at that moment corresponds to the area $S_{l_a}$ of the green triangle (see Fig.~\ref{fig case_overload_1}).
In order to find the triangle area one can find $l_x$ --- time interval  corresponding to zero VCO frequency:
\begin{equation}
  \begin{aligned}
    & l_x =\min \bigg\{
           -\frac{C}{I_p}(
             v_k
             +\frac{\omega_{\rm vco}^{\rm free}}{K_{\rm vco}}
             -I_pR
          ),
           -\tau_k
    \bigg\}.\\
  \end{aligned}
\end{equation}
Since the triangle in question ($S_{la}$) is part of the larger tringle $S$,
we get
\begin{equation}
\label{overload sla}
  \begin{aligned}
    & S = K_{\rm vco}(\tau_k+l_x)^2\frac{I_p}{2C},\\
    & S_{la}=S\mod 1.\\
  \end{aligned}
\end{equation}

Condition $\tau_{k+1}<0$ means that VCO edge triggered change of PFD state from 0 to $-I_p$.
Since phase change between two consequetive falling edges of VCO is $1$, we get
\begin{equation}
\begin{aligned}
  & S_{la}+l_b(K_{\rm vco}v_k+\omega_{\rm vco}^{\rm free})=1,\\
  & l_b = \frac{1-S_{la}}{K_{\rm vco}v_k+\omega_{\rm vco}^{\rm free}}.\\ 
  \\
\end{aligned}
\end{equation}
where $l_b$ is time during which PFD was in zero state.
If $l_b < T_{\rm ref}$, then
\begin{equation}
  \begin{aligned}
    & \tau_{k+1} = -(T_{\rm ref} - l_b).
  \end{aligned}
\end{equation}
If $l_b \geq T_{\rm ref}$
then we should proceed to the next case (O2).

\subsection*{{\bf Case O2}. $\tau_k<0$, $K_{\rm vco}v_k+\omega_{\rm vco}^{\rm free}>0$, $\tau_{k+1} \geq 0$ (equivalently $l_b \geq T_{\rm ref}$)}
\begin{figure}[H]
  \centering
    \includegraphics[width=\linewidth]{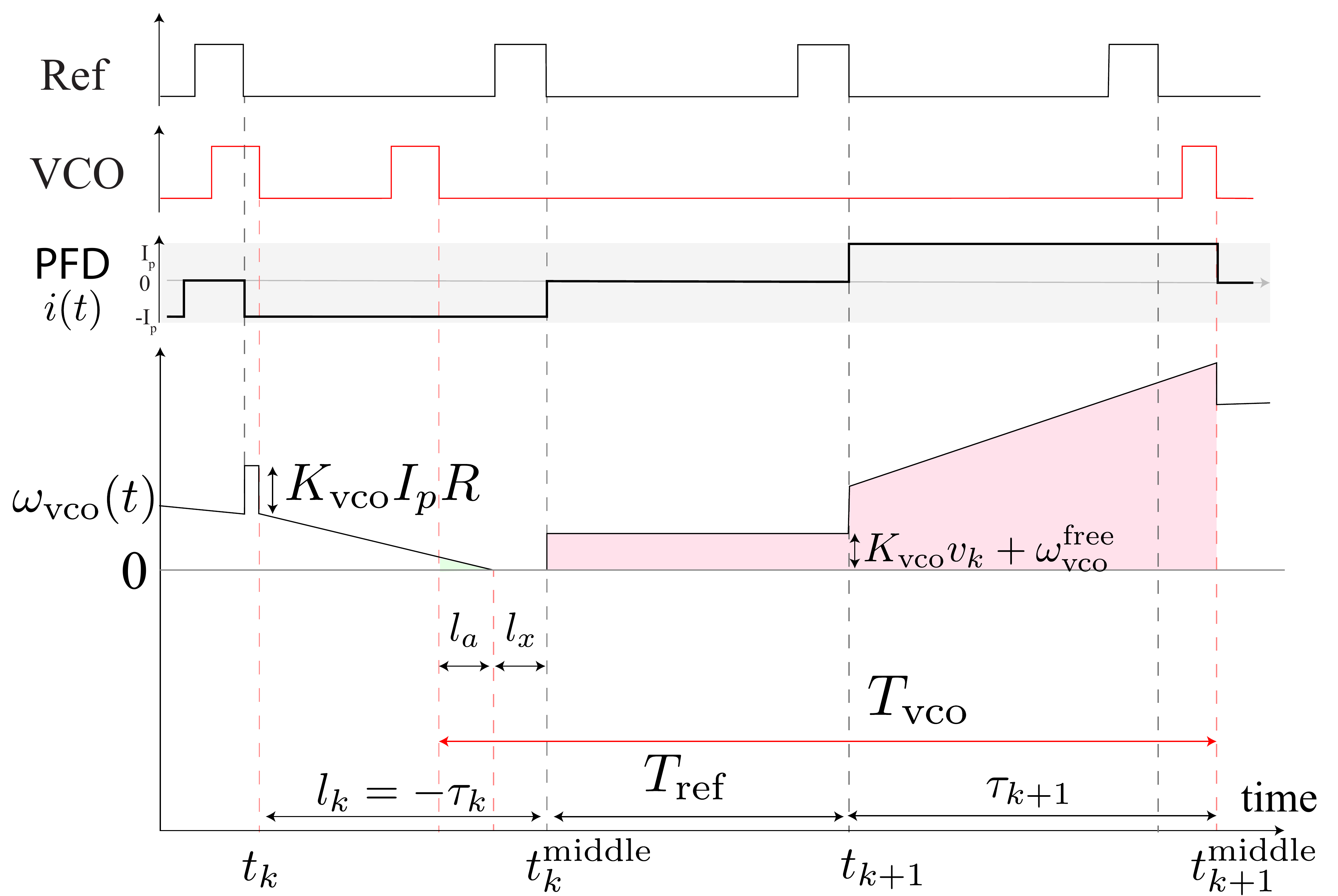}
    \caption{VCO overload case O2. $\tau_k<0$, $K_{\rm vco}v_k+\omega_{\rm vco}^{\rm free}>0$, $\tau_{k+1} \geq 0$ (equivalently $l_b \geq T_{\rm ref}$)}
    \label{fig case_overload_2}
\end{figure}
Value of $S_{la}$ is computed by \eqref{overload sla}.
Now we can compute $\tau_{k+1}$ (see Fig.~\ref{fig case_overload_2}) similarly to {\bf case 4}, \eqref{case4 solution}
\begin{equation}
\label{case4 overload solution}
\begin{aligned}
& \tau_{k+1} = \frac{-b + \sqrt{b^2 - 4ad}}{2a},
\\
& a = \frac{K_{\rm vco}I_p}{2C},
\\
&
  b = b(v_k) = \omega_{\rm vco}^{\text{free}} + K_{\rm vco}v_k
    + K_{\rm vco}I_pR,
\\
&
  d = b(\tau_k,v_k) =S_{l_a}+S_{T_{\rm ref}} - 1,
  \\
  & S_{T_{\rm ref}} = S_{T_{\rm ref}}(v_k) = T_{\rm ref}(K_{\rm vco}v_k+\omega_{\rm vco}^{\text{free}}),\\
\end{aligned}
\end{equation}


\subsection*{{\bf Case O3}. $\tau_k<0$, $K_{\rm vco}v_k+\omega_{\rm vco}^{\rm free}\leq0$, $v_k+\frac{\omega_{\rm vco}^{\rm free}}{K_{\rm VCO}}+I_pR<0$}
\begin{figure}[H]
  \centering
    \includegraphics[width=\linewidth]{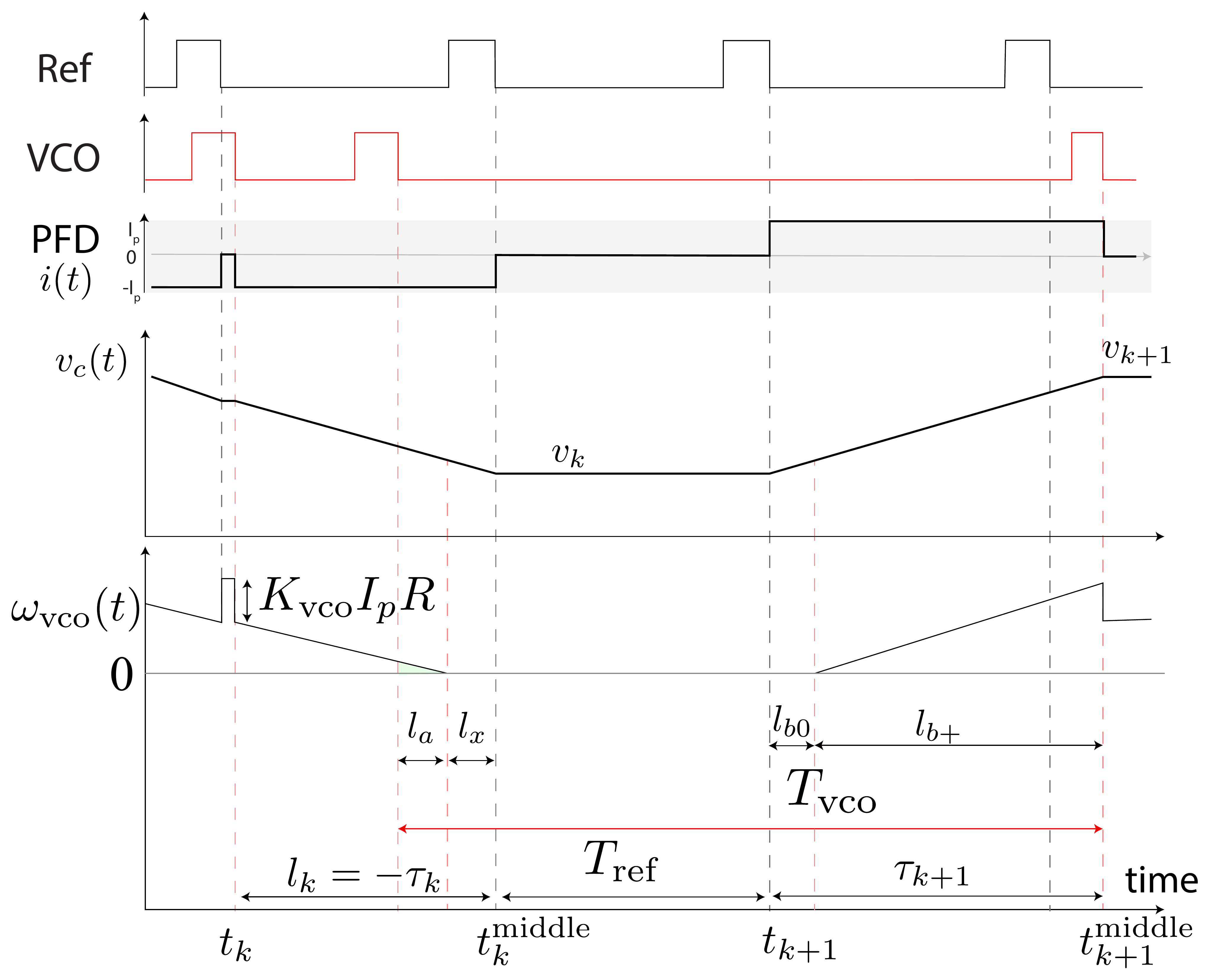}
    \caption{Overload limit, VCO cycle slipping; $\tau_k < 0, \quad \tau_{k+1} < 0$}
    \label{fig case_overload_3}
\end{figure}
Consider timing diagram on Fig.~\ref{fig case_overload_3}.
One can split interval $\tau_{k+1}$ into subintervals $l_{b0}$, $l_{b+}$ such that PFD output is zero on $l_{b0}$ and positive on $l_{b+}$:
\begin{equation}
\begin{aligned}
  & \tau_{k+1} = l_{b0} + l_{b+},\\
  & \omega_{\rm vco}^{\rm free} 
    + K_{\rm vco}(I_pR + v_k + \frac{I_p}{C}l_{b0}) 
    = 0,
  \\
  &
    l_{b0} = \frac{C}{I_p}\left(
      -\frac{\omega_{\rm vco}^{\rm free}}{K_{\rm vco}}
      -I_pR
      -v_k
      \right).
  \\
\end{aligned}
\end{equation}
Since phase difference between two consequetive falling edges of VCO is $1$, we get
\begin{equation}
\begin{aligned}
  & S_{la} + S_{lb+} = 1,
  \\
  & S_{lb+} = \frac{K_{\rm vco}I_p}{2C} l_{b+}^2,
  \\
  & l_{b+} = \sqrt{(1-S_{la})\frac{2C}{K_{\rm vco}I_p}},
  \\
\end{aligned}
\end{equation}
where $S_{la}$ and $S_{lb}$ are areas of corresponding triangles (phase difference of VCO for corresponding time intervals).

\subsection*{{\bf Case O4}. $\tau_k<0$, $K_{\rm vco}v_k+\omega_{\rm vco}^{\rm free}\leq0$, $v_k+\frac{\omega_{\rm vco}^{\rm free}}{K_{\rm VCO}}+I_pR \geq 0$}
\begin{figure}[H]
  \centering
    \includegraphics[width=\linewidth]{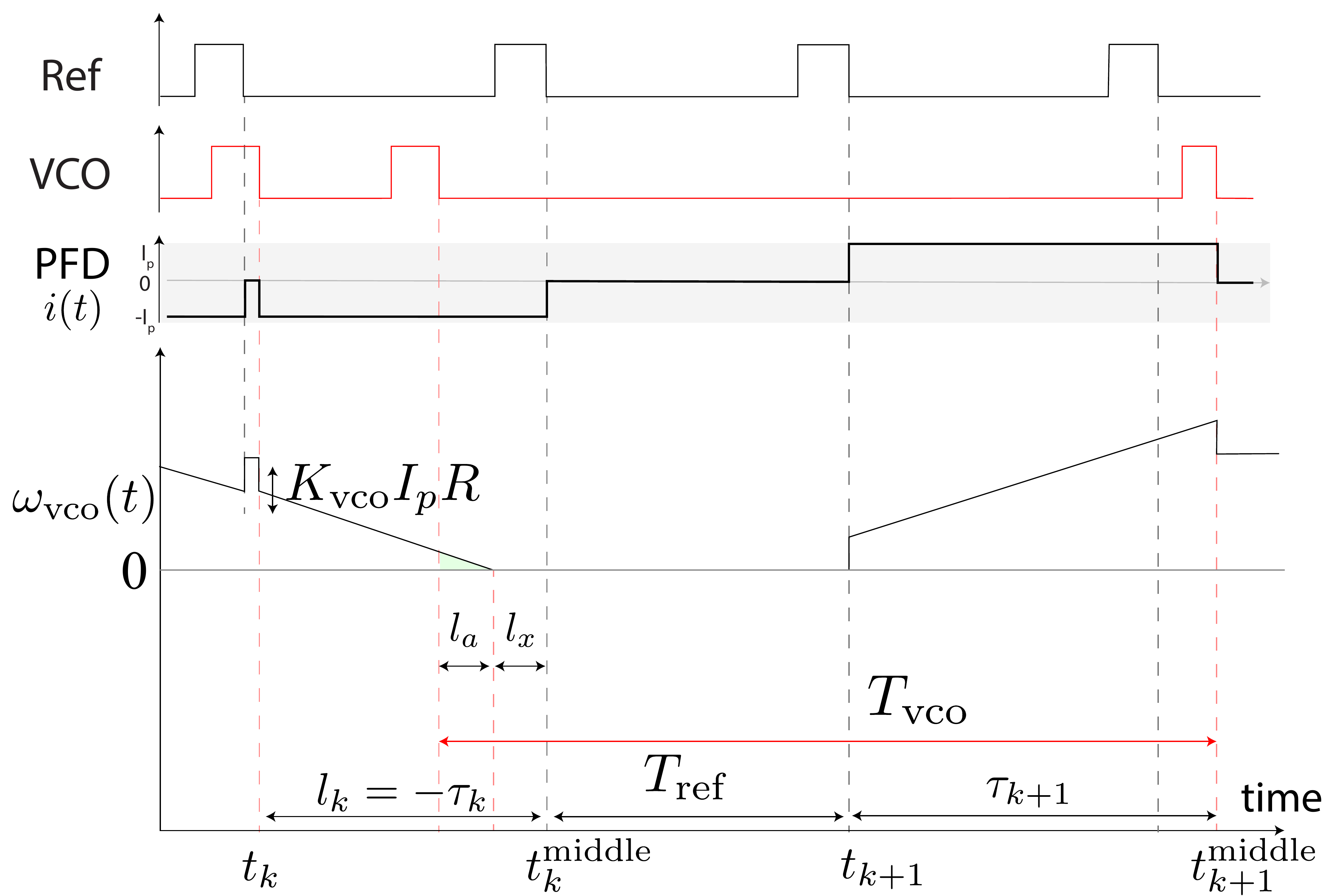}
    \caption{VCO overload case O4.$\tau_k<0$, $K_{\rm vco}v_k+\omega_{\rm vco}^{\rm free}\leq0$, $v_k+\frac{\omega_{\rm vco}^{\rm free}}{K_{\rm VCO}}+I_pR \geq 0$}
    \label{fig case_overload_4}
\end{figure}
Consider timing diagram on Fig.~\ref{fig case_overload_4}.
Since phase difference between two consequetive falling edges of VCO is $1$, we get
\begin{equation}
\begin{aligned}
  & S_{la} + S_{\tau_{k+1}} = 1,
  \\
\end{aligned}
\end{equation}
where $S_{la}$ can be computed using \eqref{overload sla}, and $S_{\tau_{k+1}}$ is phase of VCO corresponding to time interval $\tau_{k+1}$.
Then
\begin{equation}
\begin{aligned}
  & S_{\tau_{k+1}} = \frac{K_{\rm vco}I_p}{2C} \tau_{k+1}^2 
  + \tau_{k+1}(\omega_{\rm vco}^{\rm free}+K_{\rm vco}(v_k+I_pR)).
  \\
\end{aligned}
\end{equation}
Finally,
\begin{equation}
\begin{aligned}
& \tau_{k+1} = \frac{-b + \sqrt{b^2 - 4ad_{o}}}{2a},
\\
& a = \frac{K_{\rm vco}I_p}{2C},
\\
&
  b = \omega_{\rm vco}^{\text{free}} + K_{\rm vco}v_k
    + K_{\rm vco}I_pR,
\\
&
  d_{o} = -S_{\tau_{k+1}}.
\\
\end{aligned}
\end{equation}

\subsection*{{\bf Case O5}. $\tau_k \geq 0$, $K_{\rm vco}v_k+\omega_{\rm vco}^{\rm free}\leq0$, $\tau_{k+1} > 0$}
\begin{figure}[H]
  \centering
    \includegraphics[width=\linewidth]{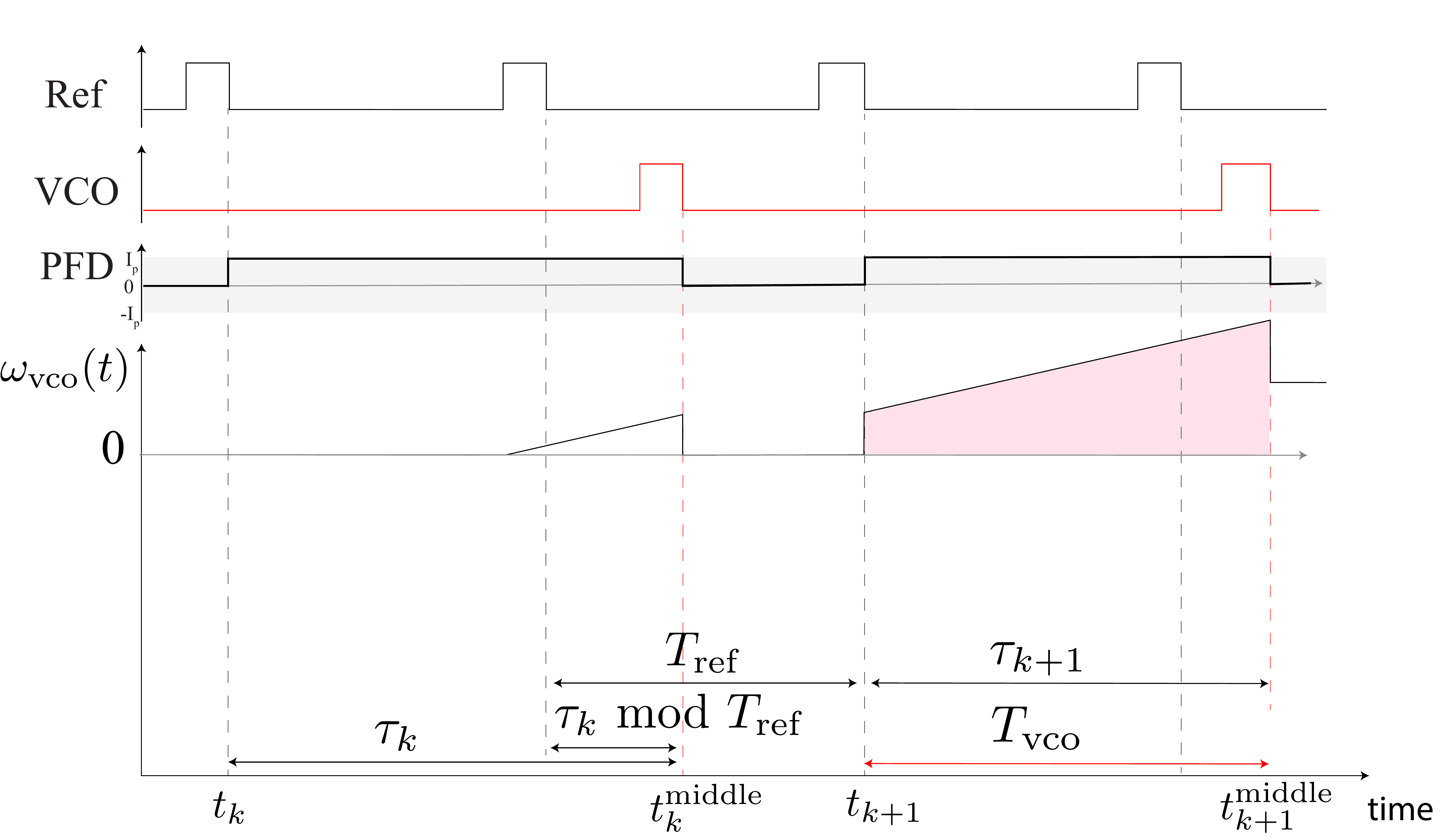}
    \caption{VCO overload case O5. $\tau_k > 0, \quad \tau_{k+1} > 0$}
    \label{fig case_overload_5}
\end{figure}
Consider timing diagram on Fig.~\ref{fig case_overload_5}.
Since phase difference between two consequetive falling edges of VCO is $1$, we get
\begin{equation}
  \begin{aligned}
    & \tau_{k+1}(\omega_{\rm vco}^{\rm free}+K_{\rm vco}(v_k+I_pR))
    +\tau_{k+1}^2\frac{K_{\rm vco}I_p}{2C}=1
  \end{aligned}
\end{equation}
Therefore
\begin{equation}
\begin{aligned}
& \tau_{k+1} = \frac{-b + \sqrt{b^2 + 4a}}{2a},
\\
& a = \frac{K_{\rm vco}I_p}{2C},
\\
&
  b = \omega_{\rm vco}^{\text{free}} + K_{\rm vco}v_k
    + K_{\rm vco}I_pR.
\\
\end{aligned}
\end{equation}

\subsection*{{\bf Case O6}. $\tau_k \geq 0$, $K_{\rm vco}v_k+\omega_{\rm vco}^{\rm free} > 0$, $\tau_{k+1} > 0$.}
\label{gz-gz-eq}
\begin{figure}[H]
  \centering
    \includegraphics[width=\linewidth]{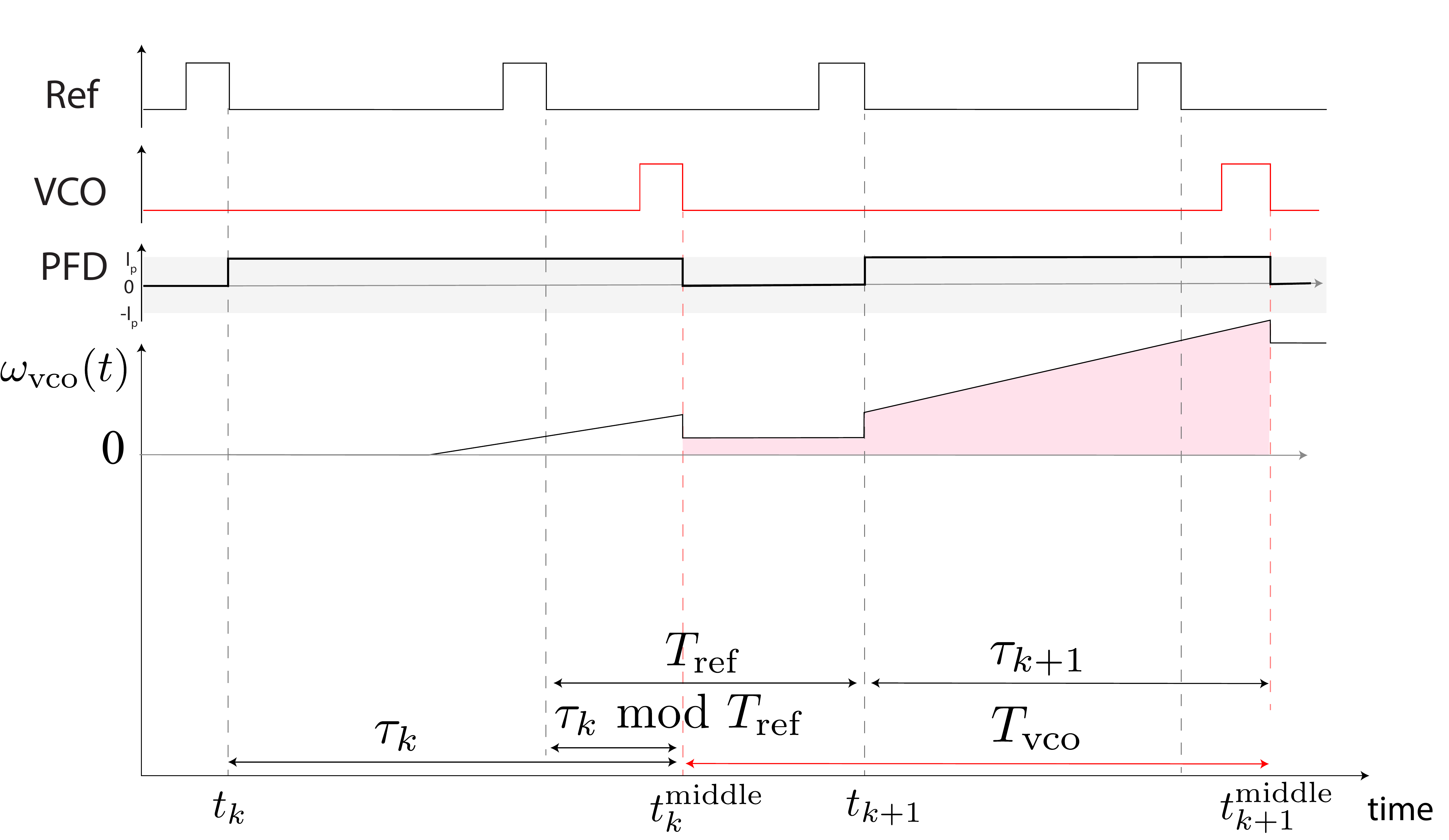}
    \caption{VCO overload case O6. $\tau_k > 0, \quad \tau_{k+1} > 0$}
    \label{fig case_overload_6}
\end{figure}

Consider timing diagram on Fig.~\ref{fig case_overload_6}.
Here $\tau_{k+1}$ can be computed using case 1 (without overload, see Fig.~\ref{fig case1_0}) 

\subsection*{{\bf Case O7}. $\tau_k \geq 0$, $K_{\rm vco}v_k+\omega_{\rm vco}^{\rm free} > 0$, $\tau_{k+1} \leq 0$}
\begin{figure}[H]
  \centering
    \includegraphics[width=\linewidth]{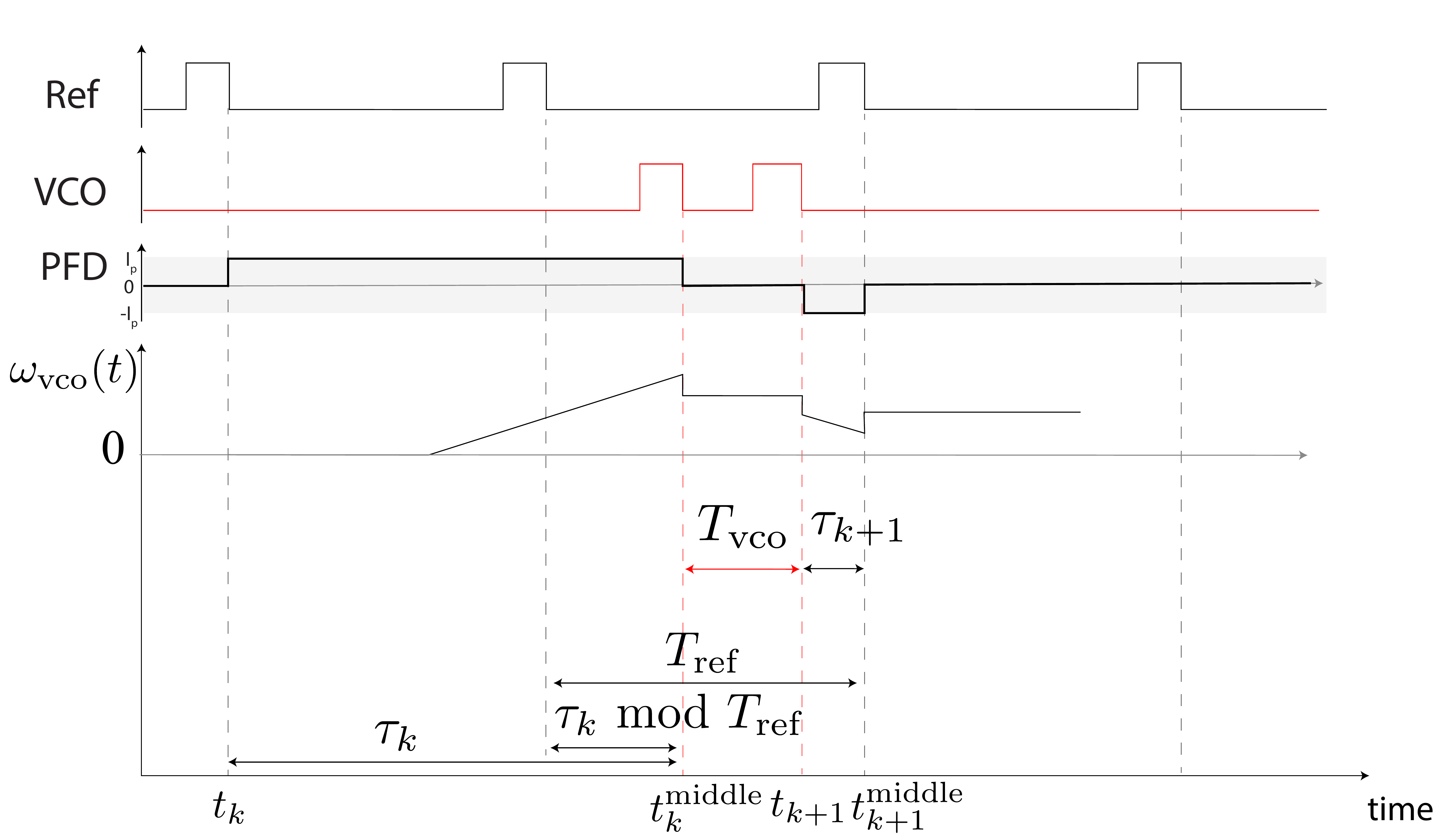}
    \caption{VCO overload case O7. $\tau_k > 0, \quad \tau_{k+1} \leq 0$}
    \label{fig case_overload_7}
\end{figure}

Consider timing diagram on Fig.~\ref{fig case_overload_7}.
Here $\tau_{k+1}$ can be computed using \eqref{case2 solution} from case 2 (without overload)

\section{Simulation of VCO overload model and comparison with Matlab Simulink}
\begin{figure*}
\centering
\includegraphics[width=0.9\linewidth]{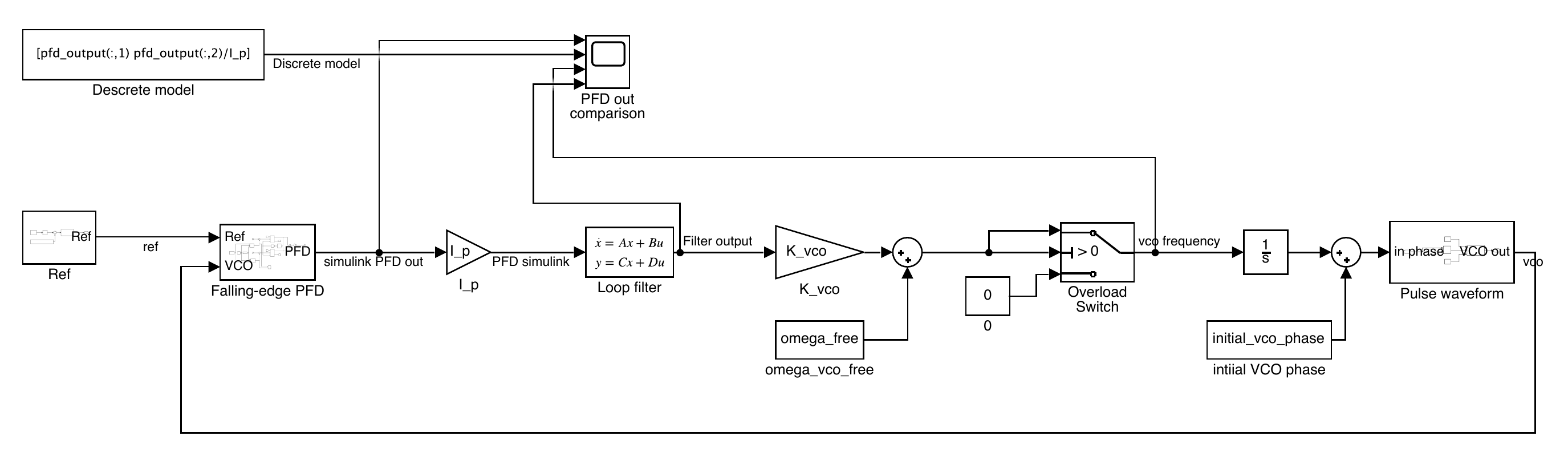}
\caption{Maltab Simulink model of CP-PLL with VCO overload}
\label{fig:cppll-simulink-model}
\end{figure*}

Simulink model is shown in Fig.~\ref{fig:cppll-simulink-model}.
Full code for simulation in Matlab and comparison with simulink model can be found in GitHub repository \url{https://github.com/mir/cppll-overload}
and in Appendix.

\begin{figure}[H]
\centering
\includegraphics[width=0.9\linewidth]{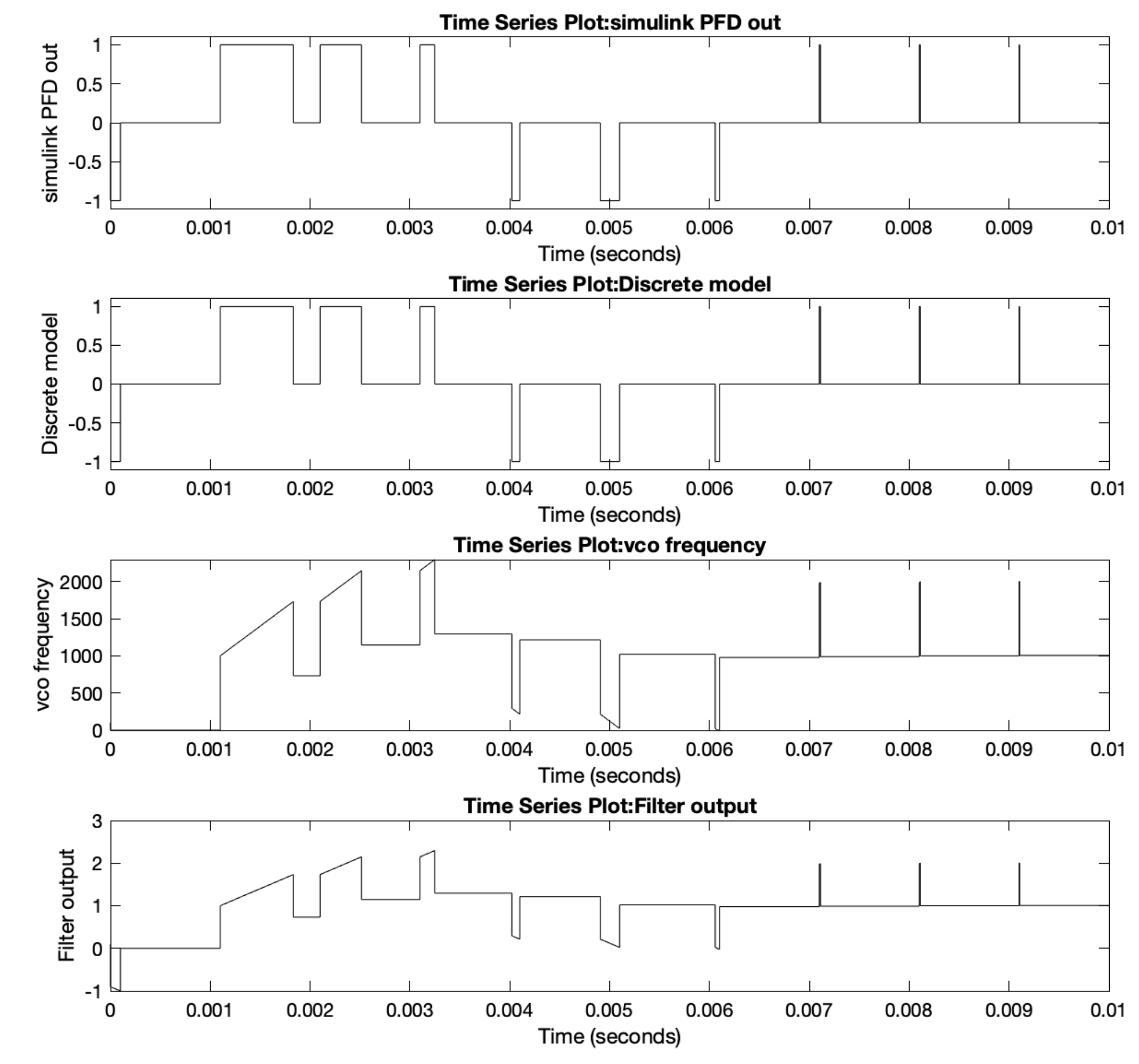}
\caption{Simulation of proposed mathematical model and comparison with Matlab Simulink.
Here {\bf Discrete model} shows output of PFD predicted by proposed mathematical model,
{\bf simulink pfd out},
{\bf vco frequency}, and {\bf Filter output} show results of Simulink simulation.
Parameters: $\omega_{\rm vco}^{\rm free} = 0;
T_{\rm ref} = 10^{-3};
R = 1000;
C = 10^{-6};
K_{\rm vco} = 1000;
I_p = 10^{-3};
v_1 = 0;
\tau_1 = -0.1 \cdot T_{\rm ref}.$}
\label{fig:cppll-overload-simulink-1}
\end{figure}

\begin{figure}[H]
\centering
\includegraphics[width=0.9\linewidth]{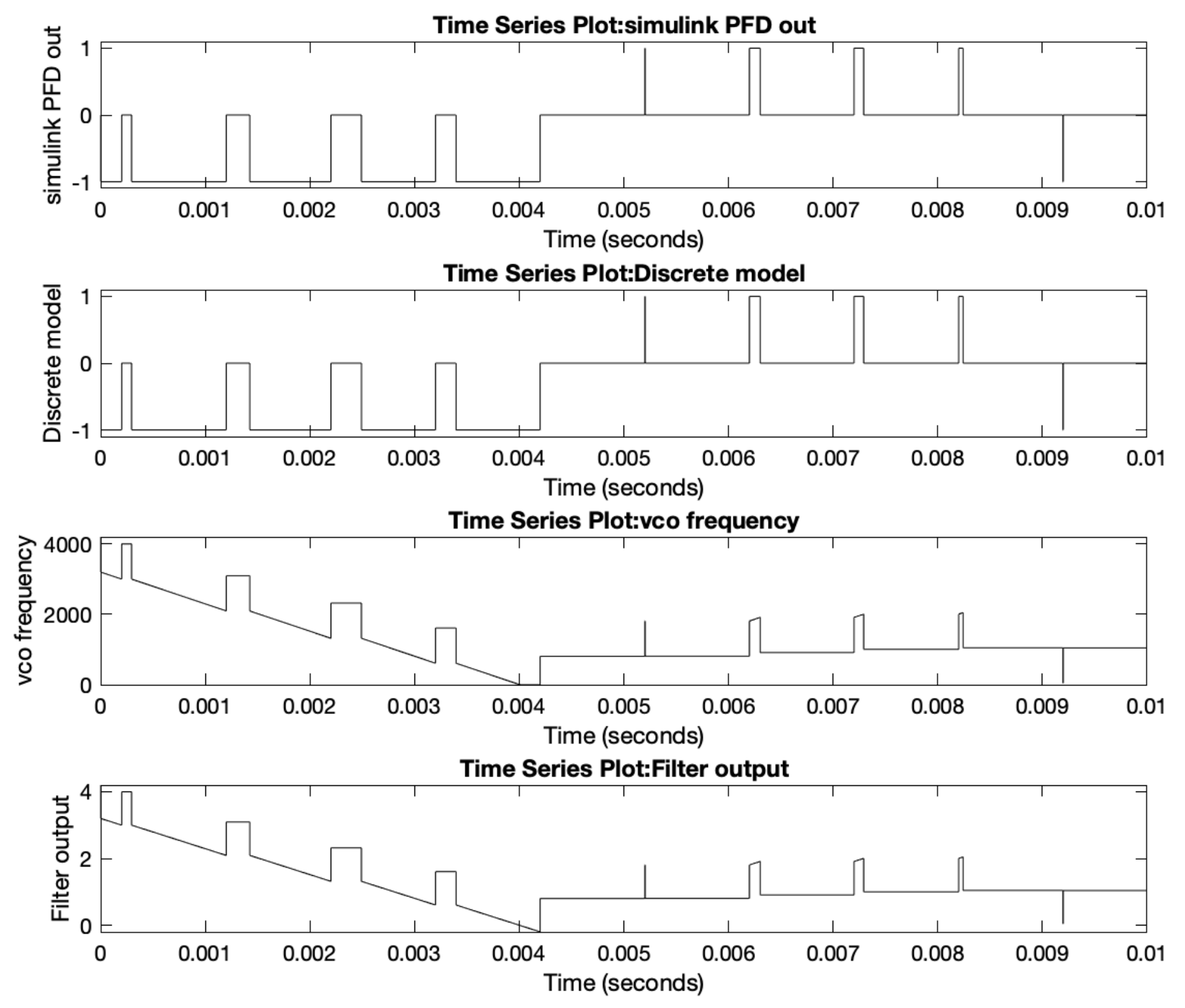}
\caption{Simulation of VCO overload model and comparison with Matlab Simulink. 
Here {\bf Discrete model} shows output of PFD predicted by proposed mathematical model,
{\bf simulink pfd out},
{\bf vco frequency}, and {\bf Filter output} show results of Simulink simulation.
Parameters: $\omega_{\rm vco}^{\rm free} = 0;
T_{\rm ref} = 10^{-3};
R = 1000;
C = 10^{-6};
K_{\rm vco} = 1000;
I_p = 10^{-3};
v_1 = 4;
\tau_1 = -0.2 \cdot T_{\rm ref};$}
\label{fig:cppll-overload-simulink-2}
\end{figure}

It is easy to see that simulation with proposed discrete model \
Simulation confirms mathematical proof and justifies proposed model.

\section*{Conclusion}
  In this paper a non-linear mathematical model of CP-PLL is rigorously derived.
  The obtained model  obviates the shortcomings in
  previously known mathematical models of CP-PLL.
  The VCO overload case initially noted in \cite{Paemel-1994} is extended to take into account new cases.
  Analysis of local stability with respect to coordinate $\tau_k$
  (partial stability, see condition \eqref{locked-state-req})
  gives us the estimation of the hold-in range.
  There is a hypothesis, which has not yet been proven rigorously,
  that the hold-in and pull-in ranges are coincide.
  For some parameters we estimate the pull-in time numerically.
  There were many attempts to generalize equations derived in \cite{Paemel-1994} for higher-order loops (see, e.g.\cite{hanumolu2004analysis,Shakhtarin-2014,bi2011linearized,Hedayat-2014-high-order,milicevic2008time,sancho2004general}), but the resulting transcendental equations can not be solved analytically without using approximations.
  Therefore proposed model is very effective and corresponding numerical
  algorithm is much faster than alternative models (e.g., Matlab Simulink).

\section*{Acknowledgements} \label{sec:acknowledgement}
The work is supported by the Russian Science Foundation project 19-41-02002
and Grant for Leading Scientific Schools of Russia (2018-2019).

Authors would like to thank Mark Van Paemel for valuable comments.
In private communication with Mark Van Paemel it turned out that some of the considered shortcoming
were avoided in FORTRAN code that he used for the simulation of the circuit.
Also Mark Van Paemel emphasized importance of VCO overload and motivated us to
develope corresponding algorithm.

\balance

\section*{Appendix: Simulink model}
\begin{figure}[H]
\centering
  \includegraphics[width=\linewidth]{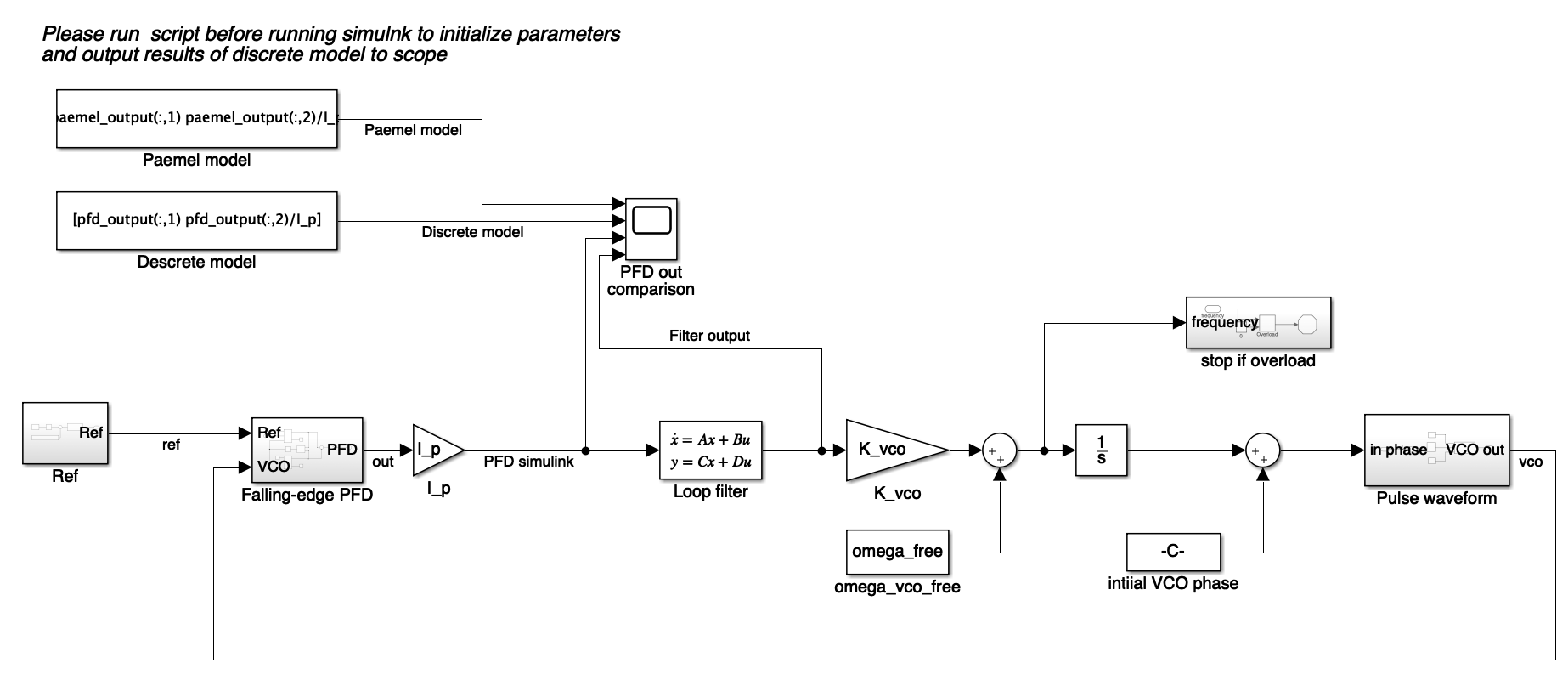}
  \caption{
  Simulink model of CP-PLL 
  }
  \label{fig:Simulink-full}
\end{figure}

\begin{figure}[H]
\centering
  \includegraphics[width=\linewidth]{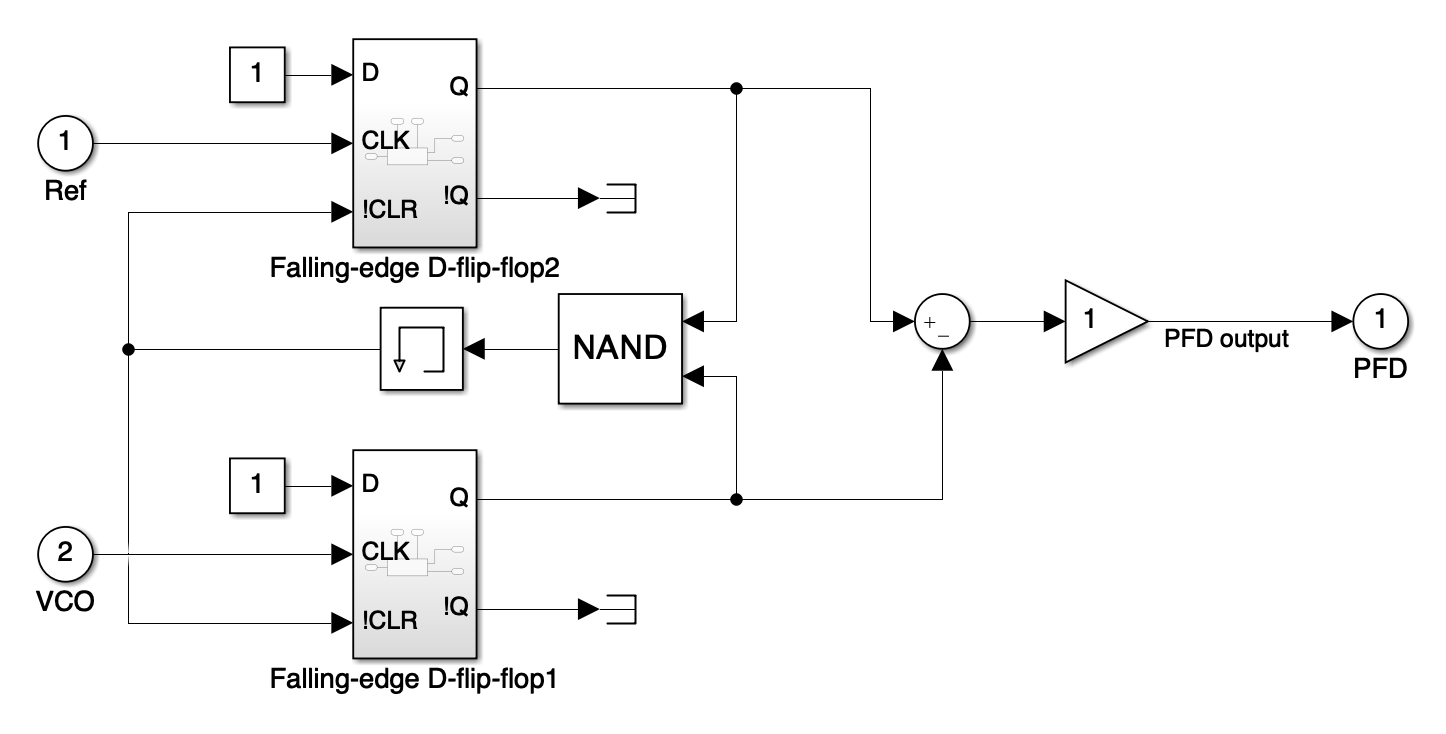}
  \caption{
  Simulink model of PFD 
  }
  \label{fig:Simulink-PFD}
\end{figure}

\begin{figure}[H]
\centering
  \includegraphics[width=\linewidth]{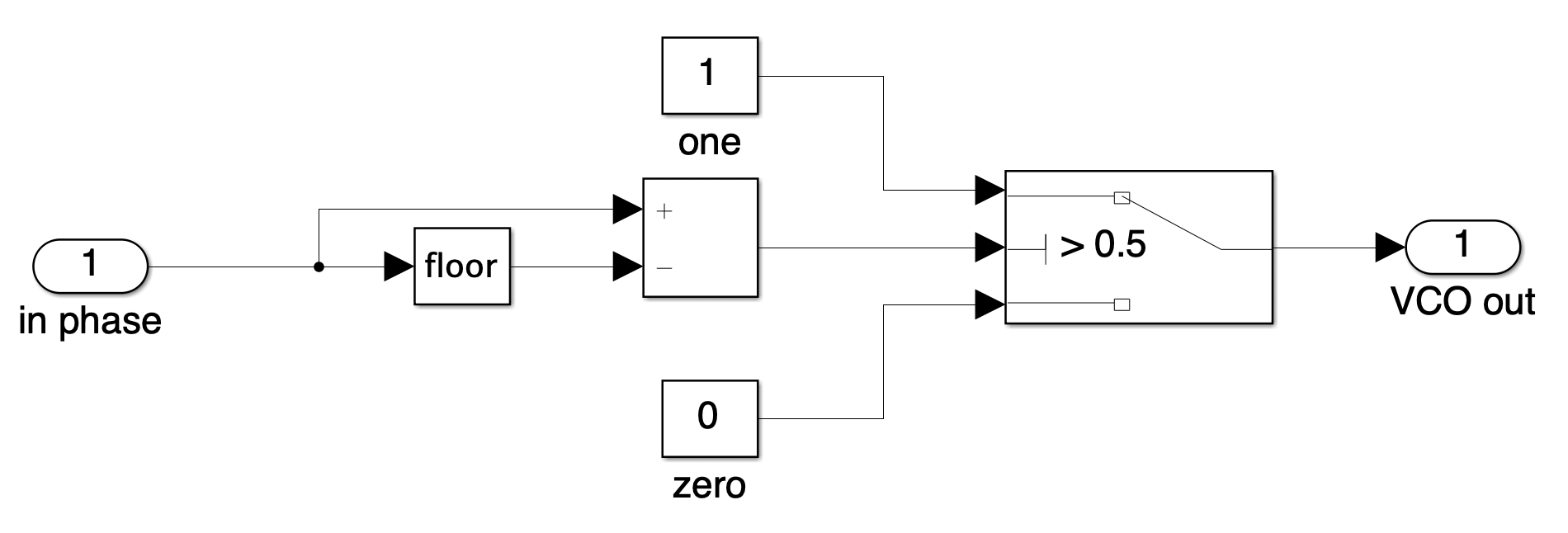}
  \caption{
  Simulink model of VCO 
  }
  \label{fig:Simulink-VCO}
\end{figure}

\begin{figure}[H]
\centering
  \includegraphics[width=0.8\linewidth]{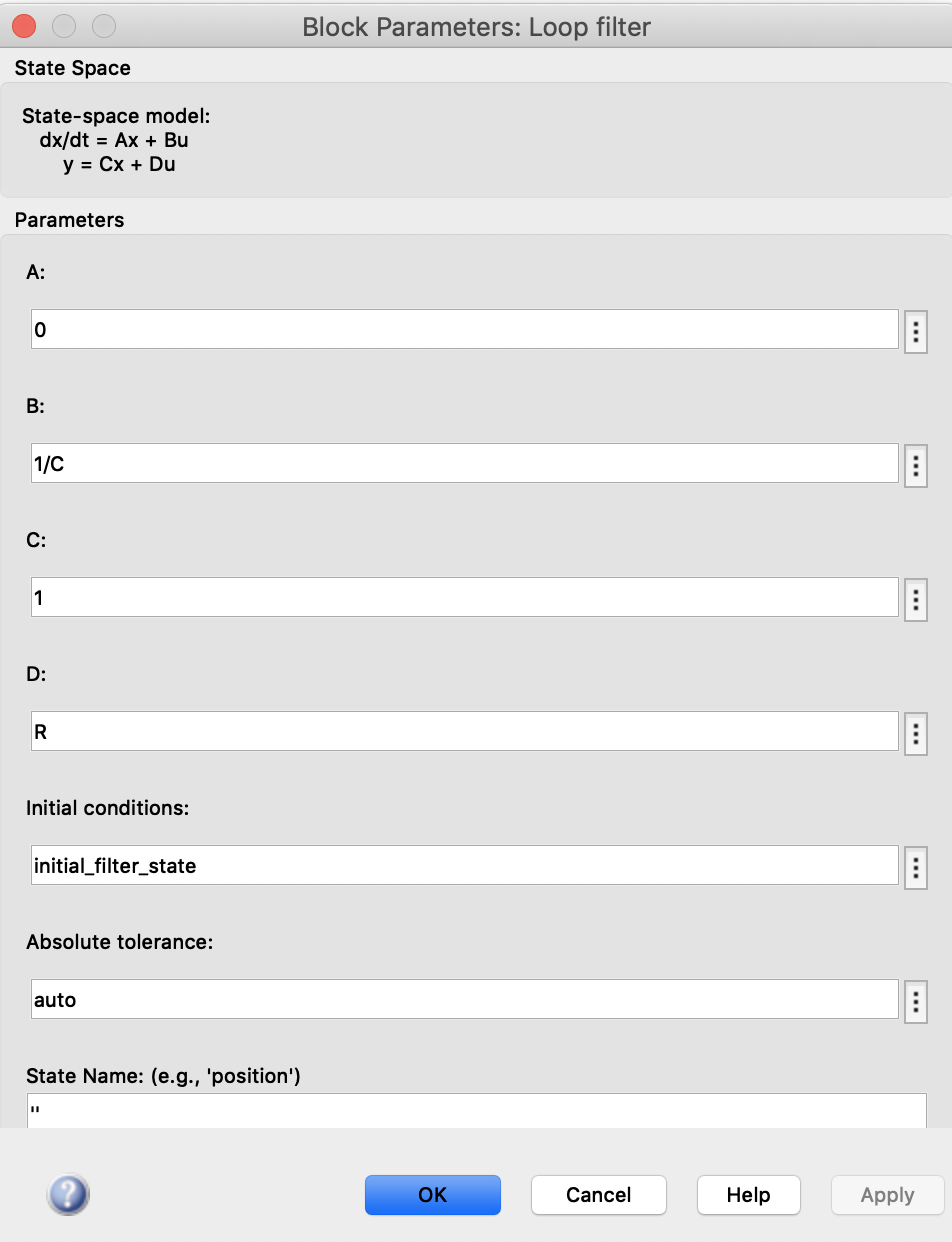}
  \caption{
  Simulink model of Loop filter 
  }
  \label{fig:Simulink-filter}
\end{figure}

\begin{lstlisting}[numbers=left, basicstyle=\tiny\ttfamily]
% parameters
omega_free = 0;
T_ref = 10^-3;
R = 1000;
C = 10^-6;
K_vco = 500;
I_p = 10^-3;

% recalculated values
tau_2N = R*C/T_ref
K_N = I_p*R*K_vco*T_ref
F_N = 1/(2*pi)*sqrt(K_N/tau_2N)
dzeta = sqrt(K_N*tau_2N)/2

% initial data
v_1 = 10;
tau_1 = T_ref*0;

% number of steps (tau_k) to simulate
max_step = 10000;

% initialize PFD output with initial data
pfd_output = zeros((max_step-1)*4,2);
pfd_output(1,:) = [0 0];
if (tau_1 >= 0)
    pfd_output(2,:) = [0 I_p];
    pfd_output(3,:) = [tau_1 I_p];
    pfd_output(4,:) = [tau_1 0];
    t_k_middle = tau_1;
    initial_vco_phase = 1 - (...
           (omega_free + K_vco*v_1)*tau_1 ...
           + K_vco*I_p/2/C*tau_1^2 ...
           - K_vco*I_p/C*tau_1^2 ...
        );        
    initial_ref_phase = 0;
else
    pfd_output(2,:) = [0 -I_p];
    pfd_output(3,:) = [-tau_1 -I_p];
    pfd_output(4,:) = [-tau_1 0];
    t_k_middle = -tau_1;
    initial_vco_phase = 0;
    initial_ref_phase = 1 + tau_1/T_ref;
end
initial_filter_state = v_1 - I_p*tau_1/C;
index = 4;


tau_v = zeros(max_step,2);
tau_v(1,:) = [tau_1 v_1];
tau_k = tau_1;
v_k = v_1;
for step = 2:(max_step - 1)  
    %check for VCO overload
    if ((tau_k > 0 ...
            && (v_k+omega_free/K_vco - I_p/C*tau_k) < 0)...
        ||...
        (tau_k < 0 ...
            && v_k+omega_free/K_vco - I_p*R < 0))       
        errordlg(sprintf('VCO overload detected at k = %0d',step-1));
        break;    
    end
    
    [tau_k1,v_k1,tau_k_zero] = righthand(tau_k,v_k ,...
                                K_vco, T_ref, I_p, C, R, omega_free);

    tau_v(step,:) = [tau_k1 v_k1];
    t_k1 = t_k_middle + tau_k_zero;
    index = index + 1;
    pfd_output(index,:) = [t_k1 0];
    
    if (tau_k1 ~= 0)
        index = index + 1;
        pfd_output(index,:) = [t_k1 I_p*sign(tau_k1)];
        
        t_k1_middle = t_k1 + abs(tau_k1);
        index = index + 1;
        pfd_output(index,:) = [t_k1_middle I_p*sign(tau_k1)];
        index = index + 1;
        pfd_output(index,:) = [t_k1_middle 0];
    end
    
    t_k = t_k1;
    t_k_middle = t_k1_middle;
    tau_k = tau_k1;
    v_k = v_k1;
end
[tau_k1,v_k1,tau_k_zero] = righthand(tau_k,v_k ,...
                                K_vco, T_ref, I_p, C, R, omega_free);
tau_v(max_step,:) = [tau_k1,v_k1];

% truncate trailing zeros
last_non_zero = find(pfd_output(:,1),1,'last');
pfd_output = pfd_output(1:last_non_zero,:);

% plot(pfd_output(:,1),pfd_output(:,2));
% ylim([-1.1*I_p 1.1*I_p]);
paemel_simulation;                            
\end{lstlisting}

\begin{lstlisting}[numbers=left, basicstyle=\tiny\ttfamily]
function [tau_k1,v_k1,tau_k_zero] = righthand( ...
    tau_k,v_k,...
    K_vco, T_ref, I_p, C, R, omega_free)
%righthandside Corrected 
if(tau_k >= 0)
    c = (T_ref - rem(tau_k,T_ref))*(omega_free+K_vco*v_k)-1;
    if (c <= 0)
%         tau(k+1) > 0, case 1) 
        a = K_vco*I_p/(2*C);
        b = omega_free + K_vco*v_k + K_vco*I_p*R;
        tau_k1 = (-b + sqrt(b^2 - 4*a*c))/(2*a);
        tau_k_zero = T_ref - rem(tau_k,T_ref);
    else
%         tau(k+1) < 0, case 2)
        tau_k1 = 1/(omega_free + K_vco*v_k) - T_ref + rem(tau_k,T_ref);
        tau_k_zero = 1/(omega_free + K_vco*v_k);
    end
else
    lk = -tau_k;
    S_lk = (K_vco*v_k-I_p*R*K_vco+omega_free)*lk+(K_vco*I_p*lk^2)/(2*C);
    S_la = rem(S_lk,1);
    S_lb = 1-S_la;
    lb = S_lb/(K_vco*v_k + omega_free);
    if lb <= T_ref
        %                 tau(k+1) < 0
        l_k1 = T_ref - lb;
        tau_k1 = -l_k1;
        tau_k_zero = lb;
    else
        %                 tau(k+1) >= 0
        S_Tref = T_ref*(K_vco*v_k +omega_free);
        c = S_la + S_Tref -1;
        b = omega_free +K_vco*v_k +K_vco*I_p*R;
        a = K_vco*I_p/(2*C);
        tau_k1 = (-b+sqrt(b^2-4*a*c))/(2*a);
        tau_k_zero = T_ref;
    end
end
v_k1 = v_k + tau_k1*I_p/C;
end
\end{lstlisting}

\begin{lstlisting}[numbers=left, basicstyle=\tiny\ttfamily]
% check that omega_free is zero
if (omega_free > 0)
    errordlg('Please set omega_free to zero');
end

% initialize PFD output with initial data
paemel_output = zeros((max_step-1)*4,2);
paemel_output(1,:) = [0 0];
if (tau_1 >= 0)
    paemel_output(2,:) = [0 I_p];
    paemel_output(3,:) = [tau_1 I_p];
    paemel_output(4,:) = [tau_1 0];
    t_k_middle = tau_1;
    initial_vco_phase = 1 - (...
           (K_vco*v_1)*tau_1 ...
           + K_vco*I_p/2/C*tau_1^2 ...
           - K_vco*I_p/C*tau_1^2 ...
        );        
    initial_ref_phase = 0;
else
    paemel_output(2,:) = [0 -I_p];
    paemel_output(3,:) = [-tau_1 -I_p];
    paemel_output(4,:) = [-tau_1 0];
    t_k_middle = -tau_1;
    initial_vco_phase = 0;
    initial_ref_phase = 1 + tau_1/T_ref;
end
initial_filter_state = v_1 - I_p*tau_1/C;
index = 4;


tau_v_paemel = zeros(max_step,2);
tau_v_paemel(1,:) = [tau_1 v_1];
tau_k = tau_1;
v_k = v_1;
for step = 2:(max_step - 1)  
    [tau_k1,v_k1,tau_k_zero] = paemel_righthand(tau_k,v_k ,...
                                K_vco, T_ref, I_p, C, R);

    % Compute case 6
    if (tau_k1 < -T_ref)
        if (step <= 2)
            errordlg('ERROR in Paemel algo case 6), no v(-1)');
            paemel_output = zeros((max_step-1)*4,2);
            break;
        end        
        v_n_1 = tau_v_paemel(step-2,2);         
        t_sum = 0;
        while (t_sum < abs(tau_k))
            t_n = (v_n_1 - I_p*R - sqrt((v_n_1 - I_p*R)^2 - 2*I_p/C/K_vco)...
              )/(I_p/C);
            t_sum = t_sum + t_n;
            v_n_prev = v_n_1;
            v_n_1 = v_n_1 - I_p/C*t_n;            
        end 
        t_sum = t_sum - t_n;
        t_a = -tau_k - t_sum;
        t_b = (1/K_vco - t_a*(v_n_prev - I_p*R) + I_p/C*t_a^2/2)/v_k;
        tau_k1 = t_b - T_ref;
        
        v_k1 = v_k + tau_k1*I_p/C;
        tau_k_zero = t_b;
    end                            
                            
    tau_v_paemel(step,:) = [tau_k1 v_k1];
    t_k1 = t_k_middle + tau_k_zero;
    index = index + 1;
    paemel_output(index,:) = [t_k1 0];
    
    if (tau_k1 ~= 0)
        index = index + 1;
        paemel_output(index,:) = [t_k1 I_p*sign(tau_k1)];
        
        t_k1_middle = t_k1 + abs(tau_k1);
        index = index + 1;
        paemel_output(index,:) = [t_k1_middle I_p*sign(tau_k1)];
        index = index + 1;
        paemel_output(index,:) = [t_k1_middle 0];
    end
    
    % case 5
    if (tau_k1 >= T_ref)
        tau_k1 = rem(tau_k1,T_ref);
    end
    
    t_k = t_k1;
    t_k_middle = t_k1_middle;
    tau_k = tau_k1;
    v_k = v_k1;
end
[tau_k1,v_k1,tau_k_zero] = righthand(tau_k,v_k ,...
                                K_vco, T_ref, I_p, C, R, omega_free);
tau_v_paemel(max_step,:) = [tau_k1,v_k1];

% truncate trailing zeros
last_non_zero = find(paemel_output(:,1),1,'last');
if (last_non_zero > 1)
    paemel_output = paemel_output(1:last_non_zero,:);
end
% plot(pfd_output(:,1),pfd_output(:,2));
% ylim([-1.1*I_p 1.1*I_p]);

\end{lstlisting}

\begin{lstlisting}[numbers=left, basicstyle=\tiny\ttfamily]
function [tau_k1,v_k1,tau_k_zero] = paemel_righthand( ...
    tau_k,v_k,...
    K_vco, T_ref, I_p, C, R)
%righthandside Paemel 
if(tau_k >= 0)
    % case 1
    a = I_p/(2*C);
    b = v_k + I_p*R;
    c = (T_ref - tau_k)*v_k-1/K_vco;
    tau_k1 = (-b + sqrt(b^2 - 4*a*c))/(2*a);
    tau_k_zero = T_ref - tau_k;
    if (tau_k1 < 0) 
        % case 3 Paemel
        tau_k1 = 1/(K_vco*v_k) - T_ref + tau_k;
        tau_k_zero = 1/(K_vco*v_k);
    end
else
    % case 2 Paemel
    tau_k1 = (...
        1/K_vco  - I_p*R*tau_k - I_p*tau_k^2/(2*C)...
        )/v_k - T_ref + tau_k;
    
    lk = -tau_k;
    S_lk = (K_vco*v_k-I_p*R*K_vco)*lk+(K_vco*I_p*lk^2)/(2*C);
    tau_k_zero = (1-rem(S_lk,1))/(K_vco*v_k);
    
    if( tau_k1 > 0)
        % case 4 Paemel
        tau_k1 = (...
            -I_p*R -v_k + sqrt((I_p*R + v_k)^2 + (2*I_p/C)*v_k*tau_k1) ...
            )/(I_p/C);
            
        tau_k_zero = T_ref;
    end
end
v_k1 = v_k + tau_k1*I_p/C;
end
\end{lstlisting}

\section*{Appendix: Matlab code for VCO overload model}

\begin{lstlisting}[numbers=left, basicstyle=\tiny\ttfamily]
% parameters
omega_free = 0;
T_ref = 10^-3;
R = 1000;
C = 10^-6;
K_vco = 1000;
I_p = 10^-3;

% recalculated values
tau_2N = R*C/T_ref
K_N = I_p*R*K_vco*T_ref
F_N = 1/(2*pi)*sqrt(K_N/tau_2N)
dzeta = sqrt(K_N*tau_2N)/2

% initial data
v_1 = 4;
tau_1 = -0.2*T_ref;
if (tau_1 < -T_ref)
    errordlg('Impossible tau_1. Can not be lower than -T_ref.');
end

% number of steps (tau_k) to simulate
max_step = 10000;

% initialize PFD output with initial data
pfd_output = zeros((max_step-1)*4,2);
pfd_output(1,:) = [0 0];
if (tau_1 >= 0)
    pfd_output(2,:) = [0 I_p];
    pfd_output(3,:) = [tau_1 I_p];
    pfd_output(4,:) = [tau_1 0];
    t_k_middle = tau_1;
    initial_vco_phase = 1 - (...
           (omega_free + K_vco*v_1 + K_vco*I_p*R - K_vco*tau_1*I_p/C)*tau_1 ...
           + K_vco*I_p/2/C*tau_1^2 ...
        );        
    % overload
    if (-K_vco*I_p/C*tau_1 + K_vco*v_1 < 0)
        if (-K_vco*I_p/C*tau_1 + K_vco*v_1 +K_vco*I_p*R < 0)
           if (omega_free + K_vco*v_1 + K_vco*I_p*R < 0)
                errordlg('Impossible initial condition. v_1 is too small.');
           end
           initial_vco_phase = 1 - (...
                (omega_free + K_vco*v_1 + K_vco*I_p*R)^2/2/(K_vco*I_p/C) ...
            );
        else
           initial_vco_phase = 1 - (...
                tau_1^2*K_vco*I_p/2/C+ ...
                tau_1*(K_vco*I_p*R + K_vco*v_1 + omega_free - tau_1*K_vco*I_p/C)...
            );
        end
    end
    if (initial_vco_phase < 0)
        errordlg('Impossible initial condition. v_1 or tau_1 is too big.');
    end
    initial_ref_phase = 0;
else
    pfd_output(2,:) = [0 -I_p];
    pfd_output(3,:) = [-tau_1 -I_p];
    pfd_output(4,:) = [-tau_1 0];
    t_k_middle = -tau_1;
    initial_vco_phase = 0;
    initial_ref_phase = 1 + tau_1/T_ref;
end
initial_filter_state = v_1 - I_p*tau_1/C;
index = 4;


tau_v = zeros(max_step,2);
tau_v(1,:) = [tau_1 v_1];
tau_k = tau_1;
v_k = v_1;
for step = 2:(max_step - 1)  
    [tau_k1,v_k1,tau_k_zero] = righthand(tau_k,v_k ,...
                                K_vco, T_ref, I_p, C, R, omega_free);

    %check for VCO overload
    if ((tau_k > 0 ...
            && (v_k+omega_free/K_vco - I_p/C*tau_k) < 0)...
        ||...
        (tau_k < 0 ...
            && v_k+omega_free/K_vco - I_p*R < 0))       
        [tau_k_o,v_k_o,tau_k_zero] = righthand_overload(tau_k,v_k ,...
                                             tau_k1,v_k1,...
                                K_vco, T_ref, I_p, C, R, omega_free);
        tau_k1 = tau_k_o;
        v_k1 = v_k_o;
    end
                            
    tau_v(step,:) = [tau_k1 v_k1];
    t_k1 = t_k_middle + tau_k_zero;
    index = index + 1;
    pfd_output(index,:) = [t_k1 0];
    
    if (tau_k1 ~= 0)
        index = index + 1;
        pfd_output(index,:) = [t_k1 I_p*sign(tau_k1)];
        
        t_k1_middle = t_k1 + abs(tau_k1);
        index = index + 1;
        pfd_output(index,:) = [t_k1_middle I_p*sign(tau_k1)];
        index = index + 1;
        pfd_output(index,:) = [t_k1_middle 0];
    end
    
    t_k = t_k1;
    t_k_middle = t_k1_middle;
    tau_k = tau_k1;
    v_k = v_k1;
end
[tau_k1,v_k1,tau_k_zero] = righthand(tau_k,v_k ,...
                                K_vco, T_ref, I_p, C, R, omega_free);
tau_v(max_step,:) = [tau_k1,v_k1];

% truncate trailing zeros
last_non_zero = find(pfd_output(:,1),1,'last');
pfd_output = pfd_output(1:last_non_zero,:);

plot(pfd_output(:,1), pfd_output(:,2)/I_p);
\end{lstlisting}

\begin{lstlisting}[numbers=left, basicstyle=\tiny\ttfamily]
function [tau_k1,v_k1,tau_k_zero] = righthand( ...
    tau_k,v_k,...
    K_vco, T_ref, I_p, C, R, omega_free)
%righthandside Corrected 
if(tau_k >= 0)
    c = (T_ref - rem(tau_k,T_ref))*(omega_free+K_vco*v_k)-1;
    if (c <= 0)
%         tau(k+1) > 0, case 1) 
        a = K_vco*I_p/(2*C);
        b = omega_free + K_vco*v_k + K_vco*I_p*R;
        tau_k1 = (-b + sqrt(b^2 - 4*a*c))/(2*a);
        tau_k_zero = T_ref - rem(tau_k,T_ref);
    else
%         tau(k+1) < 0, case 2)
        tau_k1 = 1/(omega_free + K_vco*v_k) - T_ref + rem(tau_k,T_ref);
        tau_k_zero = 1/(omega_free + K_vco*v_k);
    end
else
    lk = -tau_k;
    S_lk = (K_vco*v_k-I_p*R*K_vco+omega_free)*lk+(K_vco*I_p*lk^2)/(2*C);
    S_la = rem(S_lk,1);
    S_lb = 1-S_la;
    lb = S_lb/(K_vco*v_k + omega_free);
    if lb <= T_ref
        %                 tau(k+1) < 0
        l_k1 = T_ref - lb;
        tau_k1 = -l_k1;
        tau_k_zero = lb;
    else
        %                 tau(k+1) >= 0
        S_Tref = T_ref*(K_vco*v_k +omega_free);
        c = S_la + S_Tref -1;
        b = omega_free +K_vco*v_k +K_vco*I_p*R;
        a = K_vco*I_p/(2*C);
        tau_k1 = (-b+sqrt(b^2-4*a*c))/(2*a);
        tau_k_zero = T_ref;
    end
end
v_k1 = v_k + tau_k1*I_p/C;
end
\end{lstlisting}

\begin{lstlisting}[numbers=left, basicstyle=\tiny\ttfamily]
function [tau_k_o,v_k_o,tau_k_zero] = righthand_overload( ...
    tau_k,v_k,...
    tau_k_1,v_k_1,...
    K_vco, T_ref, I_p, C, R, omega_free)
%righthandside overload 
root2 = @(a,b,c) ((-b + sqrt(b^2 - 4*a*c))/2/a);

omega_vco = omega_free + K_vco*v_k;
a = K_vco*I_p/2/C;
b = omega_vco + K_vco*I_p*R;

if(tau_k < 0)
    l_x = min([-C/I_p*(v_k + omega_free/K_vco - I_p*R) -tau_k]);
    S = K_vco*(tau_k + l_x)^2*I_p/2/C;
    Sla = rem(S,1);
    if (omega_vco > 0)        
        if (tau_k_1 < 0)
            % O1            
            l_b = (1-Sla)/omega_vco;
            tau_k_o = -(T_ref - l_b);
            
            tau_k_zero = l_b;
        else
            % O2
            Sref = T_ref*omega_vco;
            d = Sla + Sref - 1;
            tau_k_o = root2(a,b,d);
            
            tau_k_zero = T_ref;
        end
    else
        if (v_k + omega_free/K_vco + I_p*R < 0)
            % O3
            l_b_0 = C/I_p*(-omega_free/K_vco - I_p*R - v_k);
            l_b_plus = sqrt((1 - Sla)*2*C/K_vco/I_p);
            tau_k_o = l_b_0 + l_b_plus;
            
            tau_k_zero = T_ref;
        else
            % O4            
            d_0 = Sla - 1;            
            tau_k_o = root2(a,b,d_0);
            
            tau_k_zero = T_ref;
        end
    end
else
    if (omega_vco <= 0)
        if (tau_k_1 > 0)
            % O5            
            tau_k_o = root2(a,b,-1);
            
            tau_k_zero = T_ref - rem(tau_k,T_ref);
        end
    else
        if (tau_k_1 > 0)
            % O6 equal to case 1) no overload
            c = (T_ref - rem(tau_k,T_ref))*(omega_free+K_vco*v_k)-1;
            tau_k_o = root2(a,b,c);            
            
            tau_k_zero = T_ref - rem(tau_k,T_ref);
        else
            % O7 equal to case 2) no overload
            tau_k_o = 1/(omega_free + K_vco*v_k) - T_ref + rem(tau_k,T_ref);
        
            tau_k_zero = T_ref - rem(tau_k,T_ref) + tau_k_o;
        end
    end
end
v_k_o = v_k + tau_k_o*I_p/C;
end
\end{lstlisting}

\section*{Appendix: Additional pics}
\begin{figure*}
  \begin{subfigure}[b]{0.49\textwidth}
    \includegraphics[width=\linewidth]{case1.pdf}
    \caption{Case 1\_1: general case, Reference cycle slipping; $\tau_k \geq 0$, $\tau_{k+1} \geq 0$}.
    \label{fig case1_0}
  \end{subfigure}
  \begin{subfigure}[b]{0.49\textwidth}
        \includegraphics[width=\linewidth]{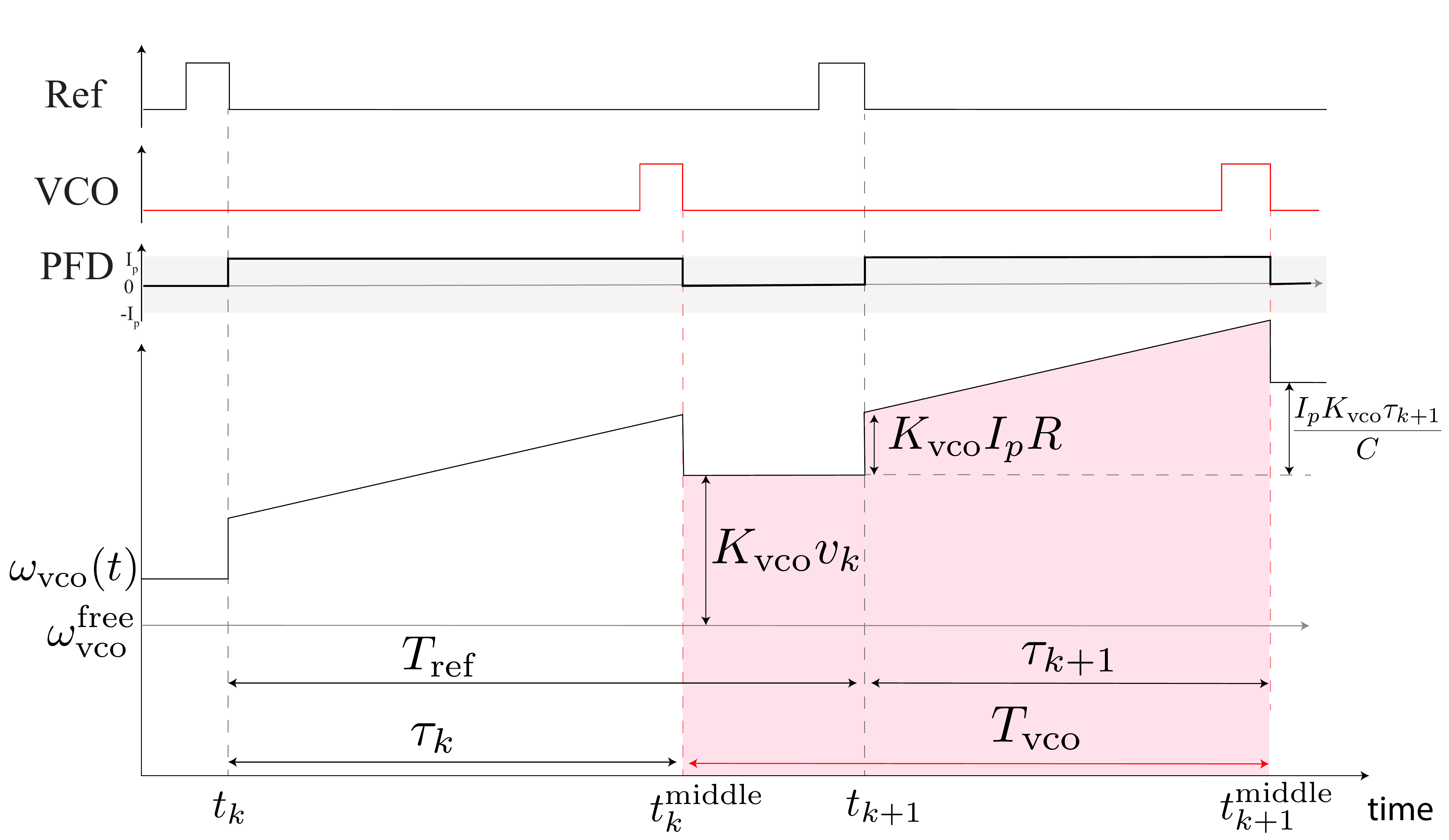}
        \caption{Case 1\_4: all VCO and Reference trailing edges happen at different time instances; $\tau_k > 0$, $\tau_{k+1} > 0$}
    \label{fig case1_3}
  \end{subfigure}
  \begin{subfigure}[b]{0.49\textwidth}
    \includegraphics[width=\linewidth]{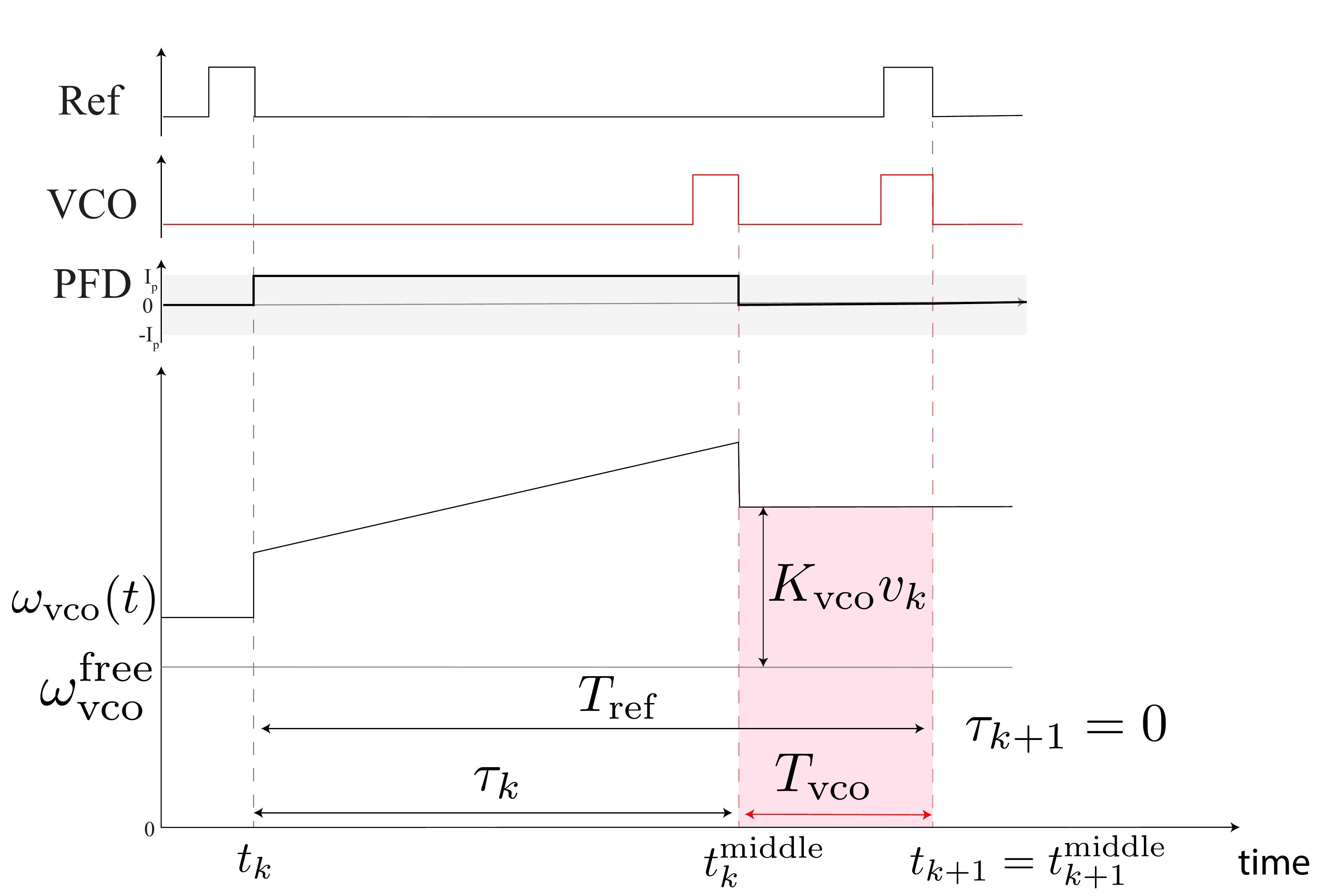}
    \caption{Case 1\_2: VCO and Reference trailing edges both happen at the same time instance $t_{k+1}$; $\tau_k > 0$, $\tau_{k+1} = 0$}
    \label{fig case1_1}
  \end{subfigure}
  \begin{subfigure}[b]{0.49\textwidth}
    \includegraphics[width=\linewidth]{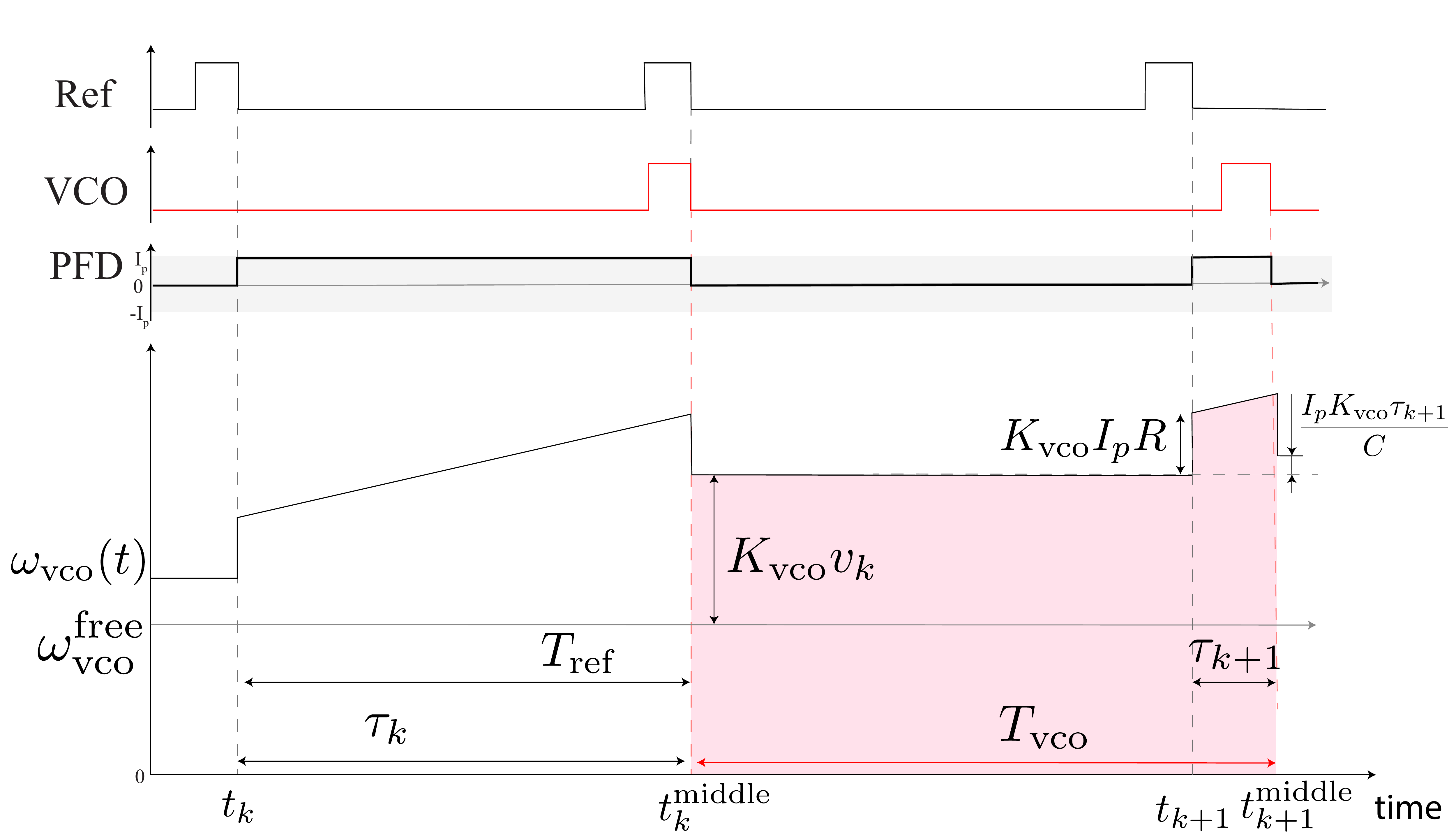}
    \caption{Case 1\_3: VCO and Reference trailing edges both happen at the same time instance $t_k^{\rm middle}$; $\tau_k \text{ mod } T_{\rm ref} = 0$, $\tau_{k+1} > 0$}
    \label{fig case1_2}
  \end{subfigure}
  \centering \begin{subfigure}[b]{0.49\textwidth}
    \includegraphics[width=\linewidth]{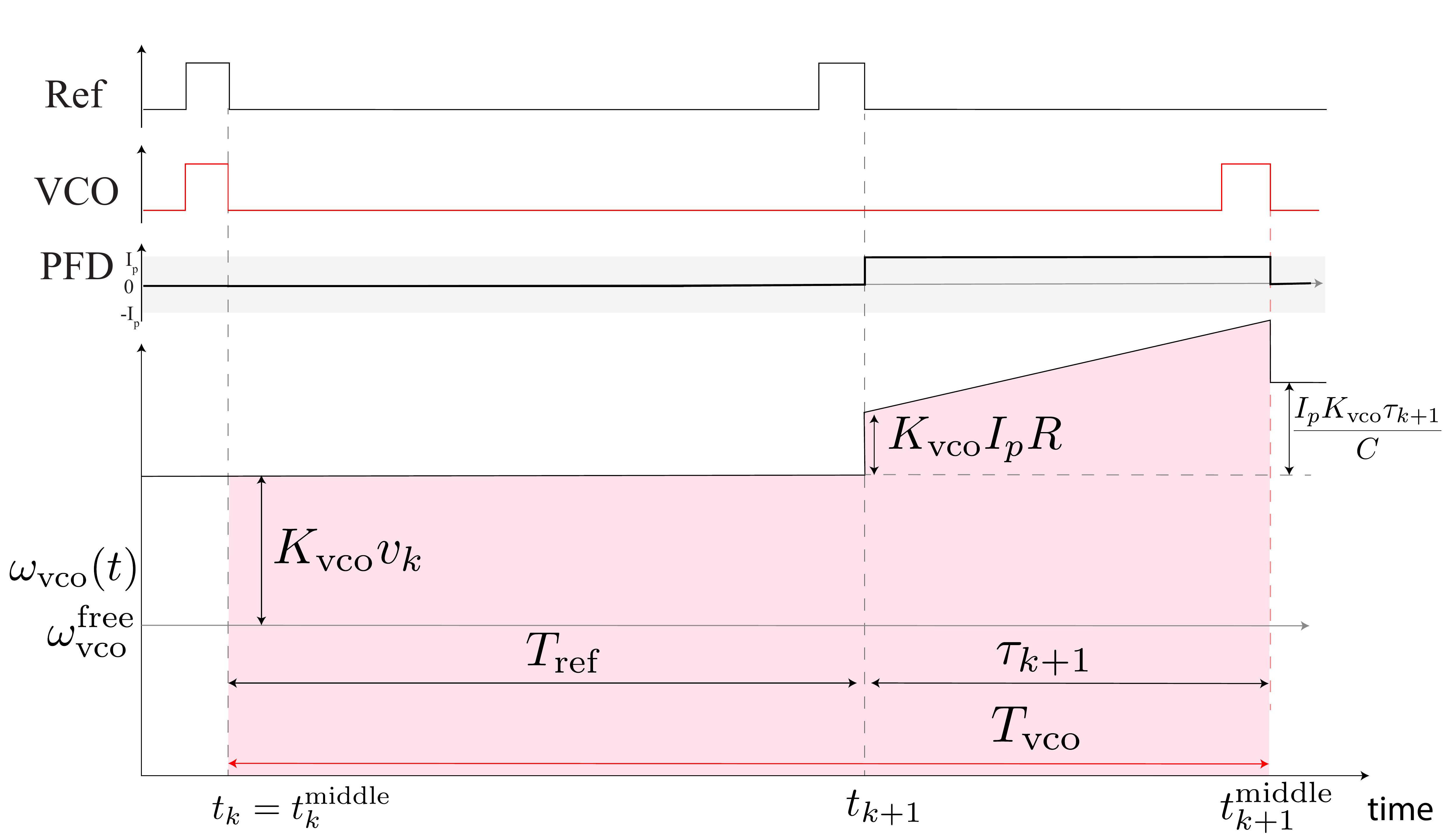}
    \caption{Case 1\_5: VCO and Reference trailing edges both happen at the same time instance $t_k$; $\tau_k = 0$, $\tau_{k+1} > 0$}
    \label{fig case1_4}
  \end{subfigure}
  \caption{
   Subcases of the Case 1. Integral of the VCO frequency $\omega_{\rm vco}$ over
   the VCO period $T_{\rm vco}$
   is pink subgraph area (grey in black/white).
  The integral is equal to $1$ according
  to the PFD switching law and definition of time intervals.
  }
  \label{fig:case1:1-5}
\end{figure*}
\begin{figure*}
  \begin{subfigure}[b]{0.49\textwidth}
    \includegraphics[width=\linewidth]{case2.pdf}
    \caption{Case 2\_1: general case, Reference and VCO cycle slipping; $\tau_k \geq 0$, $\tau_{k+1} < 0$}
    \label{fig case2_1}
  \end{subfigure}
  \begin{subfigure}[b]{0.49\textwidth}
    \includegraphics[width=\linewidth]{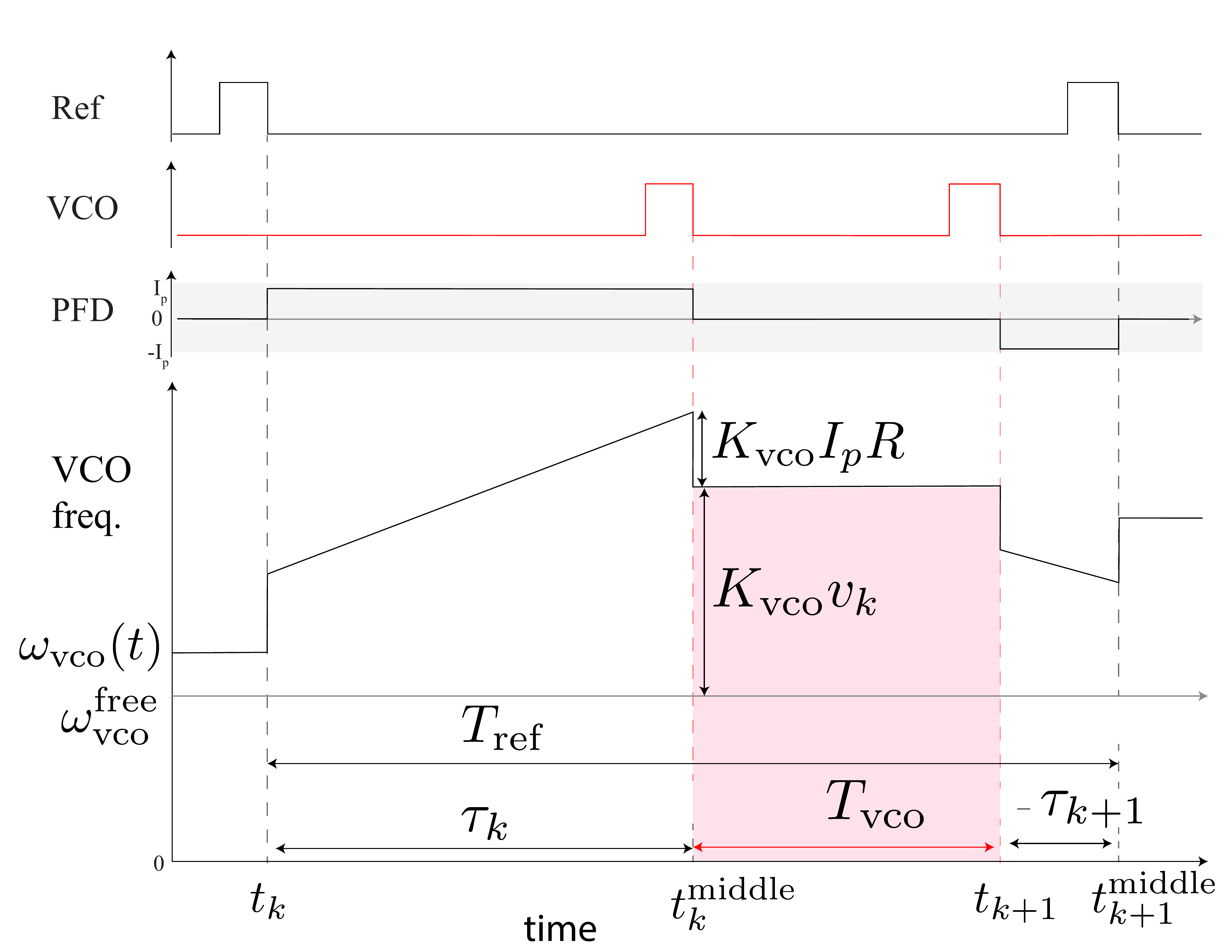}
    \caption{Case 2\_2: all VCO and Reference trailing edges happen at different time instances; $\tau_k > 0$, $\tau_{k+1} < 0$}
    \label{fig case2_2}
  \end{subfigure}
  \begin{subfigure}[b]{0.49\textwidth}
    \includegraphics[width=\linewidth]{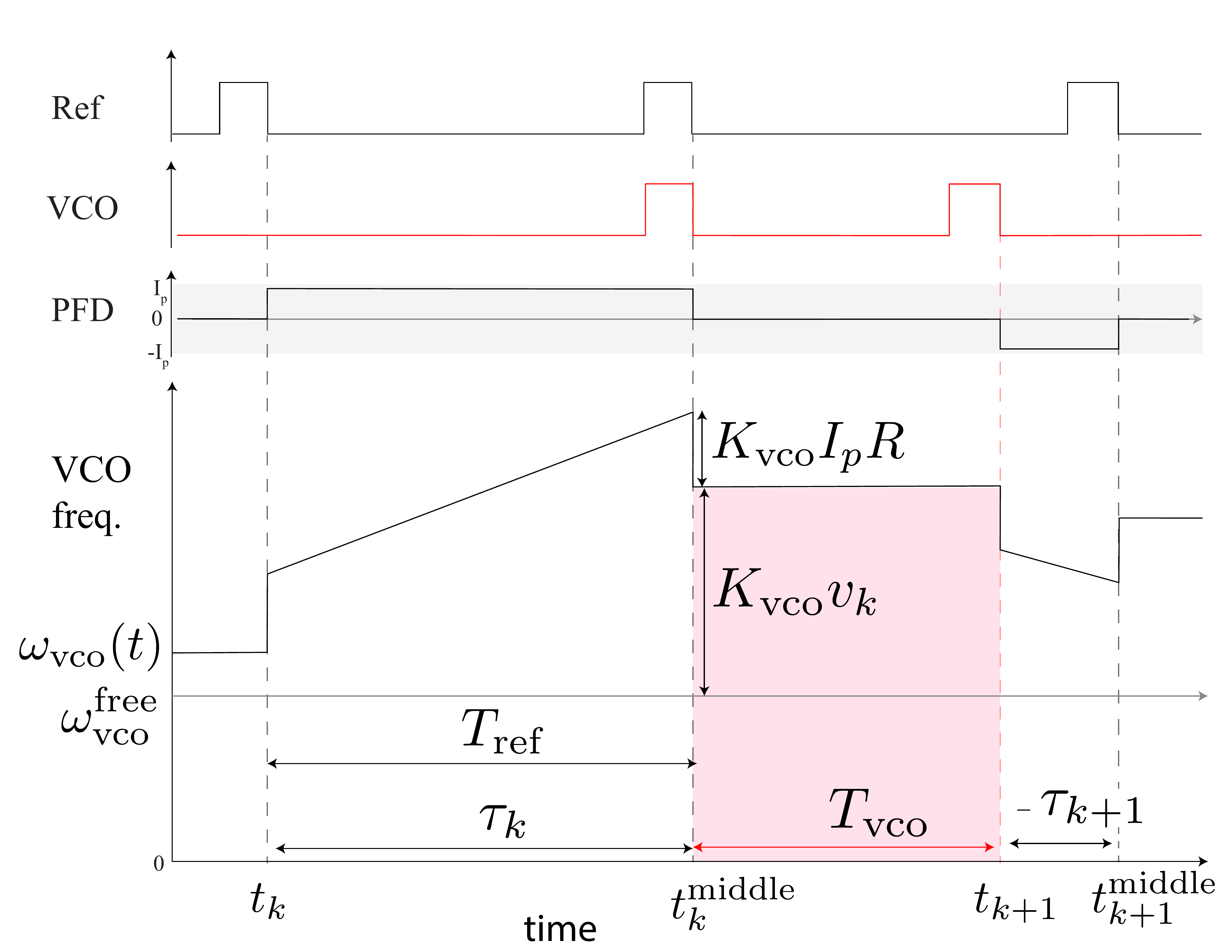}
    \caption{Case 2\_3: VCO and Reference trailing edges both happen at the same time instance $t_k^{\rm middle}$; $\tau_k \text{ mod } T_{\rm ref} = 0$, $\tau_{k+1} < 0$}
    \label{fig case2_3}
  \end{subfigure}
  \begin{subfigure}[b]{0.49\textwidth}
    \includegraphics[width=\linewidth]{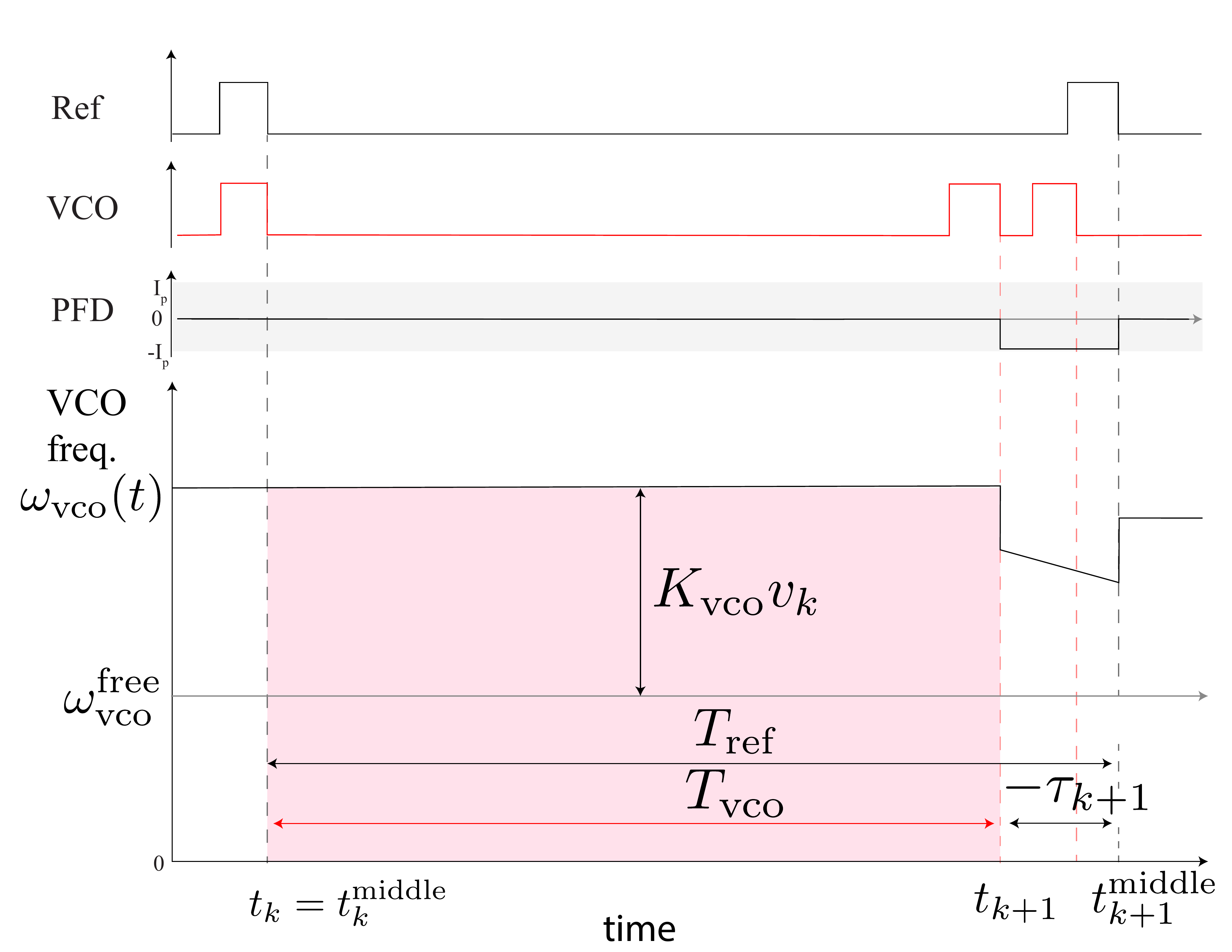}
    \caption{Case 2\_4: VCO and Reference trailing edges both happen at the same time instance $t_k$; $\tau_k = 0$, $\tau_{k+1} < 0$}
    \label{fig case2_4}
  \end{subfigure}
  \caption{
   Subcases of the Case 2.
   Integral of the VCO frequency $\omega_{\rm vco}$ over
   the VCO period $T_{\rm vco}$
   is pink subgraph area (grey in black/white).
   The integral is equal to $1$ according
   to the PFD switching law and definition of time intervals.
  }
  \label{fig:case2:1-4}
\end{figure*}

\begin{figure*}
  \begin{subfigure}[b]{0.49\textwidth}
  \centering
    \includegraphics[width=\linewidth]{case3.pdf}
    \caption{Case 3\_1: general case, VCO cycle slipping; $\tau_k < 0, \quad \tau_{k+1} < 0$}
    \label{fig case3_1}
  \end{subfigure}
  \begin{subfigure}[b]{0.49\textwidth}
  \centering
    \includegraphics[width=\linewidth]{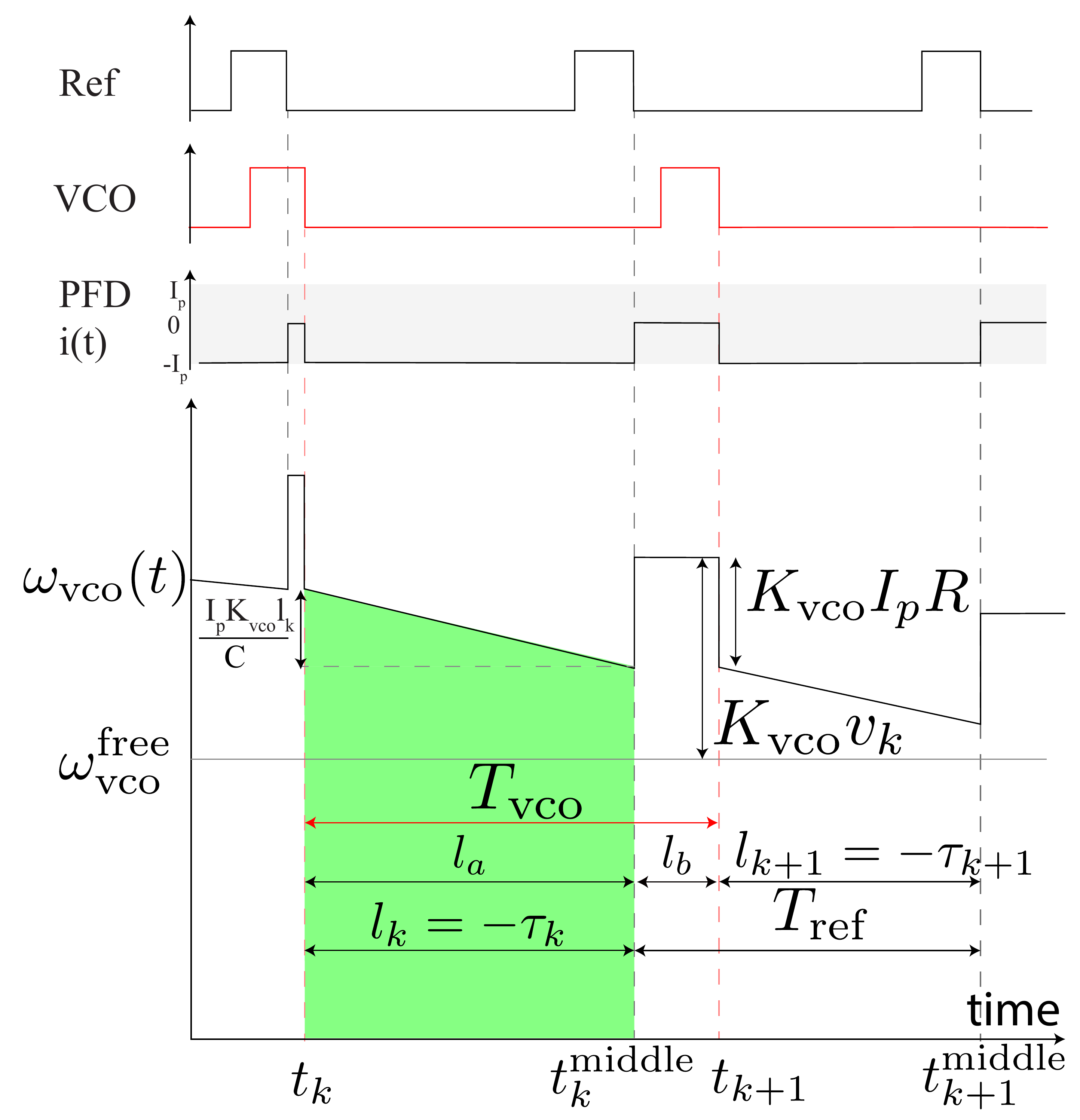}
    \caption{Case 3\_2: all VCO and Reference trailing edges happen at different time instances; $\tau_k < 0, \quad \tau_{k+1} < 0$}
    \label{fig case3_2}
  \end{subfigure}
  \begin{subfigure}[b]{0.49\textwidth}
  \centering
    \includegraphics[width=\linewidth]{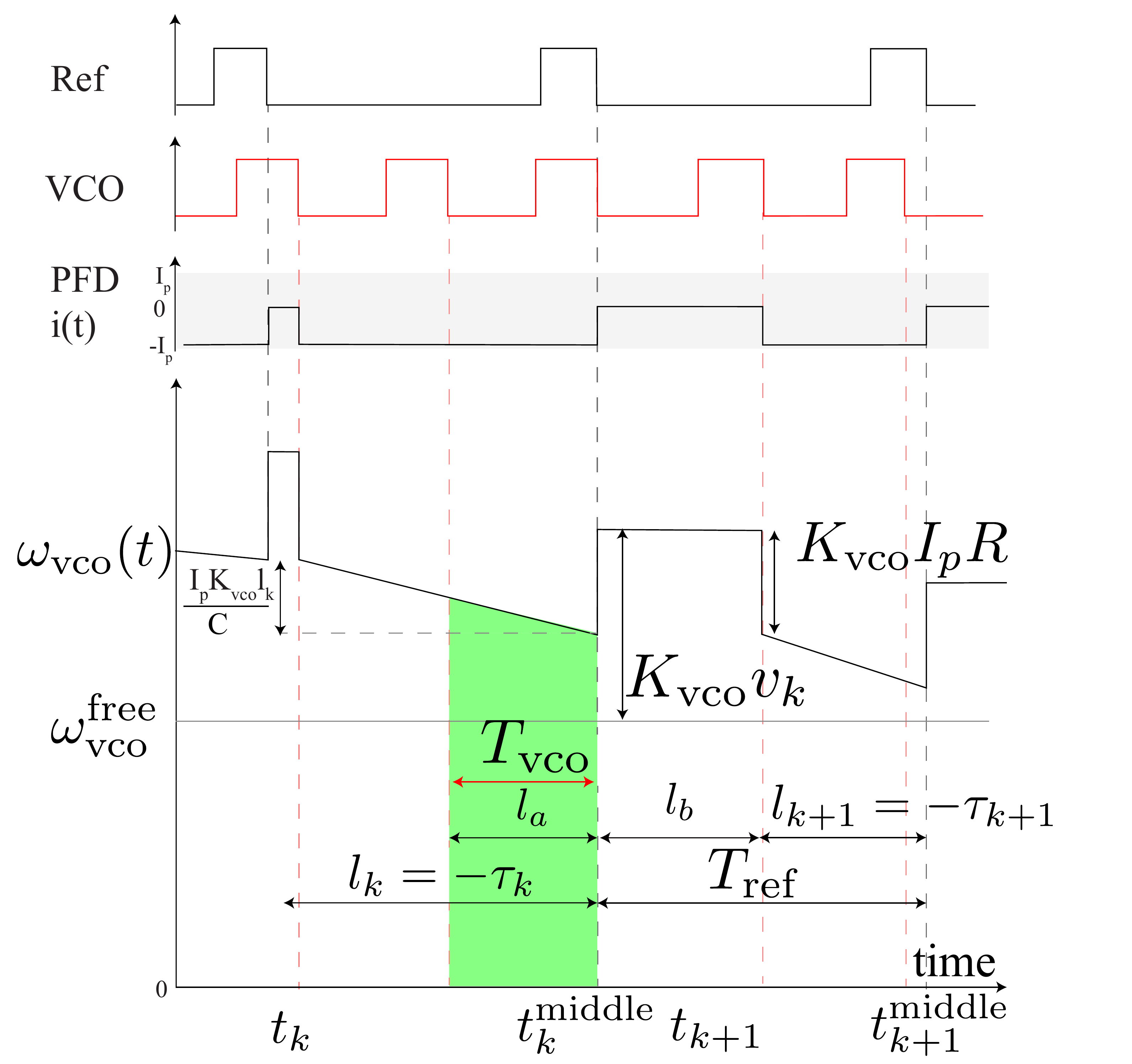}
    \caption{Case 3\_3: VCO and Reference trailing edges both happen at the same time instance $t_k^{\rm middle}$; $\tau_k < 0, \quad \tau_{k+1} < 0$}
    \label{fig case3_2}
  \end{subfigure}
  \begin{subfigure}[b]{0.49\textwidth}
  \centering
    \includegraphics[width=\linewidth]{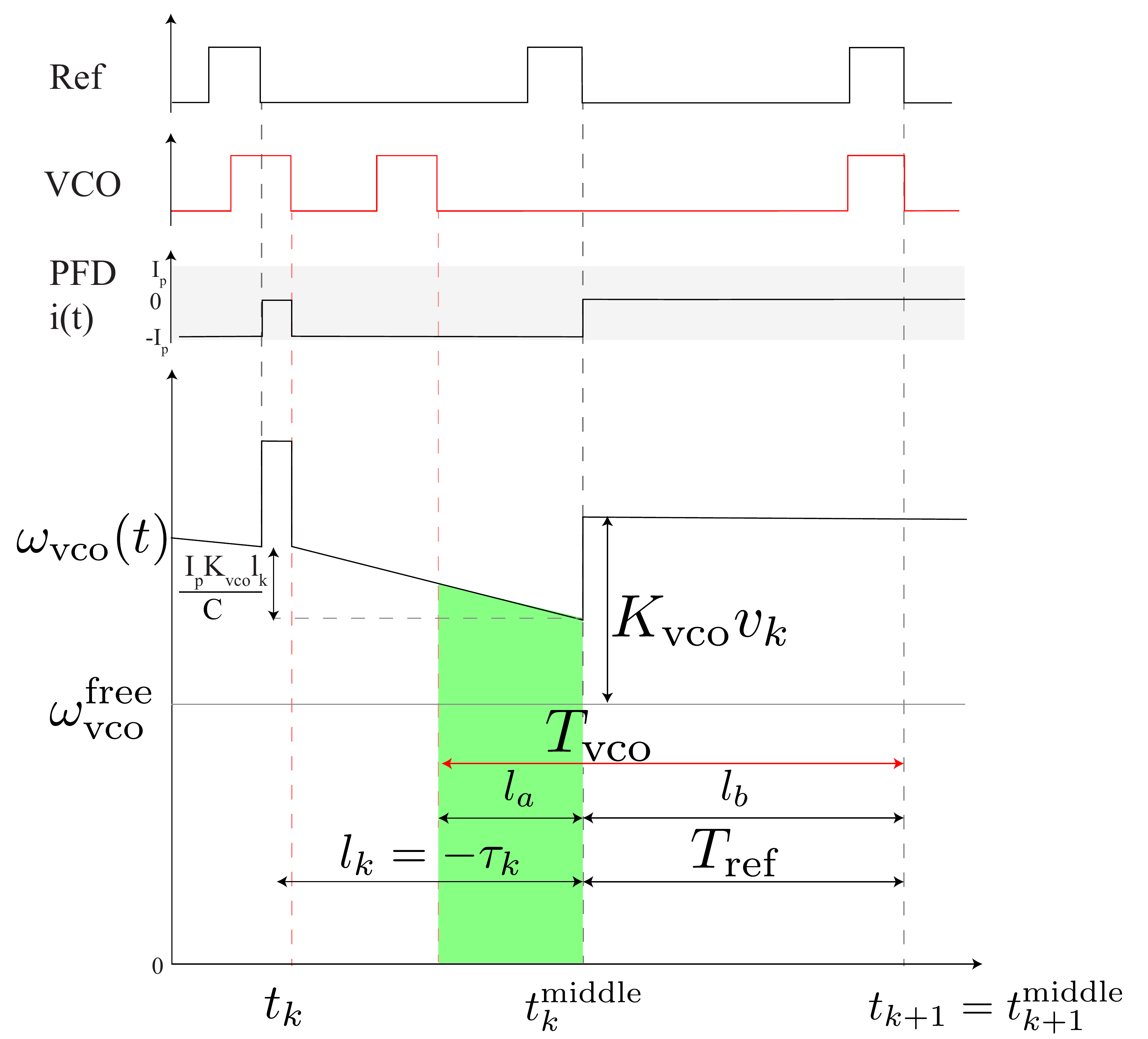}
    \caption{Case 3\_4: VCO and Reference trailing edges both happen at the same time instance $t_{k+1}$; $\tau_k < 0, \quad \tau_{k+1} = 0$}
    \label{fig case3_3}
  \end{subfigure}
  \caption{
   Subcases of the Case 3.
   Integral of the VCO frequency $\omega_{\rm vco}$ over
   the VCO period $T_{\rm vco}$
   is pink subgraph area (grey in black/white).
   The integral is equal to $1$ according
   to the PFD switching law and definition of time intervals.
  }
  \label{fig case3}
\end{figure*}

\begin{figure*}
  \begin{subfigure}[b]{0.49\textwidth}
  \centering
    \includegraphics[width=\linewidth]{case4.pdf}
    \caption{Case 4\_1: general case, VCO and Reference cycle slipping; $\tau_k < 0, \quad \tau_{k+1} > 0$}
    \label{fig case4_1}
  \end{subfigure}
  \begin{subfigure}[b]{0.49\textwidth}
  \centering
    \includegraphics[width=\linewidth]{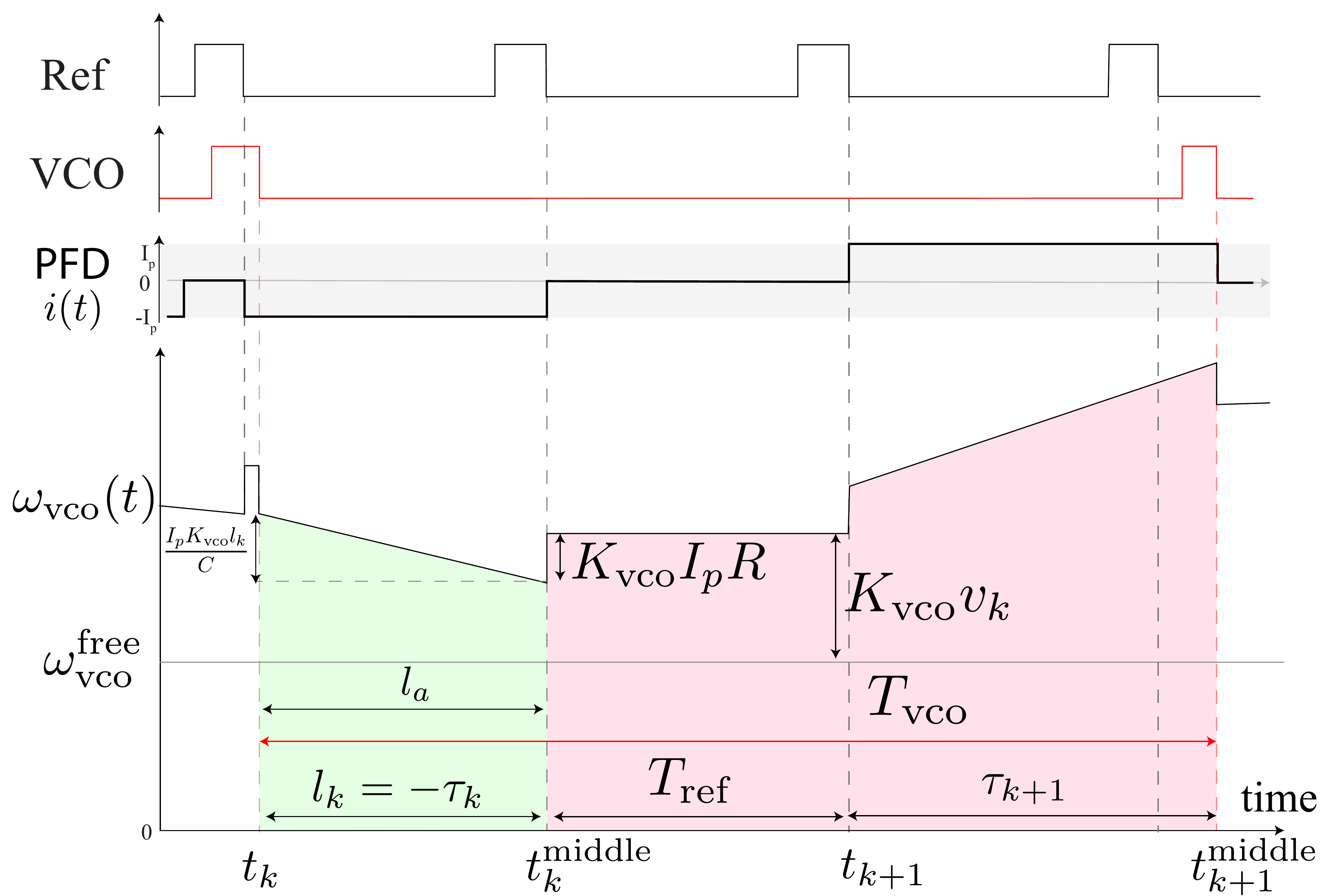}
    \caption{Case 4\_2: Reference cycle slipping; Refer all VCO and Reference trailing edges happen at different time instances;  $\tau_k < 0, \quad \tau_{k+1} > 0$}
    \label{fig case4_2}
  \end{subfigure}
  \begin{subfigure}[b]{0.49\textwidth}
  \centering
    \includegraphics[width=\linewidth]{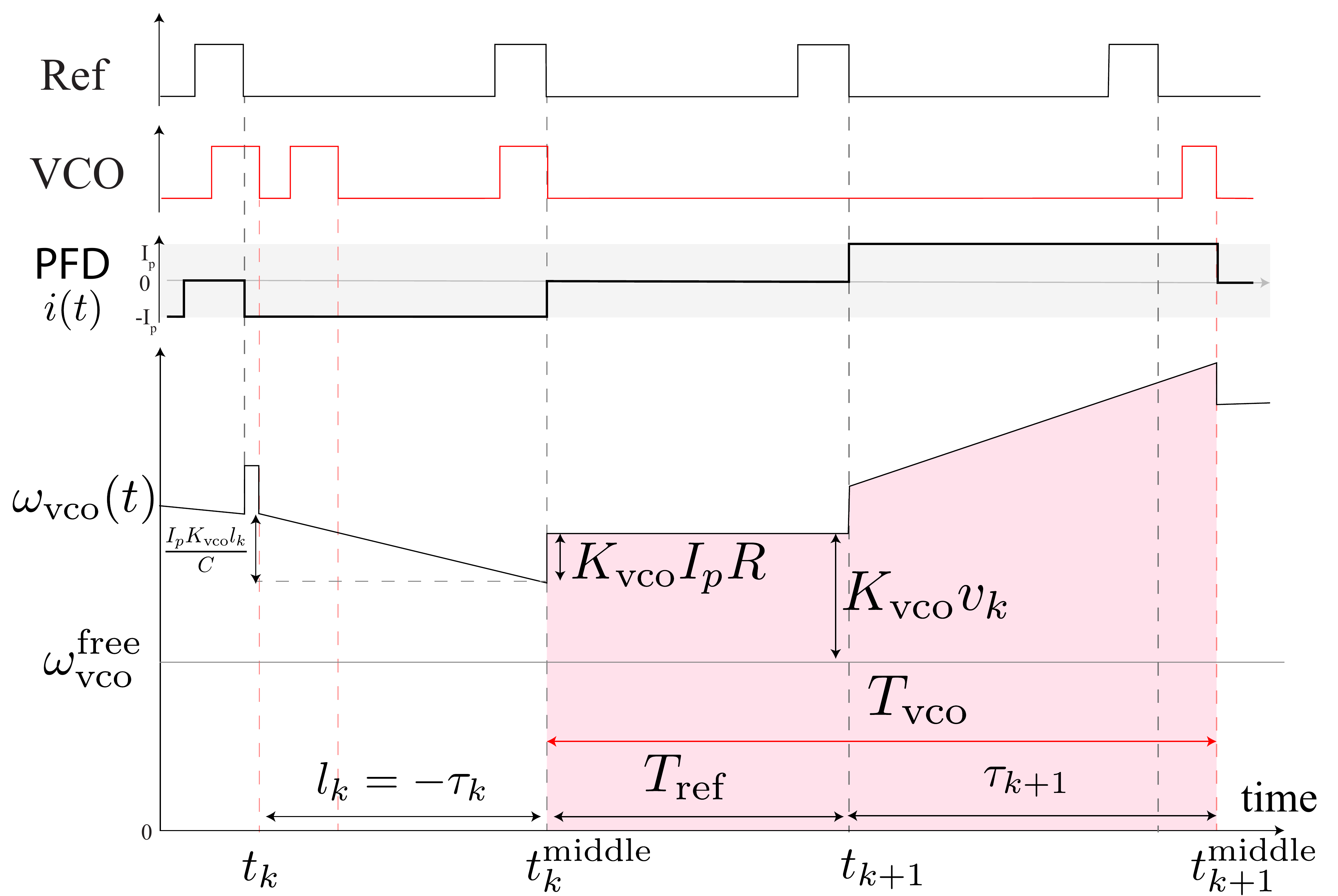}
    \caption{Case 4\_3: VCO and Reference trailing edges both happen at the same time instance $t_k^{\rm middle}$;  $\tau_k < 0, \quad \tau_{k+1} > 0$}
    \label{fig case4_3}
  \end{subfigure}
  \caption{
   Subcases of the Case 4.
   Integral of the VCO frequency $\omega_{\rm vco}$ over
   the VCO period $T_{\rm vco}$
   is pink subgraph area (grey in black/white).
   The integral is equal to $1$ according
   to the PFD switching law and definition of time intervals.
   }
  \label{fig:case4:1-3}
\end{figure*}

\end{document}